\documentclass[12pt,a5paper]{article}

\DeclareUnicodeCharacter{223C}{\textasciitilde}
\DeclareUnicodeCharacter{2009}{\,}
\usepackage{longtable}

\usepackage{textalpha}
\usepackage[T1]{fontenc}
\usepackage[utf8]{inputenc}
\usepackage[english]{babel}
\usepackage{newunicodechar}
\newunicodechar{−}{\ensuremath{-}}
\newunicodechar{≈}{\ensuremath{\approx}}
\usepackage{setspace}
\usepackage{hyperref}
\hypersetup{colorlinks=true, citecolor = magenta, linkcolor=teal, filecolor=pink, urlcolor=blue}
\usepackage{graphicx}
\usepackage{caption} 
\usepackage{subcaption}
\usepackage{chronosys}
\usepackage[strings]{underscore}
\definecolor{darkblue}{rgb}{0.0,0.0,0.55}
\usepackage{multirow}
\usepackage{booktabs}
\usepackage{tablefootnote}
\usepackage{pdfpages}
\usepackage[version=4]{mhchem}
\usepackage{amsmath} 
\usepackage{comment}
\usepackage{datetime}
\usepackage{csquotes}

\usepackage{dutchcal} % \mathcal: also small letters. \mathbcal = bold ones. 
\usepackage{multicol}

\usepackage{tikz}
\usetikzlibrary{mindmap}
\usetikzlibrary{arrows}

\newdateformat{mydate}{\THEDAY. \THEMONTH. \THEYEAR}
\usepackage[a4paper, left=2cm, right=2cm, top=1.5cm, bottom=1.8cm]{geometry}
\usepackage[sort&compress,authoryear]{natbib}
\begin{document}
\setstretch{1.5}
\begin{titlepage}

\centering
\textbf{Charles University}\\
\textbf{Faculty of Mathematics and Physics}\\
\textbf{Astronomical Institute}\\[10mm]

%%%% Replace with your study programme %%%
Study programme: Physics \break
Study branch: Theoretical Physics, Astronomy and Astrophysics (P4F1A)

\begin{figure}[h!] 
\begin{center} \includegraphics[height=7.5cm]{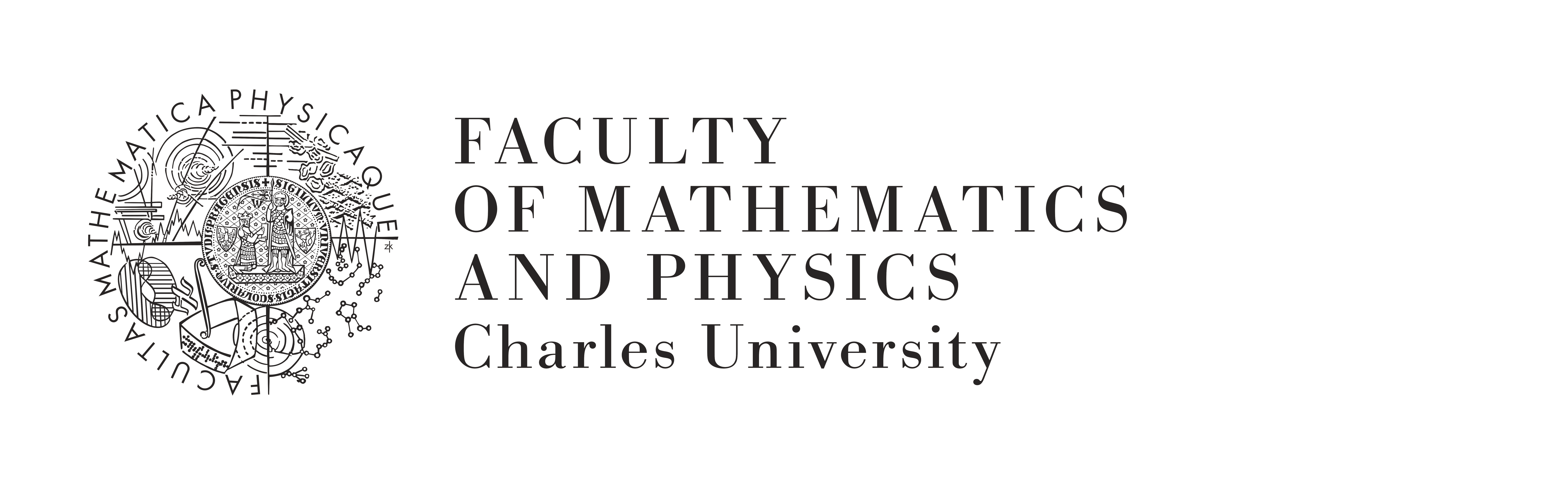}
\end{center}

\end{figure}

%%%% Replace with your name %%%
\textbf{\LARGE\href{https://sites.google.com/view/nicksamaras/home}{Nikolaos Samaras}} \\[1cm]
%%%% Replace with the thesis title in English and Czech %%%
The origin of galaxies \\[5mm]
Původ galaxií \\[1cm]

%%%% Select the type of thesis %%%
\href{https://dspace.cuni.cz/handle/20.500.11956/202014?locale-attribute=en}{Doctoral thesis}\\[1cm]

%%%% Replace with you supervisor and consultant %%%
Supervisor: \textbf{\href{https://astro.uni-bonn.de/~pavel/}{Prof. RNDr. Pavel Kroupa, Ph.D.}}\\
\null\vfill

Prague, \today

\end{titlepage}

%\pagestyle{empty}
%\null\vfill
%%% Declaration goes here (should be in Czech for Czech study programmes). 
%%% Information about declaration here: 
% https://natur.cuni.cz/fakulta/organizacni-struktura/organy-fakulty/dekan-a-kolegium/opatreni-dekana/opatreni-dekana-c-13-2023
% Example: I declare that I have independently prepared this thesis and that I have cited all the sources and literature used. Neither this thesis nor any substantial part of it has been submitted for the attainment of a different or the same academic degree.
%\noindent Declaration
%\\[2cm]
%Praha, \mydate\today \hfill Author % Add your name instead of Author
%\newpage

%\null\vfill
%%% Acknowledgement goes here %%%%
%\noindent Acknowledgment
%%%%

%\newpage
%-----------------------------------------------------------------------------------------------------------------------------------
    
\section*{Abstract}
Modified Newtonian Dynamics, or Milgromian Dynamics (MOND), has been particularly successful in predicting scaling relations for galactic systems, namely the baryonic Tully-Fisher for spirals, the Faber-Jackson for ellipticals and the Radial Acceleration Relation for late-type galaxies. Its essential tenet is the modification of the gravity law, at low accelerations. Nevertheless, despite MOND 's passed tests, the puzzle of the missing mass/gravity on galaxy clusters' scales still persist. One proposed scenario to complete this deficit and apply it further to cosmology is the so-called $\nu$HDM model. It is composed of a sterile neutrino and the MOND gravity. In this Ph.D. thesis, I have put forward this cosmological idea, conducting hydrodynamical simulations to investigate the $\nu$HDM structure formation. In the debut attempt, the $\nu$HDM model was proven capable of reproducing the $\Lambda$CDM cosmology, imitating the same expansion history ($\Omega_m \approx$ 0.3 and $H_0 \approx$ 67.6 km/s/Mpc), reproducing the CMB phenomenology, while the cosmic web is nicely formed. In the next step, I have optimized the $\nu$HDM model using Bayesian statistics, re-branding it to opt-$\nu$HDM. I managed to optimally fit the angular power spectrum of the CMB calculated by \textit{Planck}, except its fourth peak, but the opt-$\nu$HDM cosmological parameters changed significantly with respect to its precursor, resulting in a much heavier cosmos, with $\Omega_m \approx$ 0.5 and $H_0 \approx$ 55.6 km/s/Mpc. Next, I explore the opt-$\nu$HDM expansion history, its structure formation and the resulting mass function. The appearance of the early-time structures, firstly in the $\nu$HDM model, as well as in the opt-$\nu$HDM model occurs quiet late, at $z\approx4$ and at $z\approx 5.5$, respectively. Since the resolution has improved from previous studies, the $\nu$HDM variants cannot explain easily the high-redshift galaxies, observed by JWST. Furthermore, the main sequence of star forming galaxies is examined on the standard $\Lambda$CDM model cosmology. Although modern $\Lambda$CDM simulations seem to reproduce the observed relation of star formation with respect to the stellar mass in a restricted mass range, fine-tuning or unphysical interactions had to be introduced to accomplish that. Subsequent tensions are arising in the $\Lambda$CDM model, like the famous "Hubble tension" and some solutions are discussed.

%-----------------------------------------------------------------------------------------------------------------------------------

%-----------------------------------------------------------------------------------------------------------------------------------
\section*{Keywords}
%%%% Keywords go here %%%%
Cosmology – gravitation – galaxies: formation - statistics - methods: numerical\\
\newpage
%%%%%%%%%%%%%%%%%%%%%%%%%%%%%%%%%%%%%%%%%%%%%%%%%%%%%%%%%%%%%%%%%%%%%%%%%%%%%%%%%%%%%%%%%%%%%%%%%%%%%%%%%%

\section*{Acknowledgments}
There has always been one person to encourage me, from the dark times till the brightest ones, my mother. She is the strongest person I know. 

My middle brother, my one and only Hero.

My little brother, gartoll.

My father, a modern pirate.

My grandmother and my grandfather, without them, I would have not been able to survive in France. Their home is and will always be my shelter.

My dear friends Nodas and Yorgos.

And most importantly, my future wife, Vasso who believed in me. She has made my life so beautiful and fashionable. I feel so much stronger every moment she stands next to me.

Last, Prof. Pavel Kroupa, for his unique insights in physics. There are probably two kinds of physicists. Some of us who follow him and those ones who cannot.

During the last two years of my PhD, I was awarded the Charles University Grant Agency (GAUK) - 94224. I was also supported by the AKTION grant number MPC-2023-07755 for a 3-month stay at the University of Innsbruck, in the period of March-May 2024. I acknowledge support through the Deutscher Akademischer Austauschdienst (DAAD) Bonn-Prague exchange program. I have spent a great number of CPU hours, learning how to conduct numerical simulations in the cluster of CHIMERA in Prague. Last, I acknowledge the support by Ostrava Super-computing Center with the project Project OPEN-32-11 2885. 

This effort would not have been possible without the assistance of Dr. Indranil Banik for his brainy way of cosmological thinking, Dr. Ingo Thies for showing me how to perform Phantom of RAMSES simulations, Dr. Franti\v{s}ek Dinnbier, Dr. Michal \u{S}vanda, Dr. Ladislav \u{S}ubr for their help about everything at the Astronomical Institute of Charles University, Dr. Jiří Eli\'a\v{s}ek for his bright advise on the slurm scheduler system of the CHIMERA computer cluster, Dr. Moritz Haslbauer for his IGIMF ideas, Elena Asencio, Prof. Alexander Knebe for Amiga Halo Finder, Dr. Sebastian Grandis, Prof. Tim Schrabback and Prof. Francine Marleau for their hospitality in Innsbruck, Austria. My office mates, Myank Singhal and Iris Bermejo Lozano.

My ESA's Gaia experience in Brussels, at the Royal Observatory of Belgium, during 2019, had been the first sunny days for me in science. After a long period full of pain, during my M.Sc. degree in France, Dr. Ronny Blomme and Dr. Yves Fr\'emat had been very encouraging. Gaia was definitely the greatest middle step possible, towards my Ph.D. degree. 

I have become aware of MOND from the Triton station blog, maintained by Prof. Stacy McGaugh. Without his outreach articles, I could not have asked myself the difficult questions on the origin of Dark Matter (missing gravity). 

I would like to thank Prof. Thierry Jecko for his kindness, when he sold that house in Cergy, in the summer of 2018.

I would have not got into Cosmology, if I had not met Prof. Manolis Plionis, back in 2013. He was the first one, offering an opportunity to work in Astrophysics. I will never forget the excitement of my early days in Cosmology. Dr. Spyros Basilakos showed me the beauty of theoretical physics, after suggesting me a project at the Research Center for Astronomy \& Applied Mathematics of the Academy of Athens in 2014. Without them, I would have not become a Cosmologist. My good friend and colleague Dr. Christos Rantsoudis for his aid, to my baby steps on passing the first classes of my B.Sc. degree in Thessaloniki.

I would also like to express my gratitude to my first school teachers in Mathematics, Ilia Spyrou, Yannis Kastoris and Dimitris Santos. They had demonstrated to me the elegance of Mathematics.

Last but not least, I would like to say a big "thank you" to the following people: Rory Gallagher, Albert King, Stevie Ray Vaughn, Janis Joplin, Jimi Hendrix, Lynyrd Skynyrd, Dire Straits, The Doors, Pink Floyd, Rolling Stones and obviously to The Beatles for keeping me company throughout my whole life, helping me to solve some of my problems, while listening to their music.
\newpage
%%%%%%%%%%%%%%%%%%%%%%%%%%%%%%%%%%%%%%%%%%%%%%%%%%%%%%%%%%%%%%%%%%%%%%%%%%%%%%%%%%%%%%%%%%%%%%%%%%%%%%%%%%

%%%%%%%%%%%%%%%%%%%%%%%%%%%%%%%%%%%%%%%%%%%%%%%%%%%%%%%%%%%%%%%%%%%%%%%%%%%%%%%%%%%%%%%%%%%%%%%%%%%%%%%%%%
\noindent
The results of this thesis were achieved in the period of a doctoral study at the Faculty of Mathematics and Physics, Charles University in years 2021 – 2025.

\begin{longtable}{ l p{5in} } % choose suitable width for "p" column
\textbf{Student:} & Nikolaos Samaras \\ \vspace{1mm} 
\textbf{Supervisor:} &prof. RNDr. Pavel Kroupa, Ph.D.\\
\textbf{Department:} & Astronomical Institute of Charles University\\
 & V Holešovičkách 2\\
 & 180 00 Praha 8\\
\end{longtable}

\begin{longtable}{ l p{5in} }
\textbf{Referees:} & Prof. Xavier Hernandez\\
 & Instituto de Astonomia\\
 & Ciudad Universitaria\\
 & Circuito Exterior s/n\\
 & Ciudad Universitaria. Delg. Coyoacán. C.P. 04510 CDMX\\
 & Mexico\\
 & \vspace{1mm} Prof. Stacy McGaugh\\ 
 & Department of Astronomy\\
 & Case Western Reserve University\\
 & Sears Library\\
 & 2083 MLK Jr. Dr.\\
 & Cleveland, OH 44106\\
\end{longtable}

\noindent
The thesis defense will take place on \textbf{27.08.2025 at 2 p.m.} in front of a committee for thesis defense in the study programme P4F1A: Theoretical Physics, Astronomy and Astrophysics at the Faculty of Mathematics and Physics, Charles University, Ke Karlovu 3, Prague 2, room M252.\vspace{2mm}

\noindent
Chairman of study programme P4F51: Theoretical Physics, Astronomy and Astrophysics \vspace{2mm}

\noindent
\begin{longtable}{ l p{5in} }
doc. RNDr. Oldřich Semerák, DSc. & \\
Faculty of Mathematics and Physics & \\
Charles University & \\
V Holešovičkách 2 & \\
180 00 Praha 8 & \\
\end{longtable}

\noindent
The thesis is displayed at the Study Department of Doctoral Studies of the Faculty of Mathematics and Physics, Charles University, Ke Karlovu 3, Prague 2.
%%%%%%%%%%%%%%%%%%%%%%%%%%%%%%%%%%%%%%%%%%%%%%%%%%%%%%%%%%%%%%%%%%%%%%%%%%%%%%%%%%%%%%%%%%%%%%%%%%%%%%%%%%

%%%%%%%%%%%%%%%%%%%%%%%%%%%%%%%%%%%%%%%%%%%%%%%%%%%%%%%%%%%%%%%%%%%%%%%%%%%%%%%%%%%%%%%%%%%%%%%%%%%%%%%%%%
\newpage

\tableofcontents

\newpage
\scriptsize	
\section*{List of Abbreviations}
\begin{tabular}{ll}
    \textbf{s} & \textbf{s}econd\\
    \textbf{eV} & \textbf{e}lectron \textbf{V}olt\\
    \textbf{Mpc} & \textbf{M}ega \textbf{p}ar\textbf{s}ec\\
    \textbf{cMpc} & \textbf{c}omoving \textbf{M}ega \textbf{p}ar\textbf{s}ec\\
    \textbf{EW} & \textbf{E}lectro \textbf{W}eak\\
    \textbf{EM} & \textbf{E}lectro \textbf{M}agnetic\\
    \textbf{QCD} & \textbf{Q}uantum \textbf{C}hromo \textbf{D}ynamics\\
    \textbf{$\Lambda$CDM} & \textbf{$\Lambda$}ambda \textbf{C}old \textbf{D}ark \textbf{M}atter\\
    \textbf{MOND} & \textbf{Mo}dified \textbf{N}ewtonian/\textbf{M}ilgr\textbf{o}mia\textbf{n} \textbf{D}ynamics\\
    \textbf{BB} & \textbf{B}ig \textbf{B}ang \\
    \textbf{BBN} & \textbf{B}ig \textbf{B}ang \textbf{N}ucleosynthesis\\
    \textbf{EW} & \textbf{E}lectro \textbf{W}eak \\
    \textbf{CMB} & \textbf{C}osmic \textbf{M}icrowave \textbf{B}ackground radiation\\
    \textbf{BAO} & \textbf{B}aryonic \textbf{A}ccoustic \textbf{O}scillations\\
    \textbf{SNIa} & \textbf{S}uper \textbf{N}ovae \textbf{T}ype \textbf{Ia}\\
    \textbf{SNII} & \textbf{S}uper \textbf{N}ovae \textbf{T}ype \textbf{II}\\
    \textbf{DM} & \textbf{D}ark \textbf{M}atter\\
    \textbf{WDM} & \textbf{W}arm \textbf{D}ark \textbf{M}atter\\
    \textbf{HDM} & \textbf{H}ot \textbf{D}ark \textbf{M}atter\\
    \textbf{CDM} & \textbf{C}old \textbf{D}ark \textbf{M}atter\\
    \textbf{DE} & \textbf{D}ark \textbf{E}nergy\\
    \textbf{KBC} & \textbf{K}eenan \textbf{B}arger \textbf{C}owie void \\
    \textbf{H} & \textbf{H}ydrogen\\
    \textbf{yr} & \textbf{y}ear\\
    \textbf{Myr} & \textbf{M}egayear\\
    \textbf{Gyr} & \textbf{G}igayear\\
    \textbf{TNG100} & \textbf{TNG100-1}\\
    \textbf{TNG50} & \textbf{TNG50-1}\\
    \textbf{AGN} & \textbf{A}ctive \textbf{G}alactic \textbf{N}uclei\\
    \textbf{SMBH} & \textbf{S}uper \textbf{M}assive \textbf{B}lack \textbf{H}ole\\
    \textbf{EAGLE}  & \textbf{E}volution and \textbf{A}ssembly of \textbf{G}a\textbf{L}axies and their \textbf{E}nvironments\\
    \textbf{EAGLE100} & \textbf{EAGLE} RefL0\textbf{100}N1504 SFR AP30\\
    \textbf{EAGLE50} & \textbf{EAGLE} RefL00\textbf{50}N0752 SFR AP30\\
    \textbf{EAGLE25} & \textbf{EAGLE} RefL00\textbf{25}N0752 SFR AP30 \\ 
    \textbf{SPH} & \textbf{S}moothed \textbf{P}article \textbf{H}ydrodynamics\\
    \textbf{GSMF} & \textbf{G}alaxy \textbf{S}tellar \textbf{M}ass \textbf{F}unction\\ 
    \textbf{MS} & \textbf{M}ain \textbf{S}equence of galaxies\\
    \textbf{BH} & \textbf{B}lack \textbf{H}oles\\
    \textbf{SF} & \textbf{Star} \textbf{F}ormation\\
    \textbf{SFR} & \textbf{Star} \textbf{F}ormation \textbf{R}ate\\
    \textbf{SFH} & \textbf{Star} \textbf{F}ormation \textbf{H}istory\\
    \textbf{SP14} & \textbf{S}peagle et al. 2014 \\ 
    \textbf{BTFR} & \textbf{B}aryonic \textbf{T}ully \textbf{F}isher \textbf{R}elation \\
    \textbf{FB} & \textbf{F}aber \textbf{J}ackson \\
    \textbf{RAR} & \textbf{R}adial \textbf{A}cceleration \textbf{R}elation \\
\end{tabular}

\scriptsize	
\begin{tabular}{ll}
    \textbf{FLRW}&  \textbf{F}riedmann \textbf{L}emaitre \textbf{R}obertson \textbf{W}alker \\
    \textbf{MW} &  \textbf{M}ilky \textbf{W}ay galaxy \\
    \textbf{M31} & \textbf{A}ndromeda galaxy\\
    \textbf{LV} & \textbf{L}ocal \textbf{V}olume\\
    \textbf{GAMA} &  \textbf{G}alaxy \textbf{a}nd \textbf{M}ass \textbf{A}ssembly \\
    \textbf{PoR} & \textbf{P}hantom \textbf{o}f \textbf{R}AMSES\\
    \textbf{IC} & \textbf{I}nitial \textbf{C}onditions\\
    \textbf{MPI} & \textbf{M}essage \textbf{P}assing \textbf{I}nterface\\
    \textbf{AHF} & \textbf{A}miga \textbf{H}alo \textbf{F}inder\\
    \textbf{GR} & \textbf{G}eneral \textbf{R}elativity\\
    \textbf{HST} & \textbf{H}ubble \textbf{S}pace \textbf{T}elescope\\
    \textbf{JWST} & \textbf{J}ames \textbf{W}ebb \textbf{S}pace \textbf{T}elescope\\
    \textbf{BOSS}  & \textbf{B}aryon \textbf{O}scillation \textbf{S}pectroscopic \textbf{S}urvey \\
    \textbf{SDSS} & \textbf{S}loan \textbf{D}igital \textbf{S}ky \textbf{S}urvey\\
    \textbf{DESI} & \textbf{D}ark \textbf{E}nergy \textbf{S}pectroscopic \textbf{I}nstrument\\
    \textbf{TRGB} & \textbf{T}ip of the \textbf{R}ed \textbf{G}iant \textbf{B}ranch\\
    \textbf{NOVA} & \textbf{N}uMI \textbf{O}ff-axis $\nu_e$ \textbf{A}ppearance \\
    \textbf{KATRIN} & \textbf{K}arlsruhe \textbf{T}ritium \textbf{N}eutrino \textbf{E}xperiment \\
    \textbf{LSDN}  &  \textbf{L}iquid \textbf{S}cintillator \textbf{N}eutrino \textbf{D}etector \\
    \textbf{MCMC} & \textbf{M}arkov \textbf{C}hain \textbf{M}onte \textbf{C}arlo \\
    \textbf{WMAP} & \textbf{W}ilkinson \textbf{M}icrowave \textbf{A}nisotropy \textbf{P}robe \\
\end{tabular}
\normalsize	

\newpage
\noindent I, Nikolaos Samaras, declare that this thesis titled, \textquote{Origin of Galaxies} and the work presented in it are my own. I confirm that:

\begin{itemize} 
\item This work was done wholly or mainly while in candidature for a research degree at this University.
\item Where any part of this thesis has previously been submitted for a degree or any other qualification at this University or any other institution, this has been clearly stated.
\item Where I have consulted the published work of others, this is always clearly attributed.
\item Where I have quoted from the work of others, the source is always given. With the exception of such quotations, this thesis is entirely my own work.
\item I have acknowledged all main sources of help.
\item Where the thesis is based on work done by myself jointly with others, I have made clear exactly what was done by others and what I have contributed myself.\\
\end{itemize}
 
\noindent Signed:\\
\rule[0.5em]{25em}{0.5pt} % This prints a line for the signature
 
\noindent Date:\\
\rule[0.5em]{25em}{0.5pt} % This prints a line to write the date
\newpage

\listoffigures % Prints the list of figures
\newpage

\listoftables % Prints the list of tables

%-------------------------------------------------------------------------------------------------------------------------------------------------
% Chapter 1
\section{The current standard model of cosmology}
%\chapter{The current standard model of cosmology} % Main chapter title
\label{Chapter1} % For referencing the chapter elsewhere, use \ref{Chapter1} 
%-------------------------------------------------------------------------------------------------------------------------------------------------

%-------------------------------------------------------------------------------------------------------------------------------------------------
% Define some commands to keep the formatting separated from the content
%-------------------------------------------------------------------------------------------------------------------------------------------------

%-------------------------------------------------------------------------------------------------------------------------------------------------
\subsection{\texorpdfstring{The $\Lambda$CDM cosmological model}{The LCDM cosmological model}}
According to the standard cosmological model, the Universe was born approximately 13.8 Gyr ago in a BB singularity and it has been expanding ever since. Initially there was a state of very hot and dense primordial plasma in the Planck epoch (t $<{10}^{-43}$ s, T $> 10^{19}$ GeV). As the Universe was stretching, its temperature was cooling down. In the next $\approx 10^{-35}$ of the first second, the Universe was inflated in a phase of sudden accelerated expansion which made it significantly smoother. Microscopic quantum fluctuations are being developed in its density and temperature field and will consist the seeds of forthcoming cosmic structures. The electro-weak era follows at very high energies T $\sim 150$ GeV \footnote{I specifically use "$\sim$" which stands for "same order of magnitude".} (t $<10^{-12}$ s), and its phase transition (T $\sim150$ MeV) to the quark-hadron epoch, gave to quarks, leptons and gauge bosons their observed masses. At $\approx 1$ MeV, or about 1 s after the BB, the baryongenesis began. Neutrinos started decoupling from the photon-baryon fluid, too. DM has decoupled already, started forming gravitational potential wells, inside which later baryons will fall in and collapse.

\begin{table}[h]
    \begin{center}    
        \begin{tabular}{||c| c||} 
            \hline
            Phase & Energy scale\\ [0.5ex] 
            \hline\hline
            Standard Model & T $>$ 250 GeV  \\ 
            \hline
            EW transition & T $\sim$ 1 GeV  \\
            \hline
            QCD & T $\sim$ 150 MeV  \\
            \hline
            DM freeze out & ? \\
            \hline
            neutrino decoupling & T $\sim$ 1 MeV \\
            \hline
            BBN & T $\sim$ 0.1 MeV \\ 
            \hline
            Radiation-Matter equality & T $\sim$ 1eV\\
            \hline
            Recombination & T $\sim$ 0.3 eV\\
            \hline
            photon decoupling & T $\sim$ 0.27 eV\\
            \hline
            Reionization  & $\approx$ 5.0 meV \\  %100–400 Myr 11–30
            \hline
            DE-matter equality & $\approx$ 0.3 meV\\ %9 Gyr
            \hline
            Present & $\approx$ 0.2 meV \\ %13.8 Gyr
            \hline
        \end{tabular}
        \caption{$\Lambda$CDM estimates on the Universe energy scales}
        \label{energy_scales}
    \end{center}
\end{table}

It took approximately 3 minutes after the BB, when $T\sim0.1$ MeV, for the BBN to commence. Particles were being created and annihilated at a rate faster than the expansion of the Universe. Mostly photons exist, making the Universe extremely opaque. Protons and neutrons bound together to form atomic nuclei, while the Universe is still in the radiation era. The matter-radiation equality happened at $z\approx 3400 \approx$ 50,000 years after BB, at $T\sim1$ eV $\approx10^4$ K. What comes next is recombination ($z<$ 3400), at $T\sim0.3$ eV, the era when neutral H starts forming. The Universe becomes transparent. Photons start decoupling forming the CMB, at T $\approx0.27$ eV $\approx$ 380,000 yr after the BB. Photon decoupling happened right after recombination, because there were so many more photons, that some of them were still interacting with the thermal bath.

The matter-dominated era begins, after the photon decoupling, till $z\approx 0.4$. For the first Gyr, there were any, yet, formed emitting source (either stars or galaxies), and since the Universe was transparent at that particular state, the era was called "the Dark Ages" ($1100<z<20$). It is suggested that the first generation of stars were born around 400 Myr after the BB. Thereafter, this era will come to an end by the re-ionization epoch, which follows at z $\approx$ 6-8. Re-ionization is driven by the galaxy formation and evolution and neutral atoms are re-ionized, after being emitted by recently formed stars/galaxies. During the last $\approx$ 9.5 Gyr or $z<$ 0.4, DE comes into play, accelerating the expansion. At the present time, CMB is observed at at T $\approx$ 2.7 K, after 13.8 Gyr \citep{1990eaun.book.....K}.
\begin{figure*}[h] 
    \includegraphics[width=\linewidth]{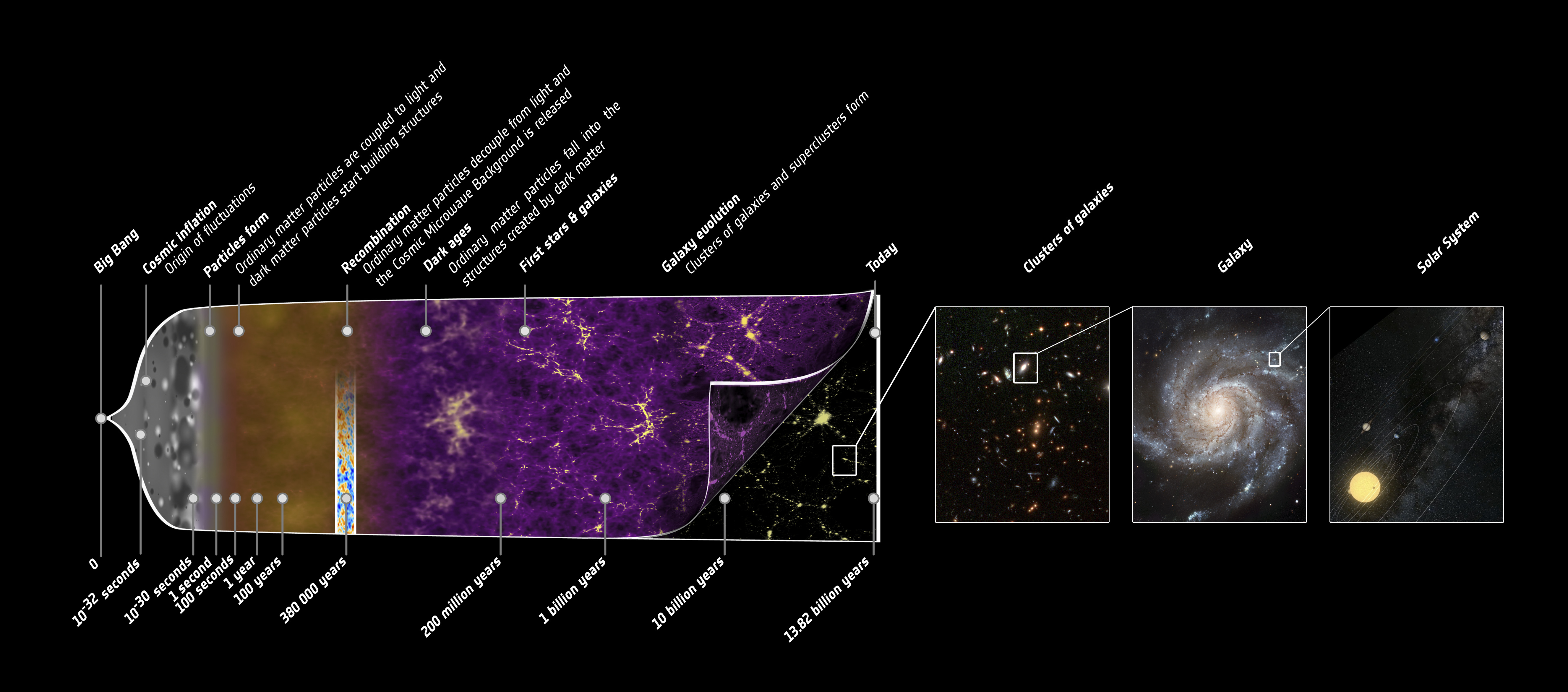}
    \caption{ESA's \textit{Planck} mission - C. Carreau}
    \label{LCDMhistory}
\end{figure*}

The two central characters operating the formation and the evolution of the gravitationally bound systems is the DM and the DE. The former is required in order to justify a number of observations, from the flatness of the rotation curves at the outskirts of galaxies, to intermediate scales of the galaxy clusters and their strong-lensing inferred mass, to the cosmological relevance of the 3rd peak of the CMB power spectrum \citep{1990Natur.348..705E, 1995Natur.377..600O}. DM is expected to be dissipationless and cold; CDM particles have low velocity dispersions and the cosmic structure proceeds hierarchically, from smaller objects merging to make bigger ones, in its distinct bottom-up cosmological scenario. On the other side, DE is proposed as the driving mechanism of the accelerated expansion of the late-time Universe \citep{Riess_1998}. It's hypothetically a form of energy, with negative pressure, acting as the opposite of gravity. Consequently, the standard model of cosmology is called $\Lambda$CDM.

Although this broadly-studied in the last decades scenario offers plenty of explanations about the emergence of the stars and galaxies, a number of auxiliary assumptions or postulates had to be formulated. The BB singularity means that all the matter and energy of the Universe is created in the beginning and there is no other source plugging mass/energy into/out of the Universe. Second in its chronological order comes the cosmological inflation. Despite the fact that it's not possible to directly probe the presumed physics before the CMB emission, inflation has been conjectured as a mathematical ansatz to solve the horizon and the flatness problems. The horizon problem refers to the observed homogeneity of causally disconnected regions of the Universe in the CMB observations. On the other hand, the flatness problem describes the coincidence that, in case of slightly different values of the matter-energy density of the Universe, it would have been non-flat, meaning geometrically non-Euclidean. In addition to the previous fine-tuning problems, the Cosmological Principle is postulated, dictating that the Universe is homogeneous and isotropic at sufficiently large scales. Wherever the observer stands, the observer sees the same distribution of matter in all directions (isotropy) and at any location (homogeneity). The FLRW metric encodes all this information, allowing to calculate distances in an expanding $\Lambda$CDM universe.

Last but not least, the cornerstone of the $\Lambda$CDM structure formation relies on the existence of exotic, DM particles, which are still undetected despite many astrophysical and particle physics experimental efforts. From the lack of observational evidence for dynamical dissipation by the DM halos, to the anisotropic distribution of satellite galaxies around the MW and the M31 and the huge underdense regions like the KBC void, the standard model of cosmology appears incomplete \citep{2023eppg.confE.231K}.

\subsection{\texorpdfstring{Cosmological $\Lambda$CDM simulations}{Cosmological LCDM simulations}}
In this dissertation, I first examine two of the most sophisticated $\Lambda$CDM hydrodynamical simulations, the Illustris TNG and the EAGLE ones. I will describe their cosmology and I will estimate the Main Sequence (MS) of galaxies and its scatter. I will explain why the MS should not be used as major prediction of the $\Lambda$CDM theory, because of their fine-tunned subgrid physics and their numerical calibration. 

%\begin{figure*}[h] 
%    \includegraphics[width=\linewidth]{Figures/gsmf.png}
%    \caption{The GSMF of $\Lambda$CDM simulations and LV observations}
%    \label{gsmf}
%\end{figure*}

The Illustris TNG Simulation project \citep{10.1093/mnras/stx3112} is a set of cosmological simulations studying the formation and the evolution of galaxies assuming a $\Lambda$CDM cosmology. It performs calculations with the moving-mesh code AREPO \citep{Spri1} based on \citet{2014A&A...571A...1P} results. More specifically, the following cosmological parameters are adopted: $\Omega_{m,0}=0.3089$, $\Omega_{\Lambda,0}=0.6911$, $\Omega_{\text{b},0}=0.0486$, $\sigma_{8} = 0.8159$, $n_{s} = 0.9667$, $H_{0} = 100 h^{-1}$ km/s/Mpc and $h = 0.6774$. The simulations begin at redshift z = 127. The evolution of DM particles, gas cells, stars, stellar wind particles and supermassive black holes (SMBHs) are followed up to redshift z = 0 \citep{Nel}.

The friends-of-friends algorithm (FOF) is used to identify the DM halos \citep{Davies}. Subhalos within halos are identified with the Subfind algorithm \citep{Spri2, Dolag} carrying a unique ID for all the snapshots. The boundary conditions of the box are periodic. The spatial resolution of the (sub)halos is defined by the minimum gravitational potential energy of the particle and the total mass by the sum of the particles connected in it.
%with a linking length of 0.2 times the mean particle separation. The minimum particle number of each halo is 32.

For the TNG100-1 simulation (hereafter TNG100) which is the highest resolution run, there are $1820^3$ DM particles, $1820^3$ gas cells in a comoving volume of ${110.7}^{3}$ cMpc\textsuperscript{3}. The actual box length is 75 Mpc/h $\approx$ 100 Mpc with a sidelength of $\approx$ 110 cMpc. The gravitational softening length for the DM particles is 1420 comoving pc. The mass of the DM particles is $7.5 \times 10^6 M_{\odot}$ and the baryonic mass is $1.4 \times 10^6 M_{\odot}$. As for the TNG50-1 simulation (hereafter TNG50)  \citep{10.1093/mnras/stz2338}, which is the smallest resolved simulation, it includes $2160^3$ DM and gas particles in 35 Mpc/h $\approx$ 50 Mpc\textsuperscript{3} volume. The initial baryonic mass is $8.5 \times 10^4 M_{\odot}$ and a DM particle mass $4.5 \times 10^5 M_{\odot}$.

\begin{figure*} 
    \includegraphics[width=\linewidth]{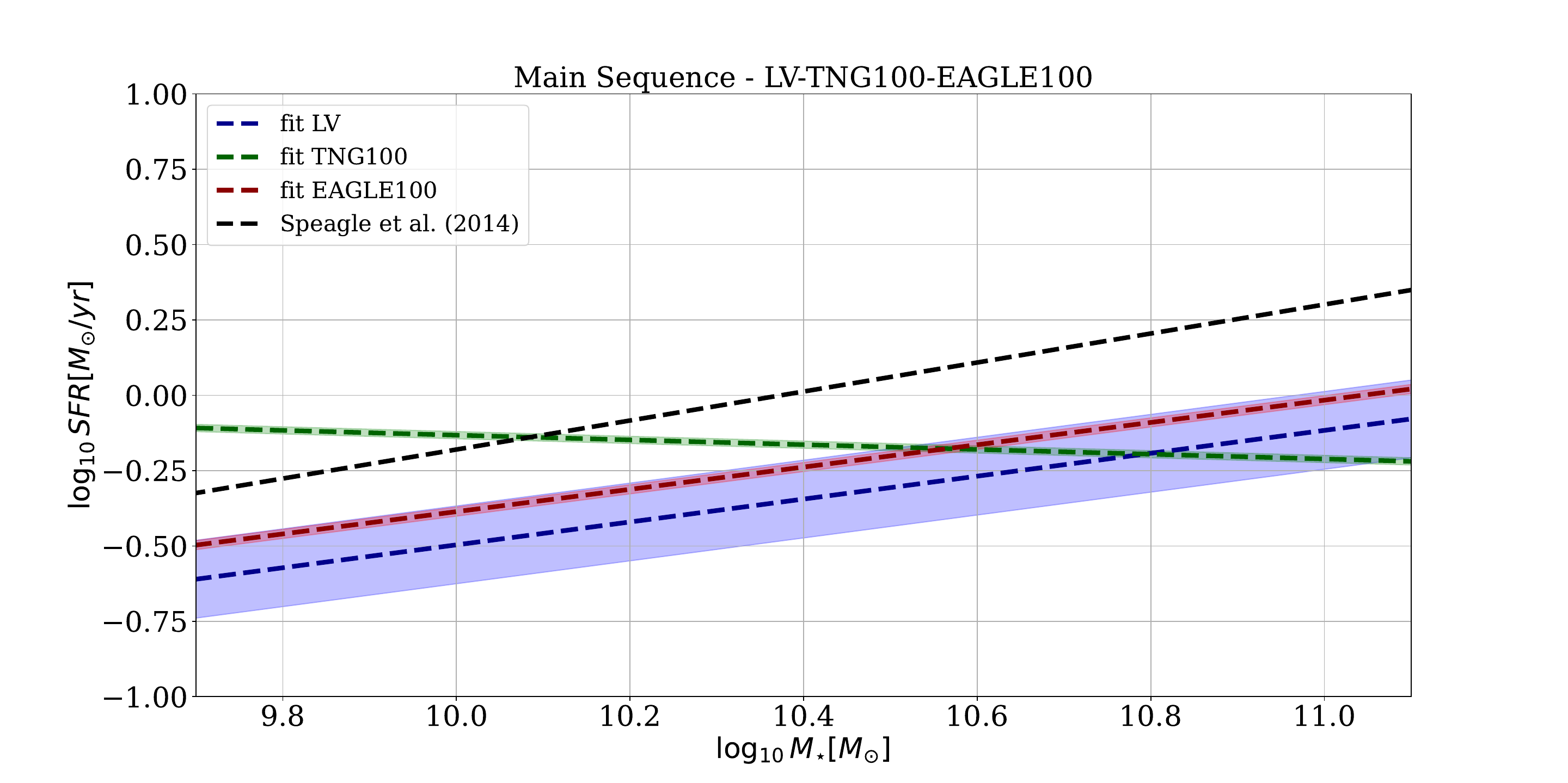}
    \caption{The MS in the Speagle range for the 100/h Mpc box.}
    \label{ms100}
\end{figure*}

The Illustris simulations include a sophisticated model of galaxy formation. The stochastic star-formation (in dense interstellar medium, gas above the threshold density $n_\text{H} \simeq$ 0.1 cm\textsuperscript{-3} forms a star) involves stellar evolution with mass loss and chemical enrichment, cooling and heating mechanisms of the interstellar medium, AGN feedback in addition to growth and evolution of SMBHs \citep{Vogel}. The reason behind the SFH stochasticity is the discreetness of SFR values at the lower resolutions runs. \citet{10.1093/mnras/stx3040} explain that a low mass galaxy on the MS with SFR $\approx$ 1$M_{\odot}$/yr will create on average $10^6M_{\odot}$ stellar particles every million years at TNG100-1 resolution. Thus, there may be no newborn star particles present even though a galaxy is actively star forming. This issue was solved by \citet{10.1093/mnras/stx1051} with a post-process re-sampling technique, eventually producing the expected populations of very young star particles.

In order to estimate the MS, I have made a further selection on the present study for a present-day (z = 0) subhalos of SFR $>10^{-3}M_{\odot}$/yr. I have also removed subhaloes with a Subhalo flag value of 0 in the TNG simulations, ensuring that only galaxies or satellites of a cosmological origin are considered. \citet{2019ComAC...6....2N} report that not all the objects are considered as "galaxies". The Subfind flag marks which satellite subhalos have been formed through baryonic processes (like disk instabilities) in already formed galaxies and the Subfind algorithm cannot distinguish them in principle. These structures, although excluded in this analysis, could generally appear as outliers in scatter plots.

On the other hand, the EAGLE suite of simulations \citep{10.1093/mnras/stu2058, Crain} employ the modified version of the Gadget-3 Tree SPH code \citep{Spri3}. I focus on two runs, in particular: the fiducial EAGLE100 and the EAGLE50. More particularly, the EAGLE100 simulation contains $2 \times 1504{^3}$ particles in total, with masses being $9.7 \times 10^6 M_{\odot}$ for DM and $1.8 \times 10^6M_{\odot}$ for baryons respectively, in a volume $L = 100$ cMpc\textsuperscript{3}. The EAGLE50 model contains $2 \times 752^3$ particles in a 50 cMpc\textsuperscript{3} volume.

EAGLE has different feedback and SF physics from the TNG simulation. Star formation is stochastic, designed to adapt the observed Kennicutt-Schmidt law \citep{SD}. The feedback processes combine radiative cooling, star formation, growth of BH via AGN activity. The aperture size is also important for the measure of stellar masses. I chose it be 30, when downloading the data. The meaning of the aperture is a spherical region of radius of 30 proper kpc, within where the galaxy stellar masses $M_{\star}$ are computed by summing up all the masses of star particles. This is also included in the radiative transfer calculation. Notably, for most of the less massive galaxies $M_{\star}< 10^{11} M_{\odot}$, the effect of the aperture is negligible \citep{10.1093/mnras/stu2058}. Nevertheless, regarding the more massive ones, the aperture size reduces the stellar masses somewhat ($\approx$ 1dex), making their metallicities and half-mass radii lower, though not affecting their SF. The larger the aperture value, the larger the mass estimate.

Similarly with the Illustris TNG simulation, the Friends-Of-Friends algorithm is used to detect haloes using the SubFind program. The EAGLE simulation is based on the cosmological parameters of $\Omega_{m,0} = 0.307$, $\Omega_{\Lambda,0} = 0.693$, $\Omega_{\text{b,0}}= 0.04825$, $\sigma_{8} = 0.8288$, $n_{s} = 0.9611$ and $ h = H_{0}/100$ km/s/Mpc$ = 0.6778 $, following \citet{2014A&A...571A...1P}. Likewise, I have filtered out galaxies with SFR $<10^{-3} M_{\odot}$/yr to examine the MS from EAGLE simulation.

\begin{table}
    \centering
    \begin{tabular}{lll}
        Data set  & $\alpha$  & $\beta$ \\ \hline
        TNG100    & -0.016    &  0.037  \\
        EAGLE100  & 0.37      & -4.085  \\ \hline
        TNG50     & 0.489     & -5.075  \\
        EAGLE50   & 0.472     & -5.098  \\ \hline
        LV        & 0.379     & -4.295  \\ \hline
        Speagle et al. & 0.48 & -4.99   \\ \hline
    \end{tabular}
    \caption{Parameters to linearly fit the MS relation.}
    \label{MSfits}
\end{table}

\begin{figure*} 
    \includegraphics[width=\linewidth]{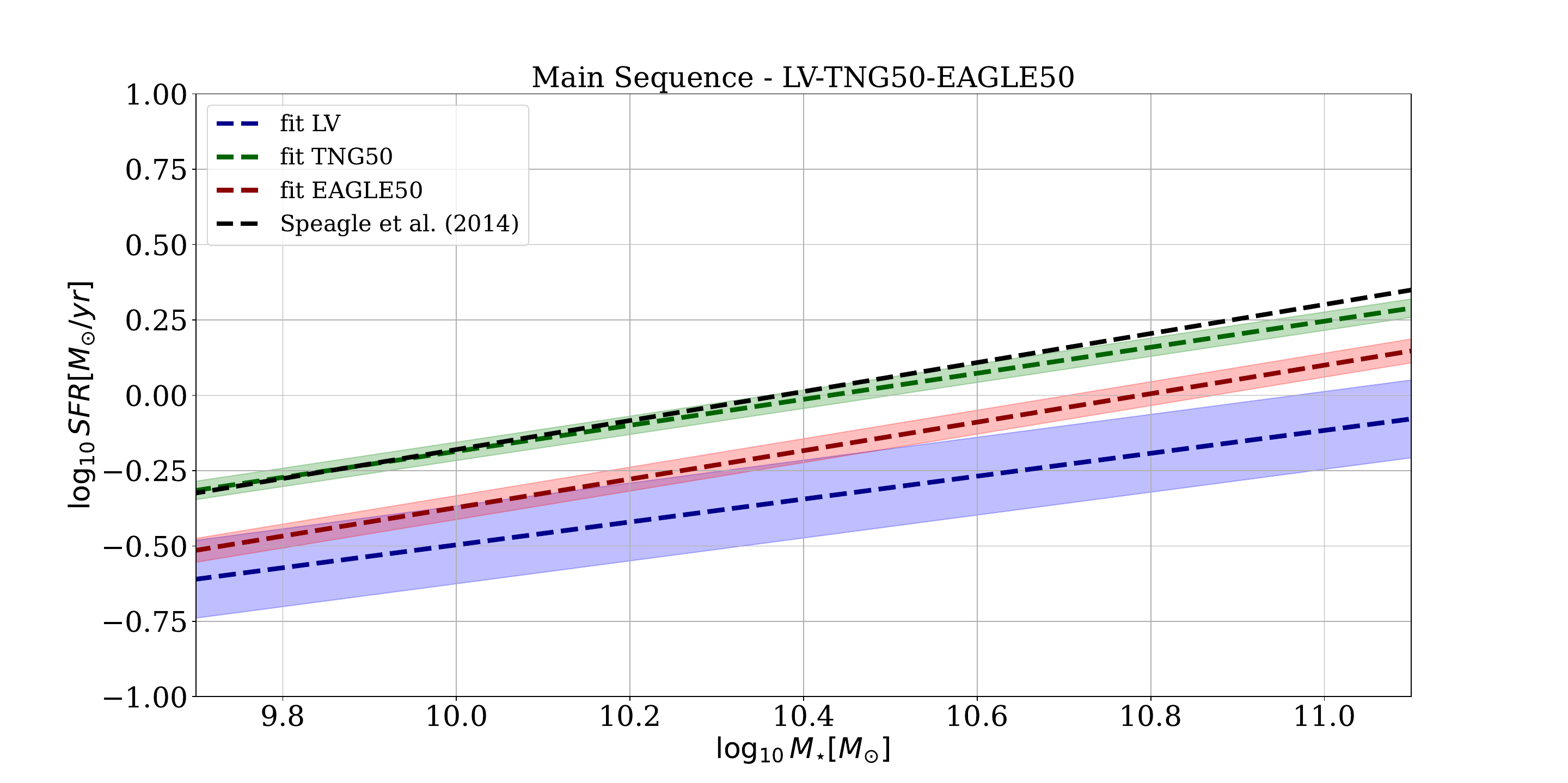}
    \caption{The MS in the Speagle range for the 50/h Mpc box.}
    \label{ms50}
\end{figure*}

\subsection{Observational Studies}
The Local cosmological Volume (hereafter LV) is defined as a sphere of radius 11 Mpc with the Milky Way at its center. I analyze the updated data from the Catalogue of Neighbouring Galaxies \citep{Karachentsev_2013} which reside in the LV. The SFRs are extracted from H$\alpha$ and Far Ultra-Violet (FUV) measurements. The catalog contains 1244 galaxies, 128 having only $\text{SFR}_{\text{H}\alpha}$, 511 only $\text{SFR}_{\text{FUV}}$, and 605 both SFR measurements. I convert the K-band luminosities to $M_{\star}$ using a mass-to-light ratio of $M_{\odot}/L_{\odot} = 0.6$ \citep{Lelli2016}. 

I have disregarded galaxies with $\text{SFR}_{\text{FUV}}$ or $\text{SFR}_{\text{H}\alpha}<10^{-3}M_{\odot}/yr$ as well as those which do not have a K-band Luminosity value (and consequently its $M_{\star}$ could not be calculated), following the same formalism as \citet{Kroupa2020, Hasl}. Hence, the resulting total sample consists of 558 galaxies,  91 of which have only H$\alpha$ measurements, 110 have only FUV measurements and 357 have both H$\alpha$ and FUV. I refer the reader to \citet{Kroupa2020} regarding the reasoning of excluding galaxies with SF activity less than $10^{-3}M_{\odot}/yr$. These are present-day SFR values. I mainly rely for this calculation on $\text{SFR}_{\text{H}\alpha}$ when they are available and in case they are not, I take the $\text{SFR}_{\text{FUV}}$ instead. In case both measurements are given, I average the value as: $\text{SFR} = \frac  {\text{SFR}_{\text{FUV}} + {\text{SFR}_{\text{H}\alpha}} } {2}$. Regarding the quiescent galaxies, I change the selection criteria appropriately, including those with SFR$<10^{-3}M_{\odot}/yr$, for stellar masses $10^{9}<M_{\star}/M_{\odot}<10^{12}$.

Additionally I compare the simulated data with observational studies, based on the compilation of \citet{Speagle_2014}. Restricting the comparison to stellar masses between $9.7<\log_{10} (M_{\star}/M_{\odot})<11.1 $ (the "Speagle range"), the time-dependent $\text{SFR}-M_{\star}$ relation is:
\begin{equation} 
    \begin{split}
        \log_{10} \text{SFR} (M_{\star}, t) & = \alpha \times \log_{10} M_{\star} \text{ + } \beta \\& = (0.84 \pm 0.02 \text{ - } 0.026 \pm 0.003 \times t) \log_{10} M_{\star} \\& \text{ - } (6.51 \pm 0.24 \text{ - } 0.11 \pm 0.03 \times t).
    \end{split}
    \label{MS}
 \end{equation}
For an age of the Universe of 13.7 Gyr, it results into a slope of  $\alpha = 0.48$ and normalization of $\beta = -4.99$. The shaded areas in Fig. \ref{ms100} and Fig. \ref{ms50}  represent Poisson statistics.

%Moreover, in order to compare the galaxy stellar mass function, I also exploit data from the GAMA team (\url{http://www.gama-survey.org/}). This projects aims to test the CDM structure formation scenario. There are $\approx$ 300,000 galaxies spectroscopically observed down to r < 19.8 mag over $\approx$ 286 $deg^{2}$, carried out by the AAOmega multi-object spectrograph on the Anglo-Australian Telescope (AAT). I consider the \citet{10.1111/j.1365-2966.2012.20340.x} paper which determines the low-redshift field galaxy stellar mass function using an area of 143 $deg^2$ and 5210 galaxies. The GSMF is plotted in Fig. \ref{gsmf} using the data from their Table. 1. I neglect the low mass end ($M_{\star}<10^{8}M_{\odot}$), since I reach the simulations' resolution limit and the comparison becomes problematic.

\begin{table}
\centering
\begin{tabular}{llll}
    Data set  & $\sigma$ & Nb of Galaxies & Total Nb of Galaxies\\ \hline
    TNG100    & 0.39     & 6884 & 337,261    \\
    EAGLE100  & 0.43     & 4393 & 325,423    \\ \hline
    TNG50     & 0.4      & 1070 & 163,267     \\
    EAGLE50   & 0.4      & 635  & 41,280      \\ \hline
    LV        & 0.61     & 60   & 1,244       \\ \hline
    Speagle et al. & 0.2   & Not Specified  & Not Specified  \\ \hline
    \end{tabular}
    \caption{MS scatter value for all the sets and their number of galaxies in the Speagle range and their total number in the set.}
    \label{scatter_table}
\end{table}

\subsection{Comparing \texorpdfstring{$\Lambda$}{L}CDM simulations and observations}
%Firstly, the galaxy stellar mass function simply counts the number of galaxies in a respective cosmological volume (Fig. \ref{gsmf}). Simulated galaxies seem to follow very well both the LV and GAMA surveys. It's only at the high-mass end where the GAMA observations are underestimated by the TNG models by a factor of 10. Generally, TNG galaxies are much more in number, compared to the EAGLE ones, hence the difference $M_{\star}>10^{11}M_{\odot}$. Curiously, there is slight decrease of the number density of LV galaxies between $10^{9.5}<M_{\odot}<10^{10.5}$, but it is within the Poisson uncertainties.

Concerning the MS plots (Fig. \ref{ms100}, \ref{ms50}), the theoretical $\Lambda$CDM model simulations are in agreement with the LV observations. The lines are linear fits to the galaxies of each set, having selected them in the constrained mass range of $9.7<\log_{10}(M_{\star}/M_{\odot})<11.1$. I summarize the fitting parameters in Table \ref{MSfits}, with $\alpha$ being the slope of the MS and $\beta$ the normalization, adopting the "Speagle" terminology. Although the MS parameters $\alpha$ and $\beta$ depend on time, I focus only at the present epoch (t = 0).

Particularly for the 100 Mpc\textsuperscript{3} box, simulations are in line with observational studies, in this limited range of $10^{9.7}<M_{\star}<10^{11.1}$ values (red and green lines of Fig. \ref{ms100}). The observational sets, the LV and the Speagle study, appear systematically parallel (black for Speagle and blue for the LV ones). Interestingly, due to their higher number of galaxies (compared to the EAGLE100), the TNG100 is the only set with a negative slope ($\alpha<0$). I note also that the TNG100 MS slope becomes even more declining ($\alpha_{\text{TNG}100}$ =  - 0.07906), if ones includes galaxies with $10^{-5}M_{\odot}/yr<\text{SFR}<10^{-3}M_{\odot}/yr$, although this is not shown in the plot. Furthermore, lower than the $10^{-5}M_{\odot}/yr$ threshold, the resolution of the simulations becomes severely poorer, thus those galaxies are not taken into consideration. Overall, $\Lambda$CDM simulations managed to reproduce the observed slope of the MS over the restricted slope of $10^{9.7}<M_{\odot}<10^{11.1}$, but the trend of TNG MS stands problematic. Less massive TNG galaxies (in the Speagle range) produce more stars per unit time than observed and more massive TNG ones produce almost what is observed. The difference on the subgrid physics between TNG and EAGLE is demonstrated in this exercise, since the EAGLE100 reproduces the LV SFR value more closely.

%Numerically, it means that most of the TNG galaxies inside this range, tend to form incorrect number of stars compared to how many real galaxies do produce.

On the other hand, the TNG50 does not exhibit such negative $\alpha$ MS slope. Both EAGLE50 and TNG50 mimic the observations fairly well. The TNG50 theoretical model fully overlaps with the Speagle relation (Eq. \ref{MS}). All the sets appear almost perfectly parallel to each other ($\alpha_{\text{TNG}50} = 0.489 \approx \alpha_{\text{EAGLE}50} = 0.472$ being very close to $\alpha_{\text{LV}}$ = 0.379), as depicted in Table \ref{MSfits}. It seems that $\Lambda$CDM captures the MS well enough at the cosmological volume of 50 Mpc/h. However, the number of the simulated galaxies is significantly smaller on a 50 Mpc box (Table \ref{scatter_table}).

For this task, the Python 3 polyfit module $(\log_{10} M_{\star}, \log_{10} \text{SFR})$ was used, which performs a least-squares polynomial fit. The degree $deg$ of the fit:\\
\overfullrule=0mm(np.polyfit(x,y,$deg$))) is set to 1.

\subsection{Scatter of the MS}
\begin{figure*} 
    \includegraphics[width=\linewidth]{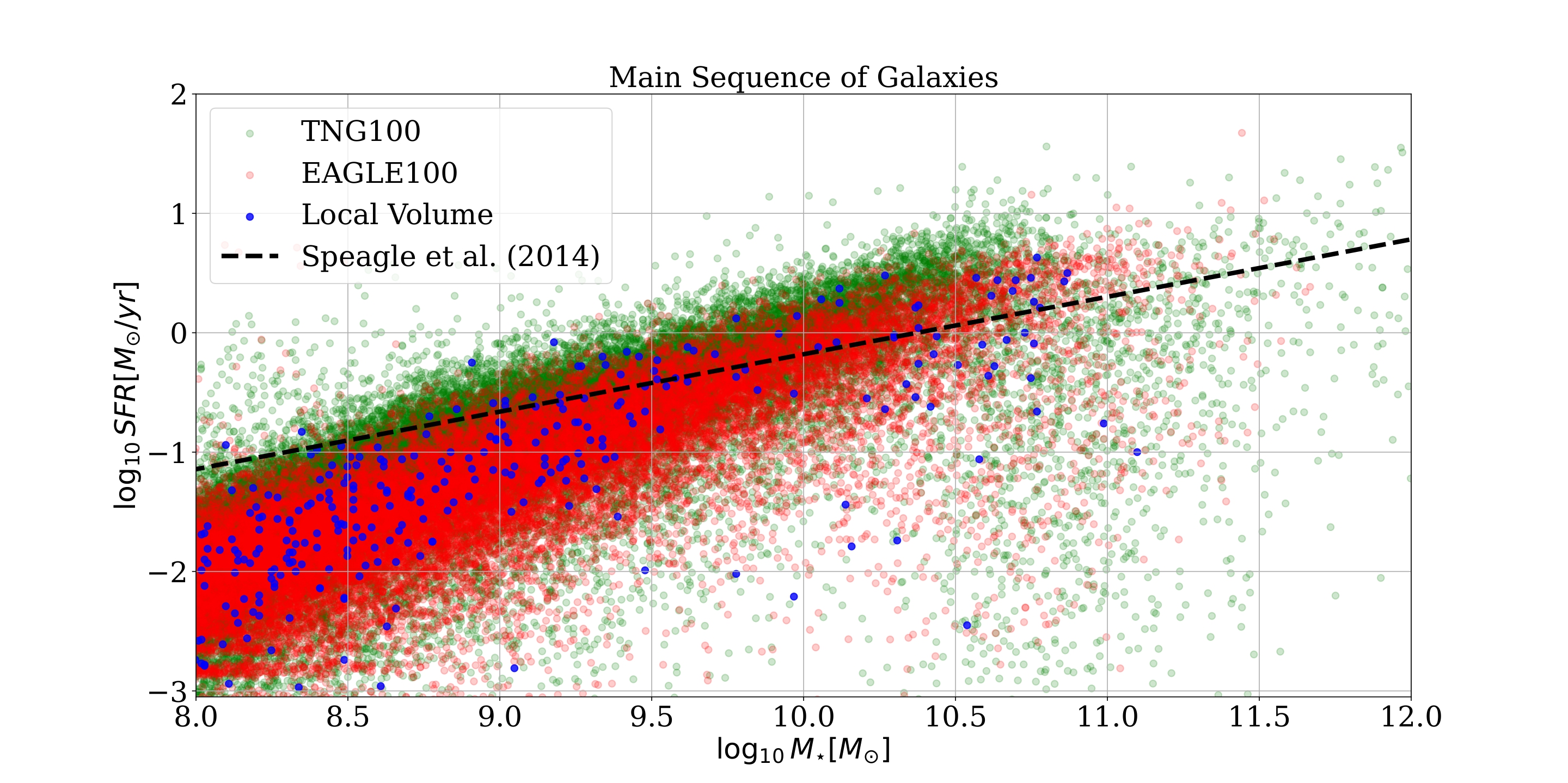}
    \caption{Main Sequence scattered plot for the $\Lambda$CDM simulations, together with the Speagle et al 2014 relation.}
    \label{MSscattered}
\end{figure*}

In this section, I quantify the MS scatter for very massive galaxies $10^{10}<M_{\star}/M_{\odot} < 10^{12}$, utilizing the Python 3 curve-fit function, as well as the numpy.histogram and matplotlib.bar functionalities. These enormous objects tend to fall off the linear line of the $\text{SFR}-M_{\star}$ relation, creating the high-mass end scatter (Fig. \ref{MSscattered}). In order to assess the stellar mass distribution, I perform the following operations:
\begin{enumerate}
    \item I bin the data in segments of 0.2 $M_{\odot}$ in the Speagle range ($9.7 <\log_{10}\frac{M_{\star}}{M_{\odot}}< 11.1$).
    \item I fit the SFR-valued normalized histograms with a Gaussian distribution.
    \item I compute the standard deviation $\sigma$.
\end{enumerate}

\begin{figure*} 
    \includegraphics[width=\linewidth]{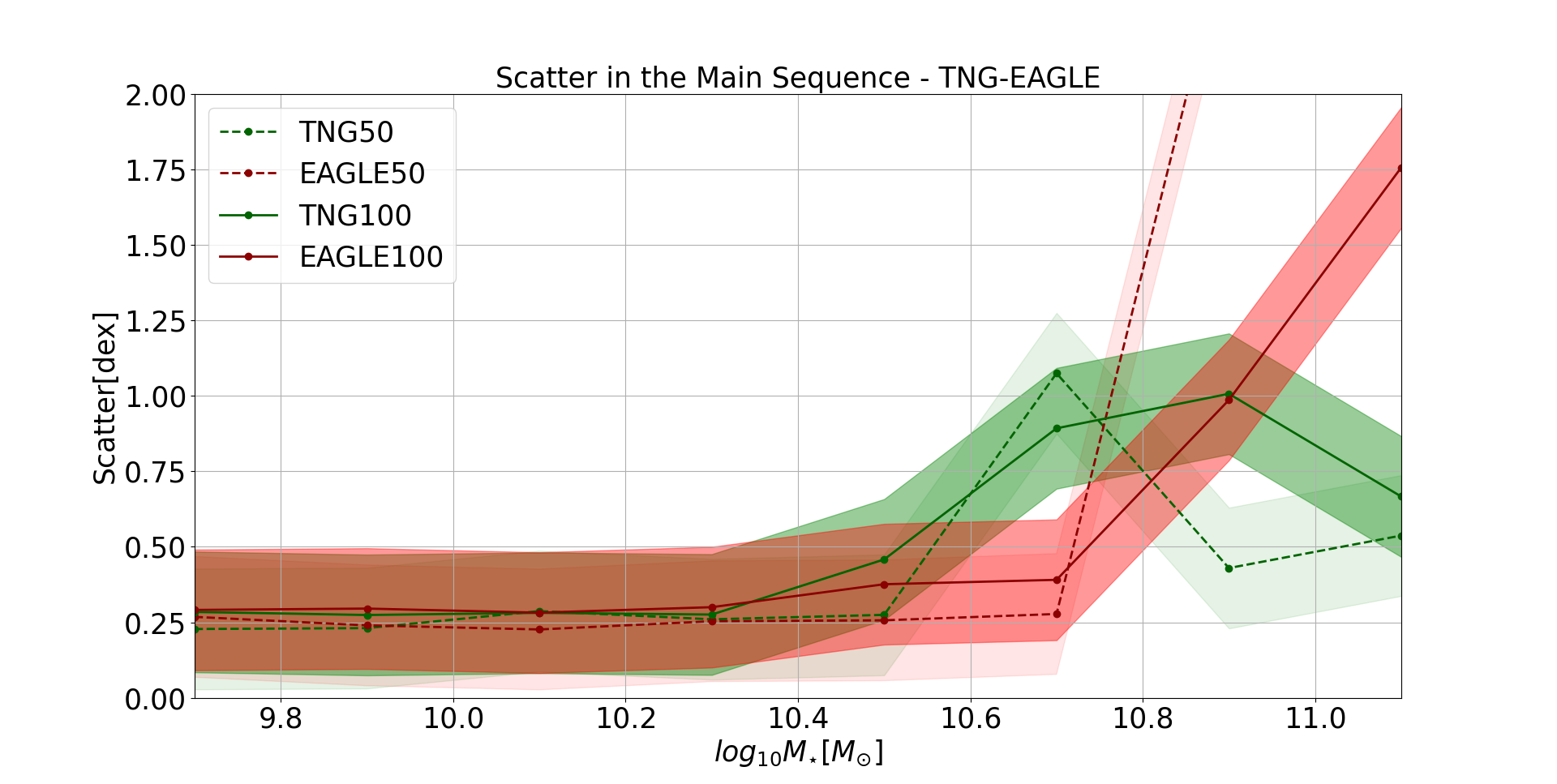}
    \caption{Estimated MS scatter for the $\Lambda$CDM simulations in the Speagle range.}
    \label{scatter}
\end{figure*}

Regarding the LV data, they give low number statistics, since there are only 60 LV galaxies in the Speagle range. I adopt the \citet[their fig. 3]{10.1093/mnras/stu2713} formalism. These authors work on a larger sample ($10^9M_{\odot} -10^{12} M_{\odot}$) than the Speagle range ($10^{9.7}<M_{\odot}<10^{11.1}$) but the binning is the same (0.2 dex). I confirm their results, showing that both TNG and EAGLE galaxies having the same trend (Fig. \ref{scatter}), of $\approx$ 0.2 MS scatter. \citet{Speagle_2014} do not provide specific information about how the selection of galaxies was achieved.

Importantly, other observational studies like \citet[their table 9]{2013ApJ...770...57B} claim a scatter around 0.21-0.39 dex, in the Speagle range.  Nevertheless, the observational uncertainties differ to the uncertainties of the simulations. The simulated models seem to agree with the data, but one should be aware that the total uncertainty is a combination of the intrinsic and the observational one, such as: 
$\sigma_{total} = \sqrt{\sigma_{obs}^2 + \sigma_{intr}^2}$. There is no observational bias in the simulations, thus the difference will be bigger.

Last but not least, there is always a selection effect. One should keep in mind that we have excluded galaxies with values of SFR<$10^{-3}M_{\odot}/yr$ (and with Subhalo Flag 0), but using a different choice criterion, the slope and the scatter will change substantially. For example, in the compilation-observational part, \citet{Speagle_2014} claim that $\text{SFR}_{\text{FUV}}$ is mostly coming from young stars while H$\alpha$ radiation is deducted from a much shorter timescale of SF activity.

One can see comprehensively the MS scatter on the Fig. \ref{MSscattered}. The $\Lambda$CDM modeled galaxies, more massive than $10^{10.4} M_{\odot}$, appear more spread. The scatter remains constant from $10^{9.7}M_{\odot}$ to $\approx 10^{10.5} M_{\odot}$ (also Fig. \ref{scatter}). Beyond $10^{10.6}M_{\odot}$, the data become much more dispersed for a given $M_{\star}$ and consequently the corresponding SFR value could vary significantly compared to the MS relation, spread almost 1 order of magnitude less then the Speagle line.

\subsection{Fine-tuning \& Calibration}
The $\Lambda$CDM simulations managed to fit the MS relation and the observed population of galaxies, but the incorporated physics of the model is debated. Regardless of the yet-undetected DM particles or the dynamical friction problem \citep{2021MNRAS.508..926R, 2024Univ...10..143O} it suffers from, the SF processes have been regulated and "corrected" in order to achieve this match with the experimental data. Notably, the adopted feedback processes of the TNG simulations introduce an interaction between the baryons of stellar winds and the velocity dispersion of DM particles, which contradicts the DM nature as a collisionless, non-electromagnetically interacting component. Motivated by \citet{Oppen}, the velocity dispersion is given by: 
\begin{equation*}
    \upsilon_{w} = k_{w}\sigma^{\text{1D}}_{\text{DM}} = k_{w} \times 110 \text{ km/sec} \times \biggl( \frac{M_{200c}}{10^{12} M_{\odot}} \biggl) ^{1/3},
\end{equation*}
where $\upsilon_{w}$ is the wind velocity and $\sigma^{\text{1D}}_{\text{DM}}$ the local, one-dimensional, velocity dispersion of the host DM halo. Note that $\sigma^{\text{1D}}_{\text{DM}}\approx$ 110 km/sec, around star - forming gas in MW-like haloes at z = 0 and $M_{200c}$ is the mass enclosed in a sphere whose mean density is 200 times the critical density of the Universe at the time the halo is considered.

It is therefore uncertain if these $\Lambda$CDM fits could be considered as major predictions of the model. \citet[their fig. 14a]{10.1093/mnras/stx3112} demonstrate the suppressed TNG100 blue-line prediction compared to the Illustris \citep{Vogel} red one at the low mass end ($M_{\star}<10^{10}M_{\odot}$), at $z = 0$ was a result of this re-calibration of the model. The well-known outcome that Illustris produced too high stellar mass fractions at $z <1$  \citep{2014Natur.509..177V} was subsequently tuned and fixed in the TNG project. \citet{10.1093/mnras/stx2656} report that a new implementation of stellar feedback via BH and via galactic wind, assist the $\Lambda$CDM model to assemble the right amount of galaxies in a volume of 100 Mpc. One could identify the disagreement with the observational data at higher redshifts too (at $z=1$ , middle row, left figure of their fig. 14) and if the galactic wind effect is not included in the feedback processes. As these authors  \citep{10.1093/mnras/stx3112} declare "better statistical sampling" eventually achieve to suppress the amplitude of the galaxy stellar mass function at the low mass end ($M_{\star}<10^{11}M_{\odot}$ for TNG project).

The EAGLE simulations are also re-calibrated to fit the MS relation. Actually, the prefix "RefL" refers to the recalibration of the simulations to match the observational data with high resolution. \citet[section 2.1]{10.1093/mnras/stu2058} and \citet[their section 2.1]{10.1093/mnras/stx2787} document that for a box of 100 cMpc side, the simulations are tailored to match the galaxy stellar mass function for a large range of stellar masses. These parameters are later used for the smaller box sizes (25cMpc - EAGLE RefL0025N0752 simulation). In Fig. \ref{ms100} and in Fig. \ref{ms50}, one can also see that both EAGLE simulations agree to each other, as well as TNG ones (TNG100-1 to TNG50-1). 

\section{The Hubble tension} \label{TheHubbleTension} 

\subsection{The Hubble tension and the local void solution}
The most notable enigma of $\Lambda$CDM cosmology nowadays is the Hubble tension, or the Hubble constant tension. It refers to the discrepancy of the measurements of the Hubble constant $H_0$ obtained by early- and late-time Universe probes. While the early-time surveys, based mainly on the \textit{Planck} satellite  \citep{2014A&A...571A...1P, planck_overview}, mapping the CMB anisotropies, provide a value of $H_0 = 67.4 \pm 0.5$ km/s/Mpc, the late-time experiments, using the distance ladder method, give a higher value of $H_0 = 73.17$ km/s/Mpc, as reported by the SH0ES team \citep{Riess_2022, 2024ApJ...973...30B}. The estimated $\approx 5\sigma$ tension signifies the importance of this disagreement in observational and theoretical cosmology, questioning the flat $\Lambda$CDM model and the systematics of the surveys.

The early Universe $H_0$ assessment is additionally accommodated by the BAO probes of the expansion history. Surveys like the BOSS \citep{2016AJ....151...44D}, or SDSS \citep{2021PhRvD.103h3533A} and DESI \citep{2025JCAP...02..021A} estimate the characteristic scale that the sound waves leave on the large-scale distribution of galaxies, obtaining a value of $H_0 = 68.5 \pm 0.6$ km/s/Mpc, supporting the \textit{Planck}-inferred calculation.

On the other hand, the late Universe measurements employ various different methods to determine the Hubble constant value by quantifying the local redshift gradient. Besides the standardize SNIa data observed by the HST, the Cepheid variable stars serve as standard candles, observed by the JWST. The Pantheon+ set of SNIa \citep{Scolnic_2022, 2022ApJ...938..110B} measured a value of $H_0 = 73.6 \pm 1.1$ km/s/Mpc, agreeing with the local result but strengthening the tension.

\begin{figure}
    \centering
    \includegraphics[scale=0.38]{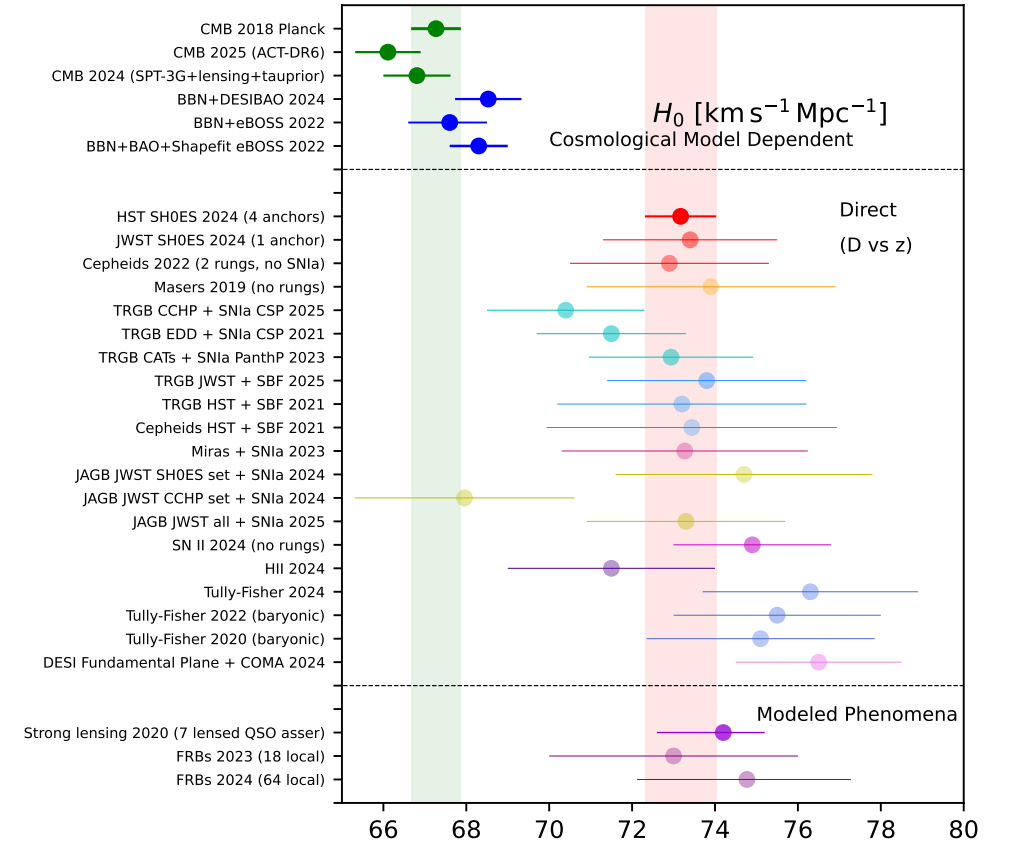}
    \caption{Summary of $H_0$ estimates from different cosmological probes with error bars smaller than 3.5 km/s/Mpc.
    Taken from The CosmoVerse White paper: Addressing observational tensions in cosmology with systematics and fundamental physics \citep{CosmoVerse}.}
    \label{WhitePaper1}
\end{figure}

There are numerous other techniques to assess the expansion rate of the Universe, especially relying on late-time physics, like the TRGB, the SNII, the surface brightness fluctuation or the BTFR, based on different stellar populations and calibration methods. An exhaustive review of the subject can be found on the CosmoVerse Paper \citep{CosmoVerse}, while \citet{2022PhR...984....1S} reviewed the proposed solutions. Fig. \ref{WhitePaper1} illustrates the distinct measurements of the Hubble constant value.

Parallely, a local under-density between 40 and 300 Mpc around the Local Group is observed \citep{2023arXiv231001473K}, the so-called KBC void \citep{Keenan_2013}. Plenty of inspections along the electromagnetic spectrum have revealed an increase in the luminosity density at a radius of 300 Mpc around the MW, following by a flattening thereafter at what is presumably the cosmic mean density, by \citet{1990MNRAS.247P...1M, 2022MNRAS.511.5742W} in optical, by \citet{10.1093/mnras/stt2024} in the near infrared wavelength, by \citet{2013A&A...555A.117R} in radio frequencies and by \citet{2015A&A...574A..26B,2020A&A...633A..19B} in X-rays. In the $\Lambda$CDM scenario, such a huge void is a paradox, given its assumed convergence to homogeneity at $\approx$ 250 Mpc \citep{2016A&A...594A..16P}.

%Although, there are counter claims that huge empty regions can arise naturally in $\Lambda$CDM \citep{2014MNRAS.441..933X}, a

A different approach is suggested by \citet{Haslbauer_KBC}. These authors proposed the $\nu$HDM cosmological model (see also Chapter \ref{nuHDMcosmogony}) which accounts and resolves the Hubble tension. The $\nu$HDM cosmology is based on the MOND gravity and a hypothetical sterile neutrino of rest-mass 11 eV/c\textsuperscript{2}. They developed a semi-analytic approach for the density and the velocity perturbations, extrapolating the MOND law to Gpc scales but keeping the expansion history identical to the standard FLRW metric. Modeling the void as a Maxwell-Boltzmann profile and taking into account the external field effect of MOND, they managed to explain the increased $H_0$, inferred with redshift gradients \citep{Sergij1}, as well as predicting the large peculiar velocities, induced potentially by this immense under-density. \citet{Sergij2} extended the study, managing to match the bulk flow observations by \citet{2023MNRAS.524.1885W}. Importantly, the faster rising predicted bulk flow is  described with a more efficient structure formation growth than in the $\Lambda$CDM paradigm on scales of tens to hundreds of Mpc. The Hubble tension emerges because an observer in the void sees the SNIa in their host galaxies falling to the sides of the void therewith appearing as a locally faster expanding Universe.

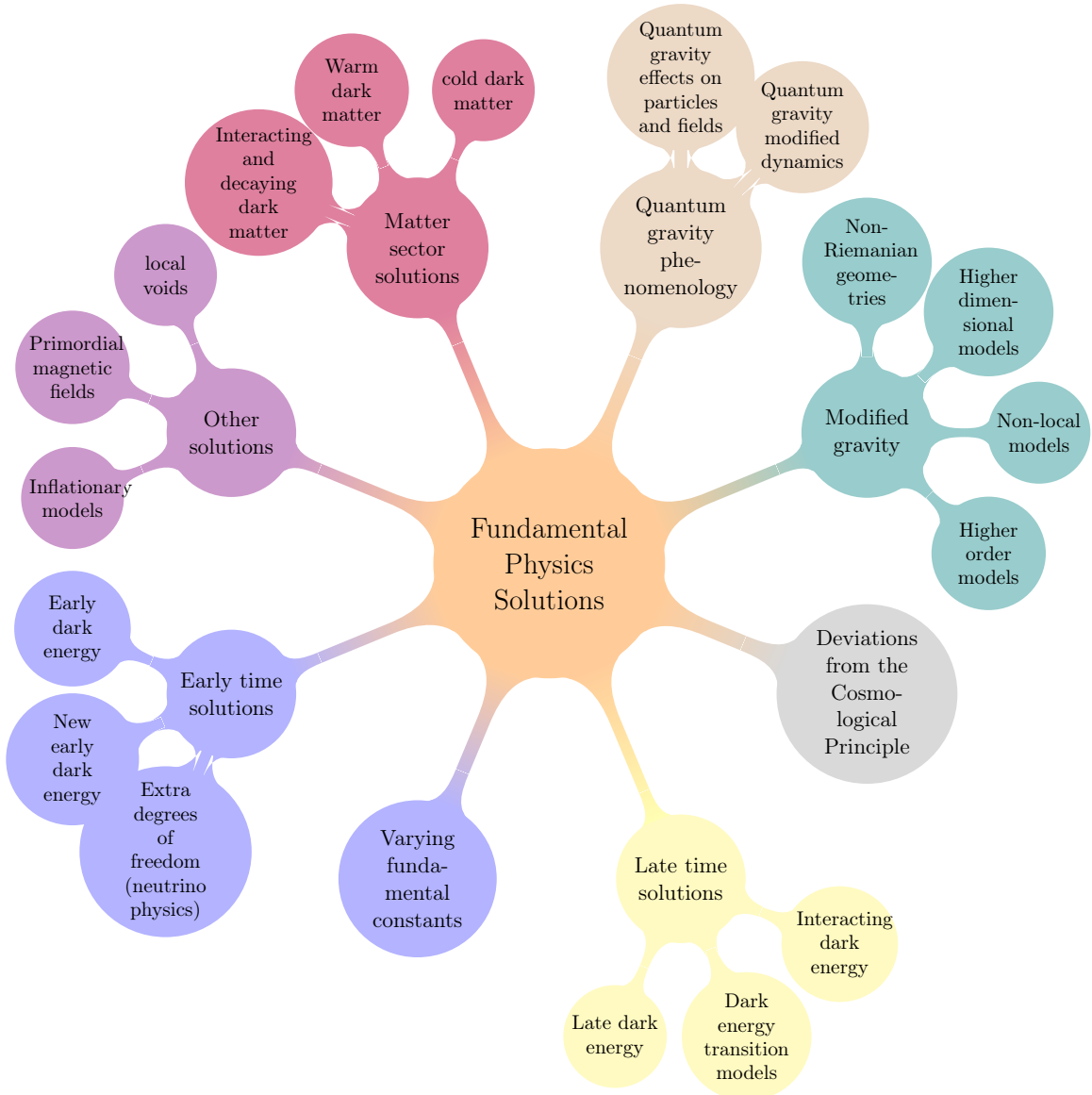
\begin{figure}
\scalebox{0.8}{
\begin{tikzpicture}[mindmap, grow cyclic, every node/.style=concept, concept color=orange!40, 
	level 1/.append style={level distance=6cm,sibling angle=360/8},
	level 2/.append style={level distance=3cm,sibling angle=45}]

\node{Fundamental Physics Solutions}
child[concept color=blue!30, minimum size=1cm]{ node {Early time solutions}
	child { node {Early dark energy}}
	child { node {New early dark energy}}
	child { node {Extra degrees of freedom (neutrino physics)}}
}
child[concept color=blue!30, minimum size=1cm]{ node {Varying fundamental constants}
}
child[concept color=yellow!30, minimum size=1cm]{ node {Late time solutions}
	child { node {Late dark energy}}
	child { node {Dark energy transition models}}
	child { node {Interacting dark energy}}
}
child[concept color=gray!30, minimum size=1cm]{ node {Deviations from the Cosmological Principle}
}
child[concept color=teal!40, minimum size=1cm]{ node {Modified gravity}
	child { node {Higher order models}}
	child { node {Non-local models}}
	child { node {Higher dimensional models}}
	child { node {Non-Riemanian geometries}}
}
child[concept color=brown!30, minimum size=1cm]{ node {Quantum gravity phenomenology}
	child { node {Quantum gravity modified dynamics}}
	child { node {Quantum gravity effects on particles and fields}}
}
child[concept color=purple!50, minimum size=1cm]{ node {Matter sector solutions}
	child { node {cold dark matter}}
	child { node {Warm dark matter}}
	child { node {Interacting and decaying dark matter}}
}
child [concept color=violet!40, minimum size=1cm]{ node {Other solutions}
            child { node {local voids} }
            child { node {Primordial magnetic fields} }
            child { node {Inflationary models} }
            };
\end{tikzpicture}}
\caption{Fundamental physics provides a number of potential solutions to address the challenge of tensions in cosmology. Taken from The CosmoVerse White Paper: Addressing observational tensions in cosmology with systematics and fundamental physics \citep{CosmoVerse}. Local void solutions are highlighted towards the lower right.}
\label{H0solutions}
\end{figure}

Nevertheless, there are plenty of other suggested solutions for the Hubble tension, each one providing a respective estimated value, as shown in Fig. \ref{WhitePaper1}. An illustrative categorization of the proposed $H_0$ explanations is provided in Fig. \ref{H0solutions}. Interestingly, the model-dependent proposed explanations yield to a low value of $\approx 67 $ km/s/Mpc, while calculations relied on surveys of the local Universe exhibit a bias upwards, to a value $\approx 74$ km/s/Mpc. In the next two sections, I will summarize my co-authored contributions on this subject, firstly why a local void scenario is reasonably justified (Section \ref{banik_paper}) and secondly why the G step model (Section \ref{desmond_paper}) could not be a viable interpretation of the Hubble tension.

\subsection{Cosmology without the CMB} \label{banik_paper}
This section recaps the argument for the late-time or the local solution of the Hubble tension, highlighted by \citet{Banik_NS}. In the $\Lambda$CDM framework, the $H_0$ and $\Omega_M$ parameters are obtained with high-precision, based on the CMB observation, by the \textit{Planck} satellite. Alternatively, they are inferred through the redshift gradients, since $z' \equiv \frac{dz}{dr} = \frac{H_0}{c}$, where $c$ is the speed of light in vacuum, by local Universe observations. This quantifies how fast the redshift $z$ increases with respect to the distance $r$ of the object. Selecting three different observational constraints which do not rely on the CMB, the aim is to evaluate the $H_0 - \Omega_M$ relation and eventually compare it with the \textit{Planck}-inferred value.

\begin{figure} 
    \centering
    \includegraphics[width=0.76\linewidth]{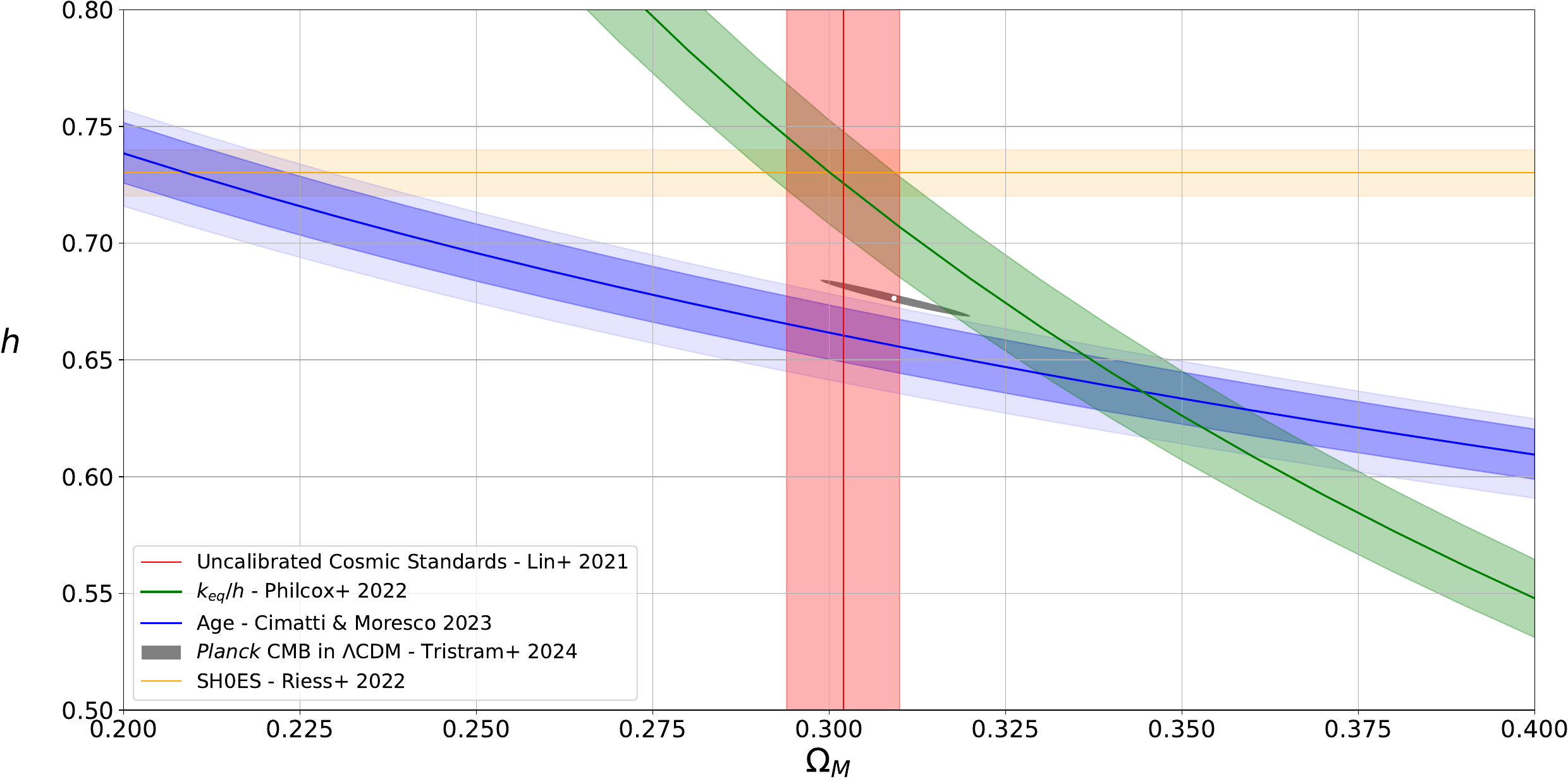}
    \caption{Same as \citet[fig. 1]{Banik_NS} for 1$\sigma$ constraints on the matter density parameter $\Omega_M$ and the Hubble constant $h \equiv H_0 /100$.}
    \label{Banik_NS_fig}
\end{figure}

The first constraint is provided by \citet{Lin_2021}. The so-called Uncalibrated Cosmic Standards (UCS) treat the absolute magnitude of SNIa and the comoving size $r_d$ of the BAO ruler as free parameters, yielding $\Omega_M = 0.302 \pm 0.008$ (red band of Fig. \ref{Banik_NS_fig}). The second constraint is coming from the characteristic peak $k_{\text{eq}}$ of the matter power spectrum, considering that matter is the dominant component of the Universe throughout its expansion history, except its early radiation-phase and the late-time take over of DE. Combining the measurement by \citet{2022PhRvD.106f3530P} of the $k_{\text{eq}}/h \left( 1.64\pm 0.05 \right) \times 10^{-2}$/Mpc, together with the relation $k_{\text{eq}} = \sqrt{2 \Omega_M H^2_0 z_{\text{eq}}}$ by \citet{Eisenstein_1998}, the most likely values are $h= 0.648$ and $\Omega_M = 0.338$ (green band of Fig. \ref{Banik_NS_fig}). Last, taking into account empirical compilations of the ages of stars, globular clusters and ultra faint dwarf galaxies in the Galactic disc and halo, based on \citet{Cimatti_2023}, the blue band puts an additional limitations on the relation $H_0 - \Omega_M$. As a consequence, a very confined region is patterned in Fig. \ref{Banik_NS_fig}. Surprisingly, the \textit{Planck} CMB value falls exactly within this range, possibly hinting that the expansion history is correctly modeled. This is a suggestion that the anomalous high local Universe value of $H_0 \approx 73.17 \pm 0.86$ km/s/Mpc is caused by outflows from a local KBC-type under-dense region \citep{Keenan_2013}, inflating the redshift gradients. 

\subsection{G step model}\label{desmond_paper}
This section argues against the G step model (GSM) \citep{PhysRevD.104.L021303, 2021PhRvD.104l3511P}, which is a potential resolution of the Hubble tension, through the sudden decrease of the Gravitational constant G around 130 Myr ago. The full length of this work is published by \citet{Banik_desmond}. In the GSM scenario, if G has indeed decreased suddenly 130 Myr ago, this would have caused more distant SNIa to look intrinsically brighter, compared to those in host galaxies with Cepheid distances (because presumably $G$ in the past was larger than today). Subsequently, their redshift would have increased more slowly with distance and this would potentially explain the discrepancy with the \textit{Planck} value. However, that also means that the Sun would have suddenly decreased in luminosity, causing an new ice age era on Earth. The Sun would have exhausted its fuel much faster and additionally solar-type stars would look younger \citep{2012PhRvD..85l3006D}.

\begin{figure} 
    \centering
    \includegraphics[width=0.76\linewidth]{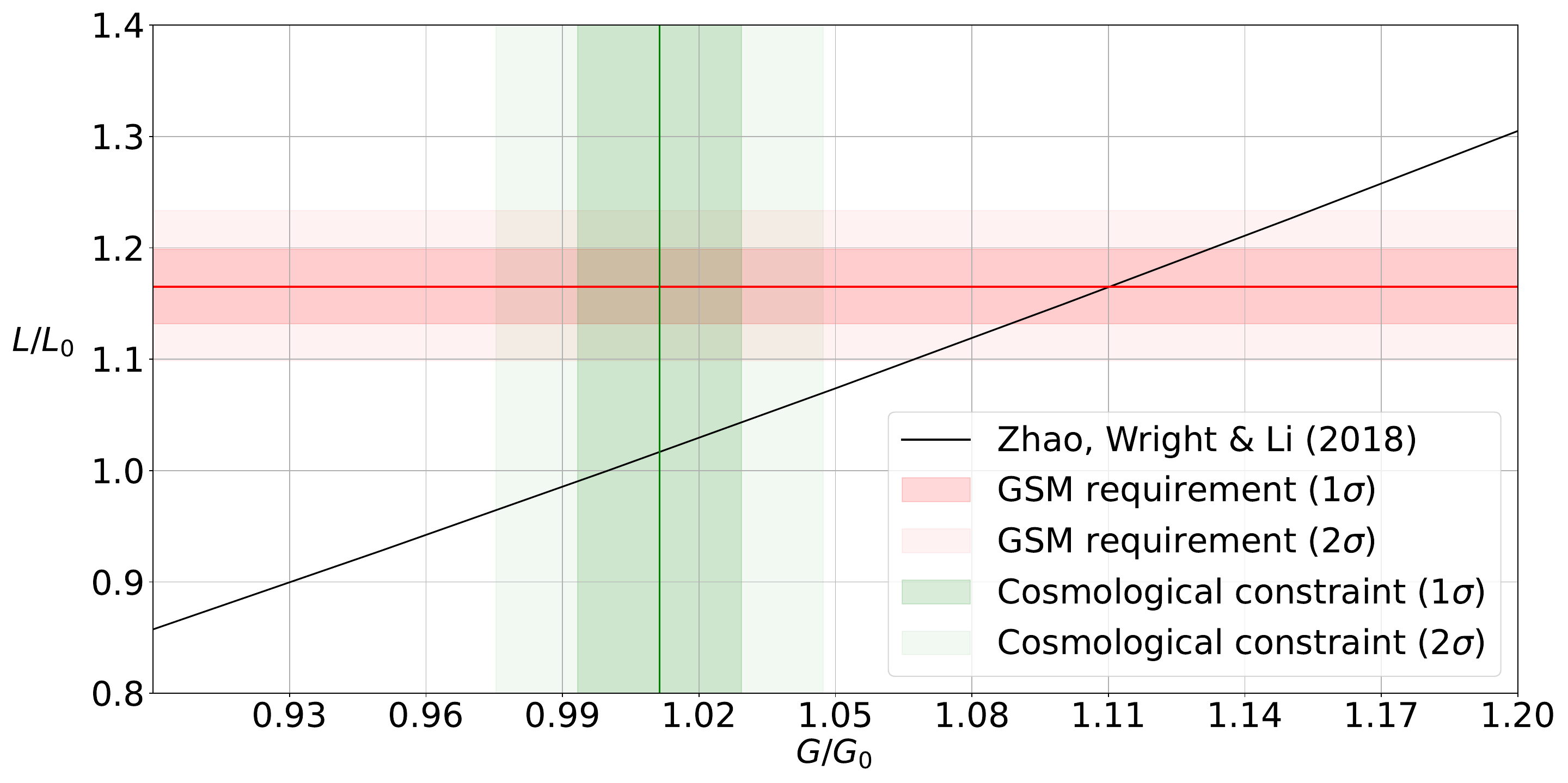}
    \caption{Same as \citet[fig. 1]{Banik_desmond} comparing the constraints on the GSM model.}
    \label{Banik_HD_fig}
\end{figure}

Nevertheless, there is no sign of a snowball phenomenon on Earth in the last 500 Myr. Most likely, they all took place $\approx$ 600 Myr ago. Even a small drop of Earth's temperature $T_{\oplus}$ can trigger planetary glaciation \citep{2020RSPSA.47600303A}. A decrease in $G$ could also expand the Earth' orbital radius, further decreasing the solar insolation. Based on the Earth's circular motion, a decrease of $G$ would result in an increase of the orbital radius $r$, in order to conserve angular momentum ($\sqrt{G M_{\odot}r}$). Furthermore, a decrease of $G$ today would imply that the luminosity of the Sun would have been $\approx$ 30\% higher in the past, since solar luminosity scales as $L \propto G^{5.6}$ with $G$ \citep{1996A&A...312..345D}. But, this would have made the Sun much more evolved by now, which is in contradiction with an estimate based on the dating of old meteorite samples using radioactivity \citep{2012Sci...338..651C}.

Fig. \ref{Banik_HD_fig} encapsulates the GSM model and its constraints. In order to explain the Hubble tension, \citet{2022ApJ...938...36R} require $\frac{L}{L_0} =\left( \frac{cz'}{H^{\textit{Planck}}_0} \right)^2$ with $H_0 \approx 73.0 \pm 1.0$ km/s/Mpc, meaning that distance of SNIa on the Hubble flow would scale as $\sqrt{L/L_0}$. So, the red line is the required value of $L$ in units of $L_0$. The vertical green band shows the cosmological constraint on $G$ using the latest CMB and baryon acoustic oscillation data \citep{2025A&A...697A.109L}. The black line is the $L\propto G^{1.46}$ scaling by \citet{PhysRevD.100.043537}, who fitted the \citet{2018PhRvD..97h3505W} SNIa data. The required enhancement to $G$ is in tension with cosmological constraints, even though the aim of the GSM is to preserve the Planck cosmology.
\section{The \texorpdfstring{$\nu$}{nu}HDM cosmological model and its variant opt-\texorpdfstring{$\nu$}{nu}HDM}
\label{nuHDMcosmogony}
%----------------------------------------------------------------------------------------
%	SECTION 1
%----------------------------------------------------------------------------------------
\subsection{Construction and history of the \texorpdfstring{$\nu$}{nu}HDM cosmology}
The $\nu$HDM cosmological model is a rather conservative idea. Fundamental postulates remain unaltered compared to the standard model of cosmology, yet crucial modifications were performed, in order to develop an alternative path towards the understanding of the emergence of galaxies. The combination of a HDM component and Milgromian gravity classifies the $\nu$HDM scenario as a \textit{hybrid} cosmological model. While plausible skepticism \citep{2015CaJPh..93..250M} diminishes any potential consolidation of MOND with DM, this could be served as a modest and numerically feasible starting point to build up a novel theory of the formation and evolution of galaxies, withdrawing the speculation of un-detected (or maybe un-detectable?) particles.

I will encapsulate the history of the $\nu$HDM model and the physics of its cosmogony, as published in \citet{2023MNRAS.523..453W}, reproducing  and enriching it with higher resolution and eventually comparing previous works with the optimized version of the model, the so-called opt-$\nu$HDM framework. The optimization of the $\nu$HDM model, through MCMC simulations, was carried out by \citet{2025MNRAS.541..784S}. 

Physically, the $\nu$HDM model begins with a BB, followed by the cosmological inflation, as postulated in the $\Lambda$CDM scenario. What differs is primarily the DM component, instead of cold dark collisionless matter, a hot light sterile neutrino is assumed, replacing the $\Omega_{\text{CDM}}$ with $\Omega_{\nu}$. The sterile neutrino is on top of the already existing 3 active ones. Essentially, the HDM contribution drives the structure formation until moderate redshifts of $z\approx 50$ \citep{Haslbauer_KBC}, when MOND gravity takes over in the simulations, assisting on the growth of the gravitationally bound systems. At late times, the DE becomes important at $z \approx 0.4$, when the Universe is accelerated to an exponential expansion. Most of the basic $\Lambda$CDM ingredients stay untouched.

The sterile neutrino has long lived as a DM candidate. In the $\Lambda$CDM framework, keV-scale sterile neutrino has been considered in the past by \citet{Bulbul_2014, 2017arXiv170208430M}, but incompatibilities can arise between constraints by detection experiments and structure formation scenarios \citep{2023APS..APRG13003A}. In the $\nu$HDM model, the sterile neutrino is a neutral fermion, with no chemical potential, with a rest-mass of about 11 eV/c\textsuperscript{2}. It was presumably relativistic until recombination, but not reheated after decoupling from the thermal bath. It contributed to the radiation in the early Universe, but behaved as matter in the subsequent era. According to \citet{Haslbauer_KBC}, its free-streaming length today is $\lambda_{\text{fs}} \approx$ 3.5 Mpc, which is about the size of a galaxy cluster. 

Various suggestions have been discussed in the literature concerning the production mechanism of the sterile neutrino. Neutrino oscillations in the early Universe \citep{Langacker:1988fp} is the most effective process, or a non-resonant transition, resulted from a primordial lepton asymmetry \citep{1994PhRvL..72...17D, PhysRevD.88.043502}. Alternatively, a singlet scalar particle interacting with the Higgs boson \citep{König_2016} is another possibility. 

Moreover, there are many particle physics experiments trying to detect (sterile) neutrinos. Recently, NOVA \citep{PhysRevLett.134.081804}, assuming a model of 3 active neutrinos and a sterile one (like in the $\nu$HDM variant models), found no evidence of the existence of a sterile component in neutrino oscillations. Nevertheless in the past, LSDN \citep{PhysRevD.64.112007} recommended that oscillations may occur in the energy range of $0.2-10$ eV\textsuperscript{2}/c\textsuperscript{4}, which potentially can induce the appearance of a sterile light neutrino. KATRIN \citep{Mohanty:2023wl} has put further constraints on the mass of a sterile neutrino to a lower limit of $0.8$ eV/c\textsuperscript{2}. Nevertheless, due to the variety of the involved systematics and the different assumed models, these experimental works are not conclusive (Samaras and Kroupa, in preparation).  

Last, a wide range of DM models has been studied the past 40 years. Generally, CDM models tend to form structures earlier than WDM/HDM or mixed DM \citep{1993ApJ...416....1K} models (combination of warm, hot and cold DM). CDM, due to its original design, builds up its cosmic evolution through mergers, in a bottom-up scenario. On the other hand, in HDM, since particles have a higher velocity dispersion, the evolution is delayed, vast over-densities form firstly, seeds are being erased during the evolution because of the free-streaming of the HDM component and structures are built up in a top-down scenario, after fragmentation processes. 

\definechronoevent{MyEvent}[textstyle=\it, barre=false, mark=false]
\setupchronology{startyear=1983,color=darkblue,stopdate=false}
\setupchronoperiode{color=green}
\setupchronoevent{textstyle=\it, markdepth=4cm}
\setupchronograduation[event]{markdepth=2cm}
\startchronology
    \chronoevent[markdepth=-1cm]{1983}{Milgrom MOND}
    \chronoevent[markdepth=1cm]{1984}{Bekenstein \& Milgrom}
    \chronoevent[markdepth=-1cm]{1998}{Sanders}
    \chronoevent[markdepth=2.8cm]{2004}{Bekenstein TeVeS}
    \chronoevent[markdepth=-1cm]{2010}{Milgrom QUMOND}
    \chronoevent[markdepth=1.9cm]{2013}{$\nu$HDM Katz et al.}
    \chronoevent[markdepth=0.3cm]{2011}{$\nu$HDM Angus \& Diaferio}
    \chronoevent[markdepth=4.5cm]{2024}{Opt-$\nu$HDM Samaras et al.}
    \chronoevent[markdepth=2.8cm]{2023}{Hydro $\nu$HDM Wittenburg et al.}
    \chronoevent[mark=true, markdepth=-2.4cm]{2025}{opt-$\nu$HDM sims in prep.}
    \chronoevent[mark=true, markdepth=-1cm]{2025}{This Ph.D. work}
\stopchronology

Numerically, volume effects are also present in the simulations, especially in HDM models, since perturbations depend on the boundary conditions (Samaras and Kroupa, in prep). Furthermore, all the aforementioned models use as IC the CMB power spectrum published by \textit{Planck} collaboration \citep{2014A&A...571A...1P}, which is $\Lambda$CDM-contaminated. Regardless, as \citet{Haslbauer_KBC} showed, the $\nu$HDM will have the same expansion history as the standard $\Lambda$CDM model, following the FLRW metric. Therefore the background evolution will be unaltered, but the gravitational collapse will adopt the MOND law. Interestingly, on the opt-$\nu$HDM model \citep{2025MNRAS.541..784S}, the cosmological parameters are significantly altered ($H_0 \approx55.6$ km/s/Mpc and $\Omega_M \approx 0.49$), by optimizing for the \textit{Planck} CMB. Therefore, the evolution of the background will be evolved according to the lowered $H_0$ value and the big $\Omega_M$ content. I will present further down a set of preliminary results of the opt-$\nu$HDM simulations, contrasting them with the "classic" $\nu$HDM progenitor cosmology. 

Historically, the $\nu$HDM scenario emerged upon MOND struggles on the galaxy clusters scale. When \citet{Sanders_1999} compared the observed mass of galaxy clusters to the Newtonian dynamical mass and to the MONDian dynamical one, he realized that MOND can reduce the discrepancy from the observations, compared to the Newtonian predictions, but a factor of 2 remains. The need of "unseen" matter was assigned firstly to an active neutrino, which was later theoretically configured to be sterile of $\approx$ 2 eV/c\textsuperscript{2} rest-mass. \citet{2010MNRAS.402..395A} used a MCMC approach to find best-fit values for the neutrino content with respect to the WMAP CMB power spectrum \citep{2009ApJS..180..306D} and $\nu$HDM was born. \citet{2013ApJ...772...10K} had investigated the bulk flows on the $\nu$HDM model, discovering the problem of the emergence of "pink elephants", that is, galaxy clusters which are unphysically massive, of  about $M _{200} \approx10^{17} M_{\odot}$ ($M_{200}$ is the mass enclosed by the viral radius where the density reaches 200 times the critical density).  A decade later, \citet{2023MNRAS.523..453W} conducted hydrodynamical simulations, confirming the previous results by \citet{2010MNRAS.402..395A} and by \citet{2013ApJ...772...10K}, demonstrating also a late onset of structure formation ($z\approx4$). 

Particularly on the galaxy scales, Milgromian Dynamics has been proven remarkably successful in predicting scaling relations for spiral, elliptical and early-type galaxies. Namely, the BTFR, the FB and the RAR \citep{2017ApJ...836..152L} relations are considered to be major achievements \citep{2012LRR....15...10F} of MOND, which was formulated over 40 years ago by \citet{1983ApJ...270..365M}. Progress on the simulation front has been also possible by \citet{2010MNRAS.403..886M}, who developed QUMOND, which is a quasi-linear form of the theory in order to effectively solve the MOND equations of motions. Numerical simulations in quasilinear Modified Newtonian dynamics have afterwards studied various topics, SF galaxies by \citet{2023MNRAS.519.5128N}, disk galaxies' interactions by \citet{2016arXiv160604942T}, or the asymmetry in the tidal tails of open star clusters by \citet{2022MNRAS.517.3613K, 2024ApJ...970...94K, 2025A&A...693A.127P} and wide binaries \citep{2024MNRAS.533..729H} confirming and extending MOND 's validity to different scales. 

Phenomenologically, MOND explains the missing mass (or gravity) problem in galaxy scales, without invoking exotic particles. \citet{1983ApJ...270..365M} originally formulated it and later \citet{1984ApJ...286....7B} developed a Lagrangian formalism conserving angular momentum for isolated systems. MOND is as an acceleration based theory, introducing a departure from standard Newtonian dynamics, possibly due to quantum vacuum effects \citep{1999PhLA..253..273M}. The transition from Newtonian to Milgromian regime occurs at very weak gravitational fields $g\ll a_0 \approx 1.2 \times 10^{-10}$m/s\textsuperscript{2}. The empirical value of $a_0$ has remained stable for last decades \citep{1991MNRAS.249..523B}. 

Mathematically, the generalized MONDian version of the  Poisson equation is defined as:
\begin{equation} \label{mond_equation}
    \nabla \cdot \Biggl[ \mu \Big( \frac{|\nabla \Phi|}{a_0} \Big) \nabla \Phi \big( \textbf{r}\big) \Biggr] = 4\pi G \rho \left( \textbf{r}\right)\equiv  \nabla \cdot  \nabla \Phi_N \left( \textbf{r}  \right),
\end{equation}
where $\Phi$ is the gravitational potential, $N$ subscript denotes the Newtonian case, \textbf{r} is the position 3D vector and $\mu$ is the interpolating function with dimensionless argument being: $\mathcal{y}_{int} \equiv \frac{|\nabla \Phi|}{a_0}$, which is the gravitational field in the units of $a_0$. The transition regime between Milgromian and Newtonian is happening through the interpolating function:
\begin{equation}
    \mu \left( \mathcal{y}_{int} \right) \rightarrow
    \begin{cases}
        \text{} \mathcal{y}_{int}, & \mbox{if } \text{g} \ll a_0  \left( \mathcal{y}_{int} \ll 1 \right), \\
        1,         & \mbox{if } \text{g}\gg a_0 \left( \mathcal{y}_{int} \gg 1 \right).
    \end{cases}
\end{equation}
In the spherical symmetric case, $\mu g \equiv g_N$. The simulations I will be analyzing further though, are based on a different and less computationally-expensive scheme. Theoretical and observational studies of MOND can be found in the review papers by \citet{2012LRR....15...10F}, \citet{2022Symm...14.1331B} and \citet{2023eppg.confE.231K}.

\subsection{Cosmological simulations on the \texorpdfstring{$\nu$}{nu}HDM framework}
Numerically, $\nu$HDM simulations are not a trivial task. One has to utilize different software to generate IC, perform the hydrodynamical simulations, apply halo finder algorithms to extract gravitationally bound structures and eventually do some post-processing analysis in order to examine the behavior of the models and the characteristics of the products. Most of the programs should be (and they have been) modified in the source codes to accommodate either the MONDian gravity or any demand occurred on the process. Note for example that the Amiga Halo Finder \citep{2009ApJS..182..608K} only computes Newtonian quantities, thus if one wants to study the MONDian virial masses or the escape velocities, then limitations come up and further developments are vital duties for the future. 

\begin{figure*} 
    \includegraphics[width=\linewidth]{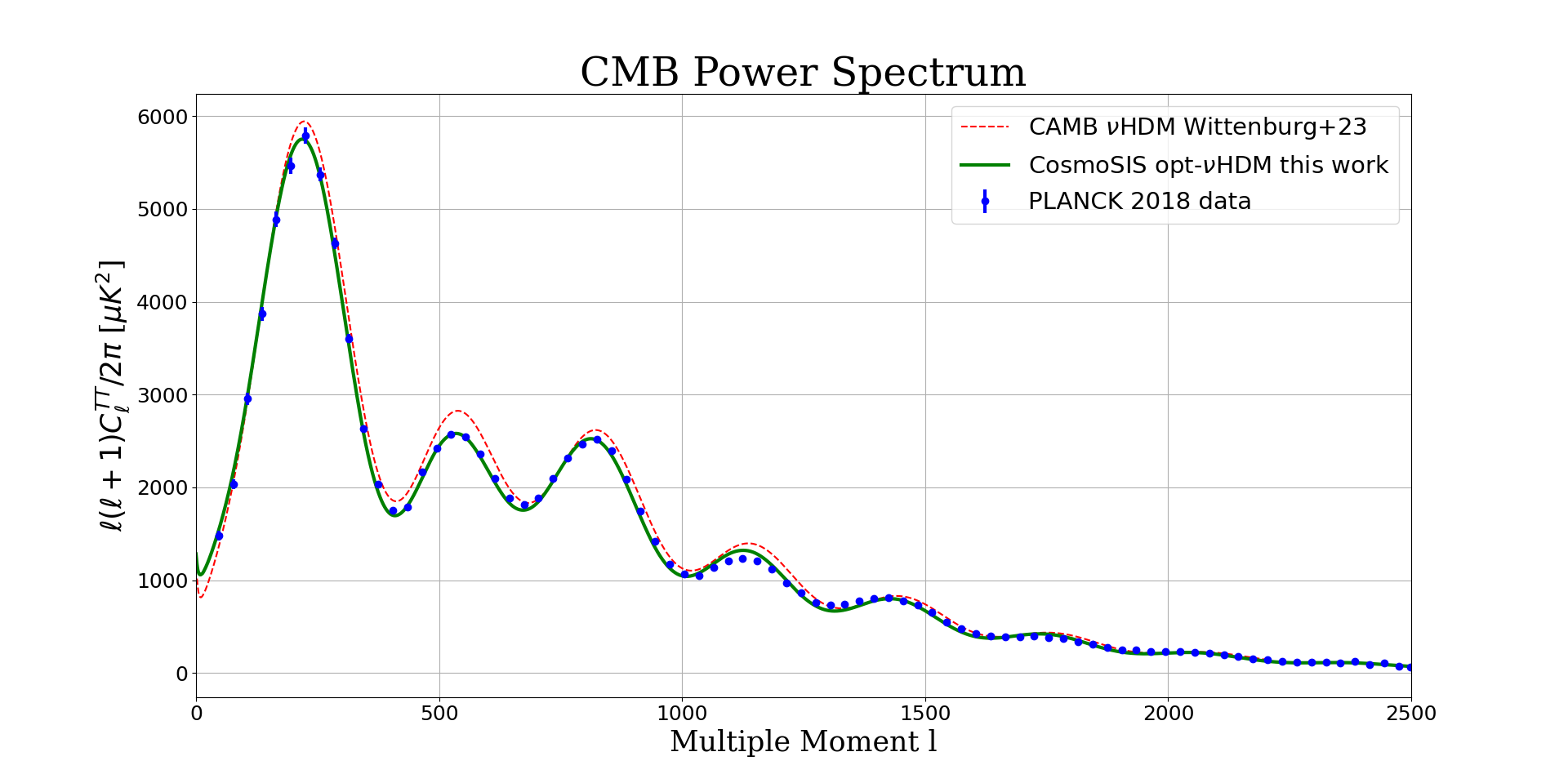}
\caption{The temperature fluctuations power spectrum of the CMB for the 2 different cosmological $\nu$HDM models and the \textit{Planck} 2018 data. Credit to \citet{2025MNRAS.541..784S}.}    
\label{cmb}
\end{figure*}

\subsubsection{Initial Conditions}
Generally, the setup of the IC and the hydro-dynamical simulations summarizes the idea of $\nu$HDM being a conservative idea. All but one (the CDM hypothesis) are kept in place, with a BB, cosmological inflation, a collisionless component to drive the 3rd acoustic peak of the CMB and DE to accelerate the Universe at late times. 

\begin{figure*} 
    \includegraphics[width=\linewidth]{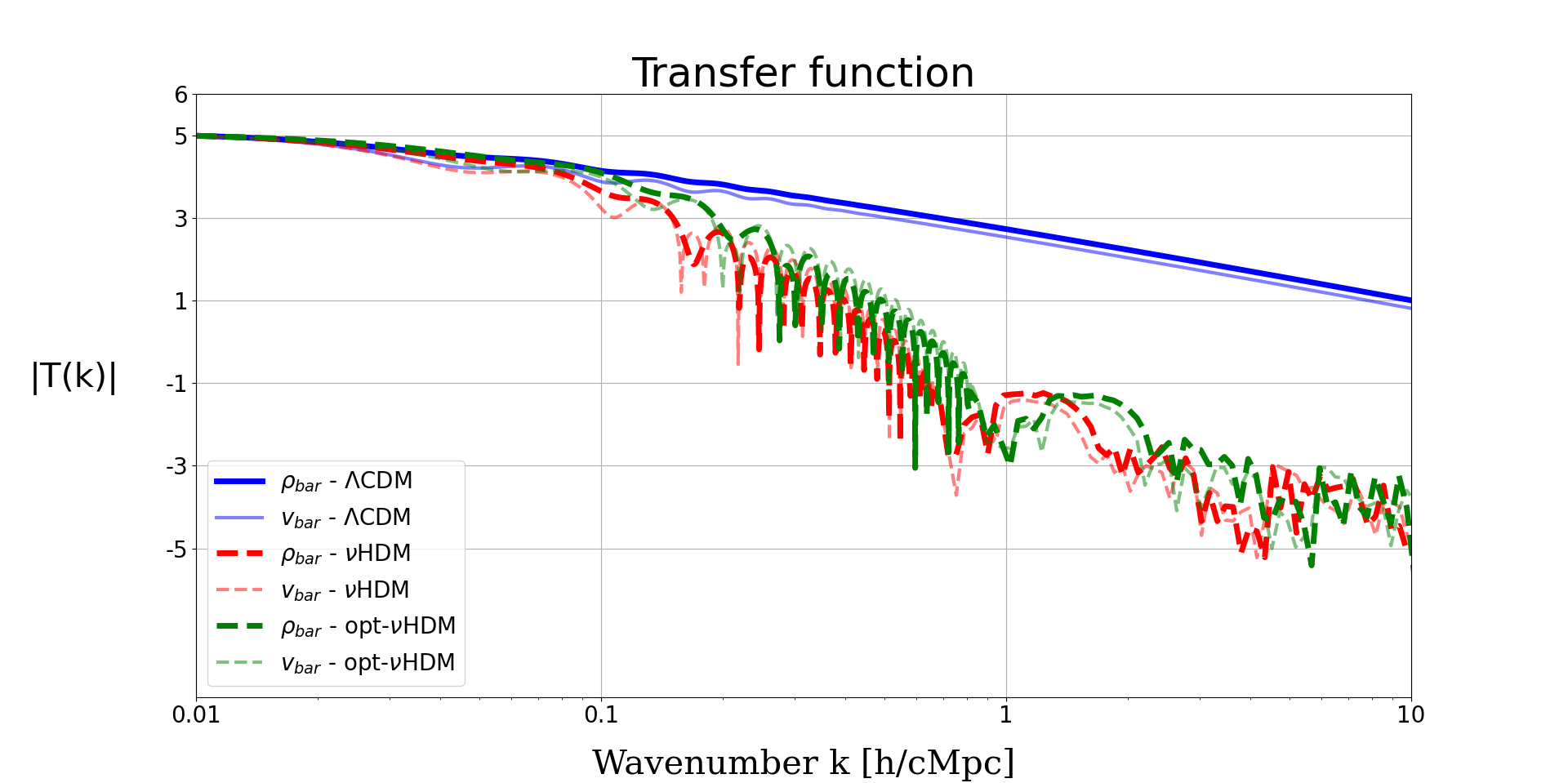}
    \caption{The density and velocity transfer functions of baryons only, for the 3 studied models by CAMB, at $z=199.0$, which is the beginning of the simulations. The solid lines are the baryonic density and the lighter ones are the baryonic velocity.}
    \label{transfer_velocity}
\end{figure*}

As \citet{2023MNRAS.523..453W} described, the IC of the $\nu$HDM were generated by CAMB \citep{Lewis_2000} and subsequently by MUSIC \citep{10.1111/j.1365-2966.2011.18820.x}. In CAMB, the $\Omega_{\text{CDM}}$ has been eliminated to 0.0001, since the program crashes at exactly 0.0 value. The neutrino contribution has been assigned the equivalent matter content with $\Omega_{\nu} = 0.26$, which was before in the $\Omega_{\text{CDM}}$. The rest of the cosmological parameters remain identical to the $\Lambda$CDM ones, except the total number of effective degrees of freedom of the neutrinos $N_{\text{eff}}$ \citep[section 2.2]{2023MNRAS.523..453W}, which was increased to $N_{\text{eff}} = 4.044$. This is the main parameter capturing the additional effects of the $\nu$HDM hypothesis, accommodating active and sterile neutrinos. Importantly, the sterile neutrino is set, in CAMB by hand, to have two eigenstates, with different mass degeneracies and mass fractions. Three active massless neutrinos are still assumed to exist together with the light but non-zero rest-mass, sterile, right-handed one. CAMB outputs the temperature fluctuations of the CMB (Fig. \ref{cmb}) as predicted by the models, as well as the transfer functions (Fig. \ref{transfer_velocity}) of all the species and the total matter power spectrum \citep[their fig. 4]{2025MNRAS.541..784S}, solving the Boltzmann equations up to the desired redshift. The simulations are set to begin at $z=199.0$, so that the perturbations have enough time to grow in the box (following the Newtonian regime) and since accelerations are expected to reach the MONDian regime much later, at $z\approx50$ \citep[section 3.1.3]{Haslbauer_KBC}. The cosmological parameters used on these simulations are summarized in Table. \ref{cosmo_params}.

The fit to the \textit{Planck} CMB power spectrum is presented in Fig. \ref{cmb}. I have used consistently the green color for the opt-$\nu$HDM model, the red one for the $\nu$HDM model of \citet{2023MNRAS.523..453W} and the blue one for the $\Lambda$CDM model or the \textit{Planck} data. The opt-$\nu$HDM model fits the data extremely well, except the 4th peak of the damping tail. These scales are very small; MONDian effects may be important due to the CMB lensing of photons, but GR extensions connected with MOND are still under investigation \citep{2021PhRvL.127p1302S}. However, Samaras and Kroupa (in prep.) may be able to estimate the matter power spectrum of the opt-$\nu$HDM simulations at $z=0$ and calculate the difference with the Newtonian one, estimating how much power the Milgromian potential should provide in order to obtain a reasonable distribution of groups of galaxies.

\citet{2023MNRAS.523..453W} and \citet{2025MNRAS.541..784S} focused on the transfer function of the density of the most dominant species of the models, since this is the essential contributor driving the structure formation (CDM for $\Lambda$CDM and sterile neutrino for $\nu$HDM variants). In this work, I also investigate the velocity components, presenting (Fig. \ref{transfer_velocity}) the density (lighter lines) and the velocity transfer functions (solid lines) of the baryons only of each of the cosmological models at $z=199.0$ (which is the beginning of the hydro-simulations). The baryonic component obeys the gravitational potential prescribed by the leading DM element of the models. The velocity component responds to the density one, systematically demonstrating the same behavior with slightly greater values. Eventually, the momentum of the opt-$\nu$HDM baryonic part is expected to be higher than the $\nu$HDM one, due to the marginally more massive sterile neutrino (See Table \ref{cosmo_params}). The systematic excess of power on the absolute value of the transfer functions (either density or velocity) of the baryons of the opt-$\nu$HDM compared to the $\nu$HDM, at scales of $10^{-1}<k<1$ $h$/cMpc, is subsequently a major result of \citet{2025MNRAS.541..784S} study, distinguishing the optimized model, from its progenitor $\nu$HDM model. The absolute value is necessary to capture the change of sign of the forming cosmic structures, since many of them become under-dense while being over-dense and vice-versa during the process of the comic structure formation via Milgromian gravitational collapse. 

\begin{table*} 
    \centering
    \begin{tabular}{l|l|l|l}
        Model parameters      & $\Lambda$CDM         & $\nu$HDM      & opt-$\nu$HDM \\ \hline \hline
        $\omega_{\text{$\nu$}} = \Omega_{\text{$\nu$}}h^2 $          &  0.00065 \textcolor{red}  & 0.1205  & $0.1307 \pm 0.001$ \\ \hline
        $\omega_{\text{c}} = \Omega_{\text{c}}h^2$   & $0.1200 \pm 0.0012$       &  0.000645     & 0.00001                      \\ \hline
        $\omega_{\text{b}} = \Omega_{\text{b}}h^2$   & $0.02237 \pm 0.00015$     &  0.022383     & 0.022383                     \\ \hline
        $\Omega_{m}$          & $0.3153 \pm 0.0073$  &  0.3          &    $0.495 \pm 0.009$      \\ \hline
        $\Omega_{\Lambda}$    &    0.690194          &  0.690194     &    0.505694               \\ \hline
        $\Omega_{K}$          &    0.0               & 0.0           &    0.0                    \\ \hline
        $H_0$ [km/s/Mpc]      & $67.64 \pm 0.54$     &  67.64        &    $55.64 \pm 0.32$       \\ \hline
        $A_{\text{Pl}}$       & $0.997 \pm 0.031$    &  1.0          &    $1.0\pm 0.0025 $       \\ \hline
        $n_s$                 & $0.9649 \pm 0.0042$  &  0.966        &    $0.872 \pm  0.0037$    \\ \hline
        $\tau$                & $0.0544 \pm 0.0073$  &  0.0543       &    $0.0375\pm 0.0069$     \\ \hline
        $\ln(10^{10}A_s)$     & $3.044 \pm 0.014$    &  3.0448       &    $2.987\pm 0.015$       \\ \hline
    \end{tabular}
    \caption{Cosmological parameters for the 3 studied flat cosmologies. For $\Lambda$CDM, these are the best fits from \citet[their Table 7]{2020A&A...641A...1P}. For $\nu$HDM, the model parameters are the \citet{2023MNRAS.523..453W} values. For the opt-$\nu$HDM, they are taken from \citet{2025MNRAS.541..784S}. Notice the minimal contribution of the CDM component in the $\nu$HDM variant models. Except from the opt-$\nu$HDM $\Omega_M$, those parameters with errors are the ones coming out from the posterior MCMC approach.}
    \label{cosmo_params}
\end{table*}

The main bend on the transfer function plot (Fig. \ref{transfer_velocity}) is the M\'ez\'aros effect, where in the early Universe, perturbations cannot grow in the horizon due to the radiation effects. Most importantly, at intermediate ($10^{-1}<k<1$ $h$/cMpc) and smaller scales (higher $k$, lower wavelength $\lambda$), the acoustic oscillations are present at redshift $z=199.0$. These are caused by the baryonic drag into the potential wells created by the sterile neutrino (or the CDM in $\Lambda$CDM). Although not shown in the transfer function plot, sterile neutrinos are also expected to show oscillations \citep[their fig. 1]{2023MNRAS.523..453W} and \citep[their fig. 5]{2025MNRAS.541..784S}, like baryons, unlike to the CDM component, whose collisionless nature generates no such effects. The redshift-dependent free-streaming length of the sterile neutrino determines this phenomenon. The small-scale inhomogeneities will eventually designate the infall of baryonic gas and dictate the structure formation and ensuing SF activity. However, this very point is still under investigation, since MOND non-linear gravity might perplex the cosmic evolution in a flat expanding Universe \citep{10.1046/j.1365-8711.1998.01459.x}. Last but not least, it's the transfer function which technically enters the MUSIC package \citep{2011MNRAS.415.2101H} as an input to compute density and velocity fields, and are later used for IC of the N-body solver, Phantom of RAMSES \citep{2015CaJPh..93..232L}. 

After the power spectrum is delivered by CAMB, its output is used as an input for MUSIC in order to sample it with a random probability. MUSIC has also been configured to append negative values of the transfer function, which was not its original design, in order to include the effects of the sterile neutrinos. All the aforementioned packages are described in my \href{https://github.com/NickSam121}{github} page for the $\nu$HDM simulations. 

Characteristically, the IC of the opt-$\nu$HDM differ from its precursor, the $\nu$HDM model (Fig. \ref{cmb}). The optimization process was based on Bayesian statistics, performed by \citet{2025MNRAS.541..784S} and altered the cosmological parameters intriguingly. The MCMC approach searches, in a multi-dimensional space, the optimal values of the cosmological parameters, which were selected to vary, in order to fit the \textit{Planck} CMB data \citep{2020A&A...641A...5P}. \citet{2025MNRAS.541..784S} described the procedure analytically, exemplifying the new expansion history that the model will follow. 

\begin{figure*} 
    \includegraphics[width=\linewidth]{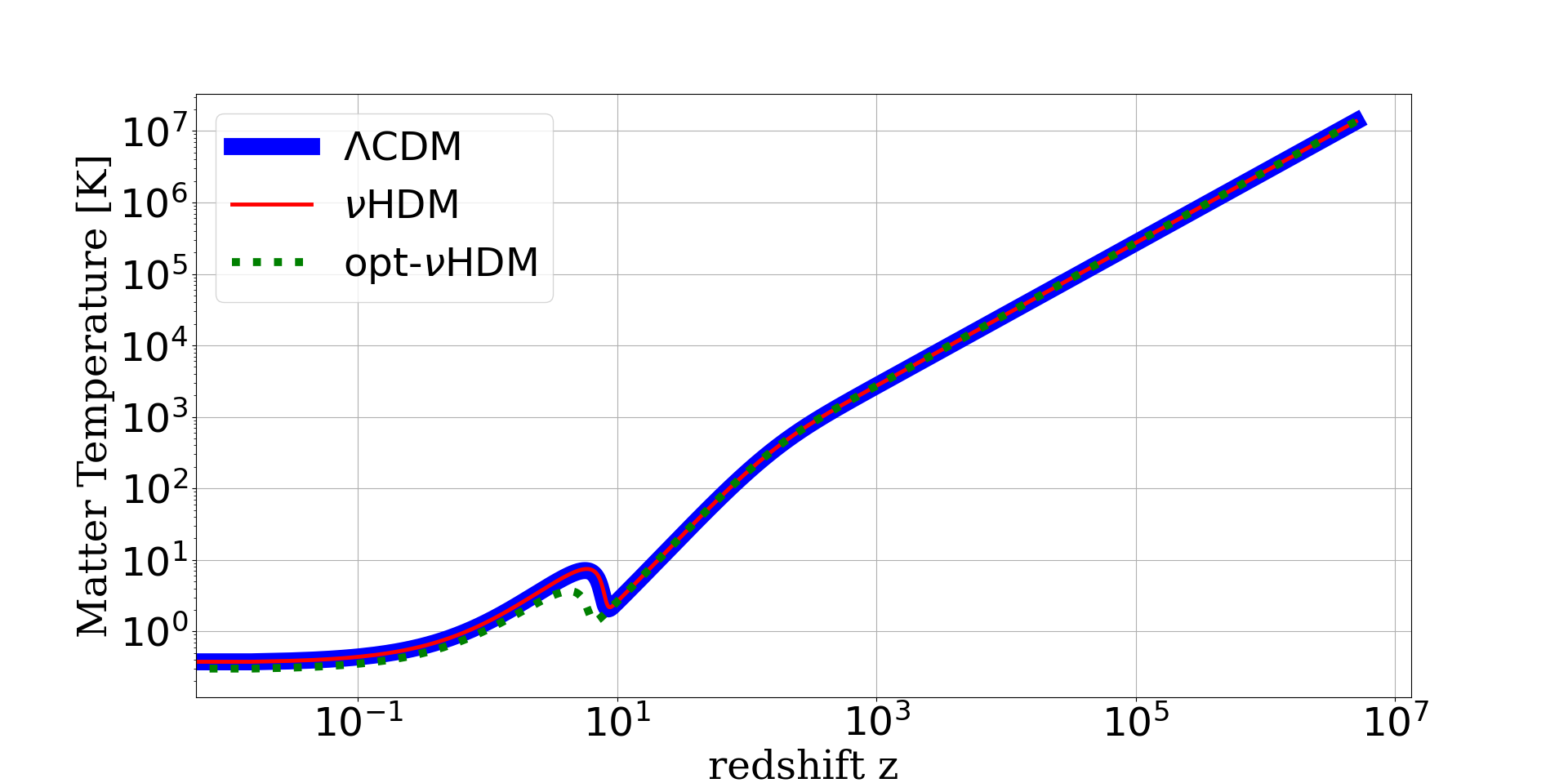}
    \caption{An estimate of the total matter temperature for the 3 studied models with CLASS.}
    \label{reionization}
\end{figure*}

Analytically, the sterile neutrino of the opt-$\nu$HDM model is slightly more massive than the one in $\nu$HDM model, $m^{\text{opt-$\nu$HDM}}_{\nu_s} \approx 13$ eV/c\textsuperscript{2} instead of $m^{\text{$\nu$HDM}}_{\nu_s} \approx 11$ eV/c\textsuperscript{2}. The opt-$\nu$HDM model obeys the FLRW metric by construction, but the matter and the DE eras no longer last the same. The "Dark Ages" and the re-ionization happen at dissimilar times, while the age of the opt-$\nu$HDM universe is estimated to be $\approx$ 0.8 Gyr older than the $\nu$HDM one (which is the same as $\Lambda$CDM age estimate of $\approx13.7$ Gyr). The later onset (than in $\Lambda$CDM) of the reionization is depicted in Fig. \ref{reionization}, where the total matter temperature is plotted against redshift. The total matter temperature is increased by the presumed reionization process at later stages of $z\approx 5.9$ for the opt-$\nu$HDM model, later than $z\approx7$ of the standard expansion history that both $\Lambda$CDM and $\nu$HDM follow. I have complementarily utilized the CLASS package \citep{2011JCAP...07..034B} to demonstrate the delayed reheating, because of the increase rest-mass of the neutrino. This is potentially a worsening factor for the anticipated delayed commencement of the opt-$\nu$HDM cosmic structure formation. Numerical issues also arise when the opt-$\nu$HDM model is treated in CAMB and CosmoSIS, but this will be explained further in Samaras and Kroupa (in preparation). Last, regarding the older age of the opt-$\nu$HDM universe, I will show in the last section that observational studies on the age estimate based on stars in the Galactic disc could be compatible with this uncommon cosmological concept.

%The transfer function, at $z=199.0$ plotted in Fig. \ref{transfer_velocity}. The velocity follows the pattern of the density ones for the baryonic components of each model. Obviously, the sterile neutrino affects the baryons, causing them to oscillate with a more pronounced manner than in CDM halos The green opt-$\nu$HDM baryonic behavior is slightly more powerful than its respective red $\nu$HDM line, since in the opt-$\nu$HDM model, the sterile neutrino is enhanced in mass to $\approx13$ eV and the total matter budget is almost double to $\Omega_M\approx$0.5. As \citet{2023MNRAS.523..453W} and \ns{Samaras et al. 2025} have explained, the absolute value is plotted since over-densities are turned into under-densities and vice versa during structure formation. 

Table \ref{cosmo_params} consists a summary of the cosmological parameters of the two $\nu$HDM variant models and the standard $\Lambda$CDM one. Both $\nu$HDM variant cosmologies are flat but with different expansion rates. The slower opt-$\nu$HDM model, with $H_0 \approx 55.64$ km/s/Mpc, has consequently a higher age of $t_0\approx14.6$ Gyr. The lower spectral index $n_s^{\text{opt-}\nu\text{HDM}}\approx 0.86<n^{\nu\text{HDM}}_s\approx 0.96$ classifies the model as a \textit{tilted} universe with significant implications for its assumed inflation; see \citet{2025MNRAS.541..784S} for a discussion on the cosmological parameters of the opt-$\nu$HDM model.

\subsubsection{Hydrodynamical simulations with Phantom of RAMSES}
The PoR code has been customized \citep{2015CaJPh..93..232L} in order to solve the QUMOND equations of motion. \citet{2023MNRAS.523..453W} have described the procedure comprehensively and I will obey their guidance to employ the flexible N-body adaptive mesh refinement (AMR) code RAMSES. The Milgromian potential will be expressed as an extra effective phantom density distribution. This is not composed of real particles; it is rather a mathematical invention, as: $\rho_{\text{eff}} \equiv \rho + {\rho}_p$, where ${\rho}_p$ is the phantom dark matter density, such that $\nabla^2 \Phi = 4 \pi G \rho_{\text{eff}}$. This allows to solve the following Poisson-like equation, instead of the complicated equation \ref{mond_equation}:
\begin{equation}
    \nabla^2 \Phi \left( \textbf{r} \right) = 4 \pi G \left[ \rho \left( \textbf{r} \right) + {\rho}_p \left( \textbf{r} \right)  \right] = 4 \pi G \rho \left( \textbf{r} \right) + \nabla \cdot \left[ \tilde\upsilon \left( \frac{|\nabla \Phi_N|}{a_0} \right) \nabla \Phi_N \left( \textbf{r} \right)  \right],
\end{equation}
where $\tilde\upsilon = \upsilon-1$ and $\upsilon$ is the QUMOND interpolating function, defined as $\mu \upsilon =1$ in spherical symmetry. The interpolating function $\upsilon$ is asymptotically:
\begin{equation}
    \upsilon\left( \mathcal{y}_{int} \right) \rightarrow     
    \begin{cases}
        \text{} \frac{1}{\sqrt{\mathcal{y}_{int}}}, & \mbox{if } g_N \ll a_0  \left( \mathcal{y}_{int} \ll 1 \right), \\
        1,         & \mbox{if } g_N\gg a_0 \left( \mathcal{y}_{int} \gg 1 \right).
    \end{cases}
\end{equation}
As such, $g = \upsilon g_N$ with $\upsilon$ taking the argument $\mathcal{y}_{int}\equiv\frac{|\nabla {\Phi}_{N}|}{a_0}$. The simple interpolating function \citep{2005MNRAS.363..603F} is used:
\begin{equation}
    \upsilon = \frac{1}{2} + \sqrt{\frac{1}{4}+\frac{1}{\mathcal{y_{int}}}} = \frac{1}{2} + \sqrt{\frac{1}{4}+ \frac{a_0}{g_N}},
\end{equation}
which is precisely the \citet{2023MNRAS.523..453W} protocol. The Poisson equation is thus solved twice, first to get the $\Phi_N$ and its derivative and second to compute the source term:
\begin{equation}
    \nabla \cdot \left[ \tilde\upsilon \left( \frac{|\nabla \Phi_N|}{a_0} \right) \nabla \Phi_N \left( r \right) \right].
\end{equation}
Furthermore, \citet{2021CaJPh..99..607N} published a PoR user guide, but numerical issues may arise and readers are advised to follow modern versions of the code \newline
(See also my \href{https://github.com/NickSam121}{https://github.com/NickSam121/nuHDM-cosmo-sims} page).

RAMSES \citep[section 2.3]{2002A&A...385..337T} solves the hydrodynamical equations for the evolution of the gas using a second-order Godunov scheme:
\begin{align} 
    &\frac{\partial{\rho}}{\partial t} + \nabla \cdot\ \left(  \rho  u\right) = 0, \\
    &\frac{\partial}{\partial t}\left(  \rho u\right) +\nabla \cdot \left(  \rho u \otimes u\right) + \nabla p = -\rho \nabla\Phi, \\
    &\frac{\partial}{\partial t}\left( \rho e\right) + \nabla \cdot \left[ \rho u \left(  e + \frac{p}{\rho} \right)  \right] = - \rho u\cdot\nabla \Phi , 
\end{align}
where $\rho$ is the mass density, $u$ is the fluid velocity, $\otimes$ is the tensor product, $e$ is the specific total energy, $\Phi$ is the Newtonian gravitational potential and p is the thermal pressure, defined as: 
\begin{equation}
    p = \left( \gamma -1 \right) \rho \left( e - \frac{1}{2}u^2 \right)
\end{equation}
\citet{2023MNRAS.523..453W} had not activated SF physics, just simple cooling and heating scheme. In the calculations that I will analyze further on, I have chosen to work specifically on the cosmological volume of 200 Mpc/h, which however, differs in physical distance because of the variant $H_0$ value in the opt-$\nu$HDM model.  Those simulations by \citet{2023MNRAS.523..453W} contain $(2^7)^3 = 2,097,152$ number of particles and a highest refinement level of $2^{10}$ which yields to $\approx$ 0.3 Mpc spatial resolution. I have increased the number of particles to $(2^{8})^3 = 16,777,216$ in order to catch more hydrodynamical effects. The refinement level is the same, since computational costs were so high, even for supercomputers. Indicatively, RAMSES \footnote{https://ramses-organisation.readthedocs.io/en/latest/wiki/Amr.html} needs approximately, with MPI, memory requirements of:
\begin{equation}
    1.4 \left( \textbf{ngridmax}/10^6 \right) + 0.7 \left( \textbf{npartmax}/10^7     \right) \text{ Gb per CPU,}
\end{equation}
for N-body and hydro runs. My simulation runs reached a total of 36 Gb per CPU for 16 tasks and 4 nodes, making the memory requirements almost 600 GB for a run. The parameter \textbf{ngridmax} denotes the maximum number of grids (or octs) that can be allocated during the run within each MPI process and the \textbf{npartmax} is the maximum number of particles of all types that can be allocated during the run within each MPI process. Last, both the CHIMERA\footnote{https://gitlab.mff.cuni.cz/mff/hpc/clusters} cluster (Prague, CZ) and the KAROLINA supercomputing\footnote{https://docs.it4i.cz/} center (Ostrava, CZ) utilize the SLURM queue system\footnote{https://docs.it4i.cz/general/karolina-slurm/} for scheduling users' demands. Technically, the PoR simulations run for a typical amount of time set by the cluster limit. Then the user must restart it from the last snapshot. This makes the task cumbersome enough and eventually it takes approximately 3 (human) days for a simulation to finish at $z= 0.0$, from an initial snapshot of $z =199.0$. 

\begin{figure*}
     \centering
     \begin{subfigure}[b]{0.32\textwidth}
         \centering
         \includegraphics[width=\linewidth, height=5cm]{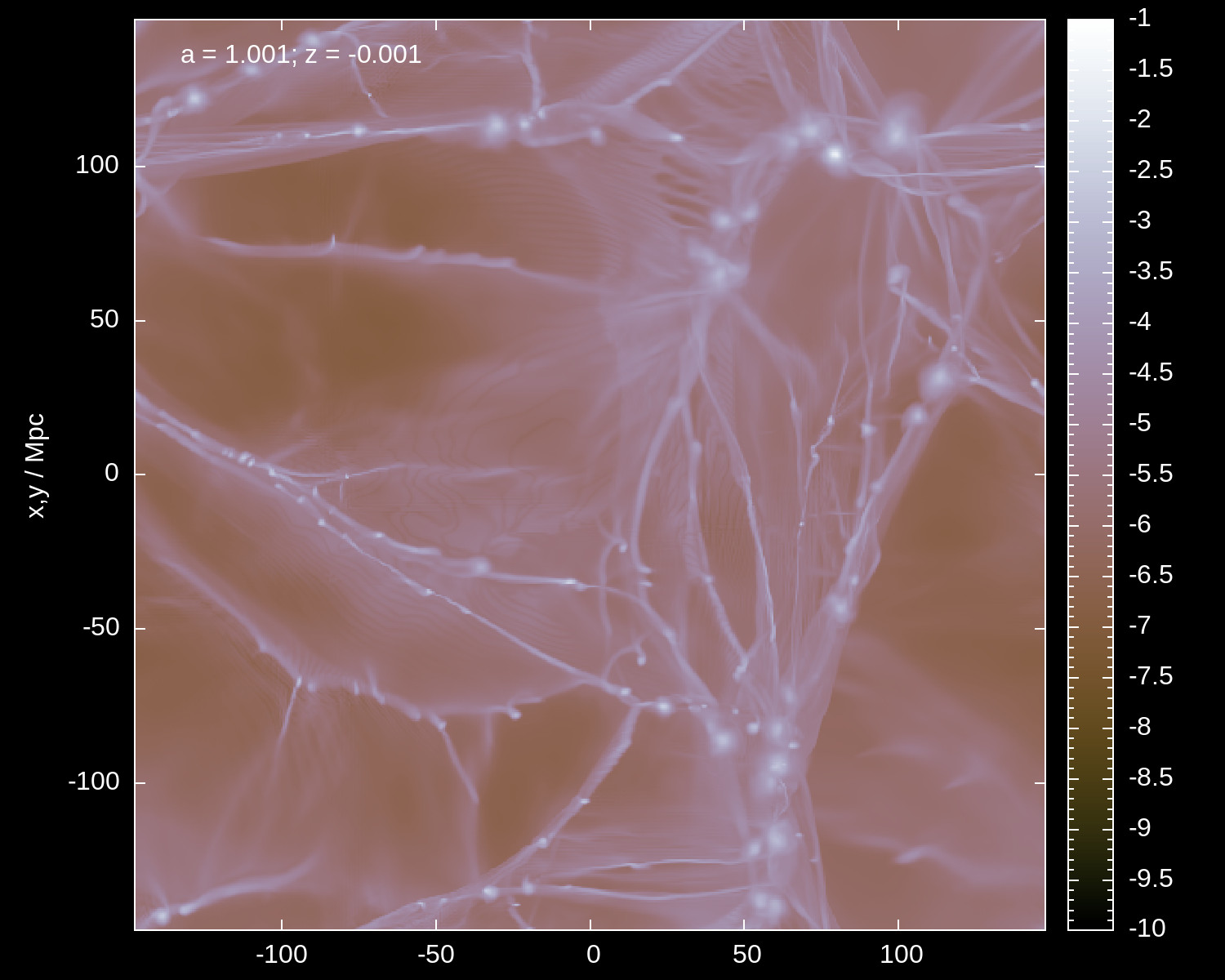}
     \end{subfigure}
     \begin{subfigure}[b]{0.32\textwidth}
         \centering
         \includegraphics[width=\linewidth, height=5cm]{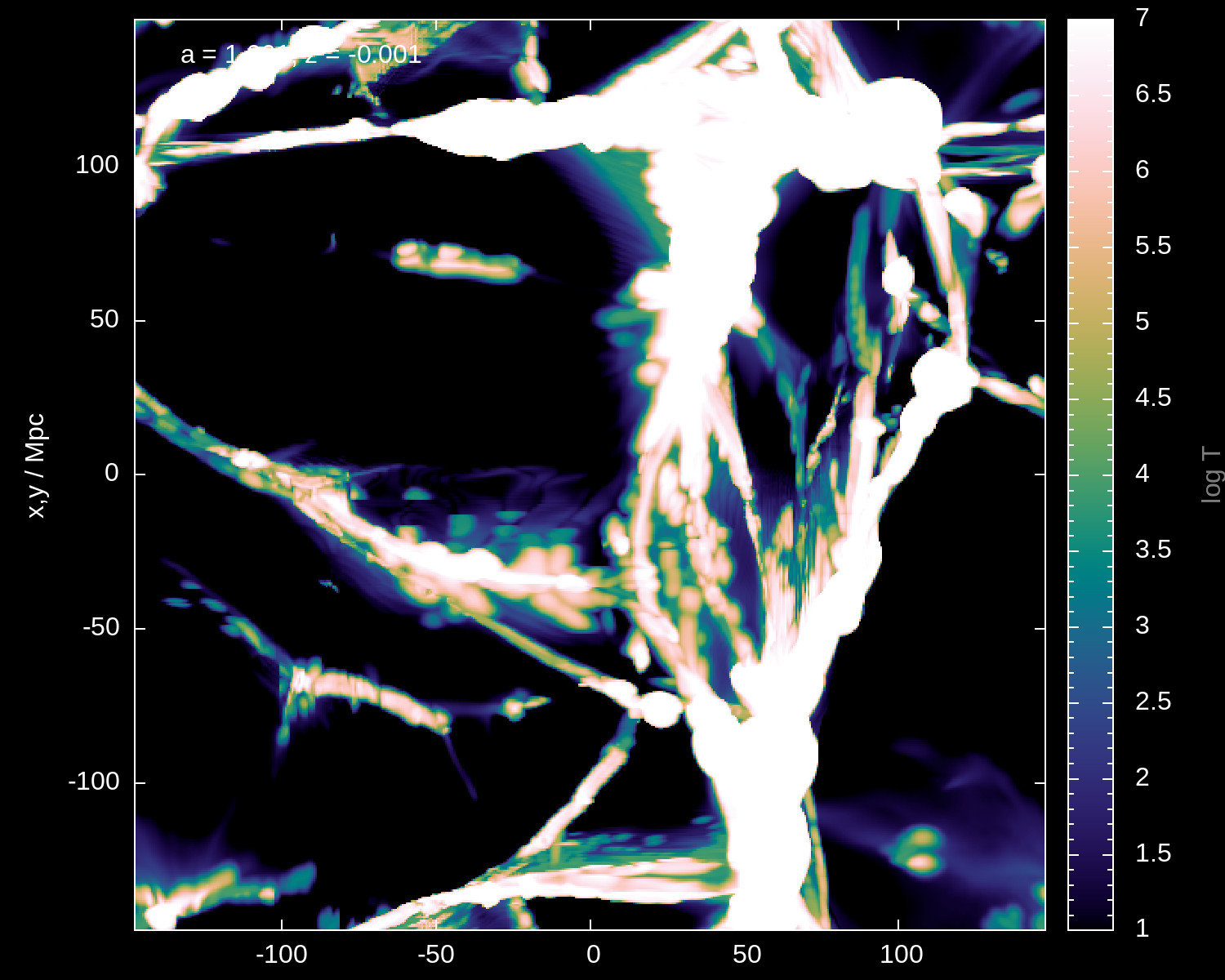}
     \end{subfigure}
     \begin{subfigure}[b]{0.32\textwidth}
         \centering
         \includegraphics[width=\linewidth, height=5cm]{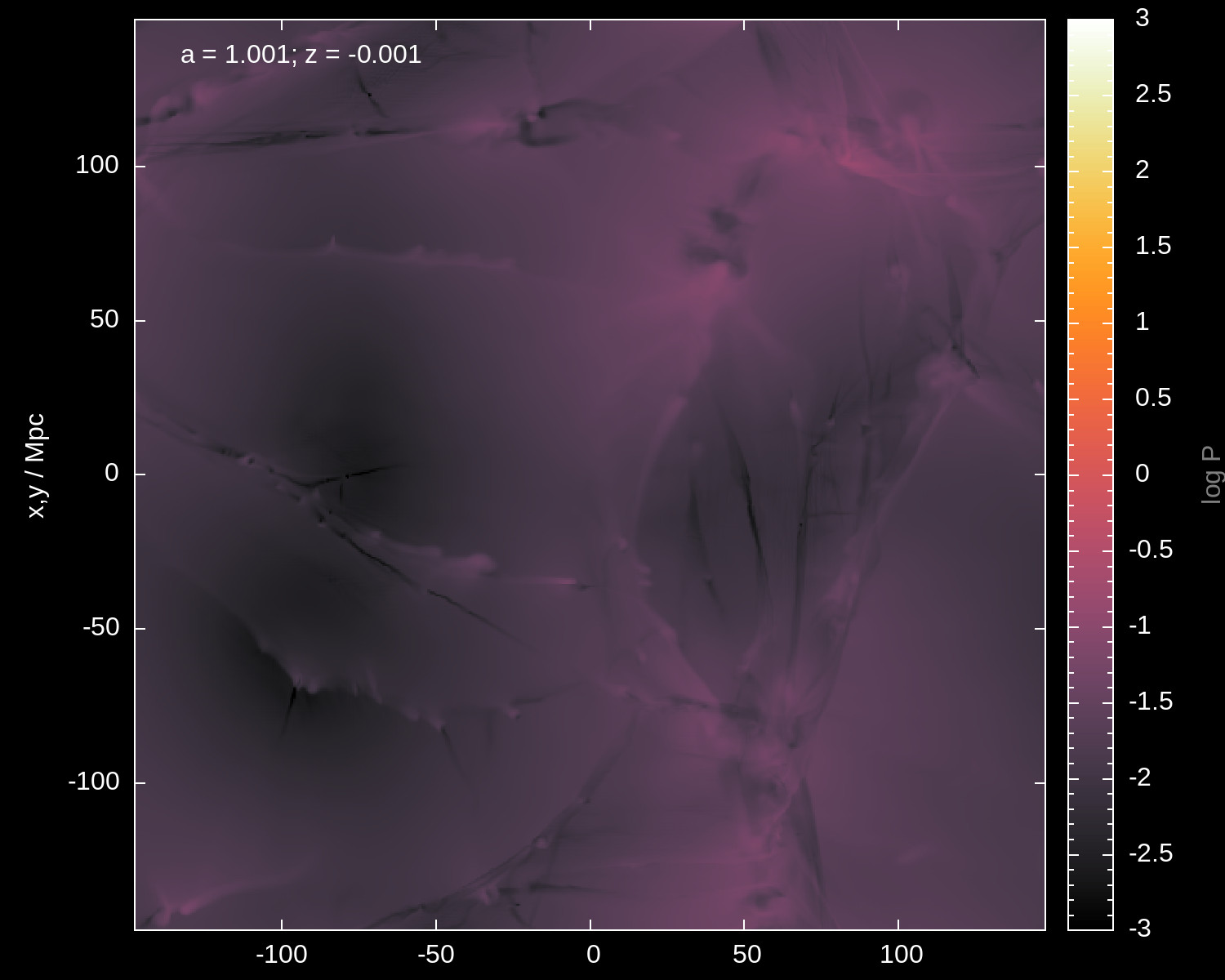}
     \end{subfigure}
    \begin{subfigure}[b]{0.32\textwidth}
         \centering
         \includegraphics[width=\linewidth, height=5cm]{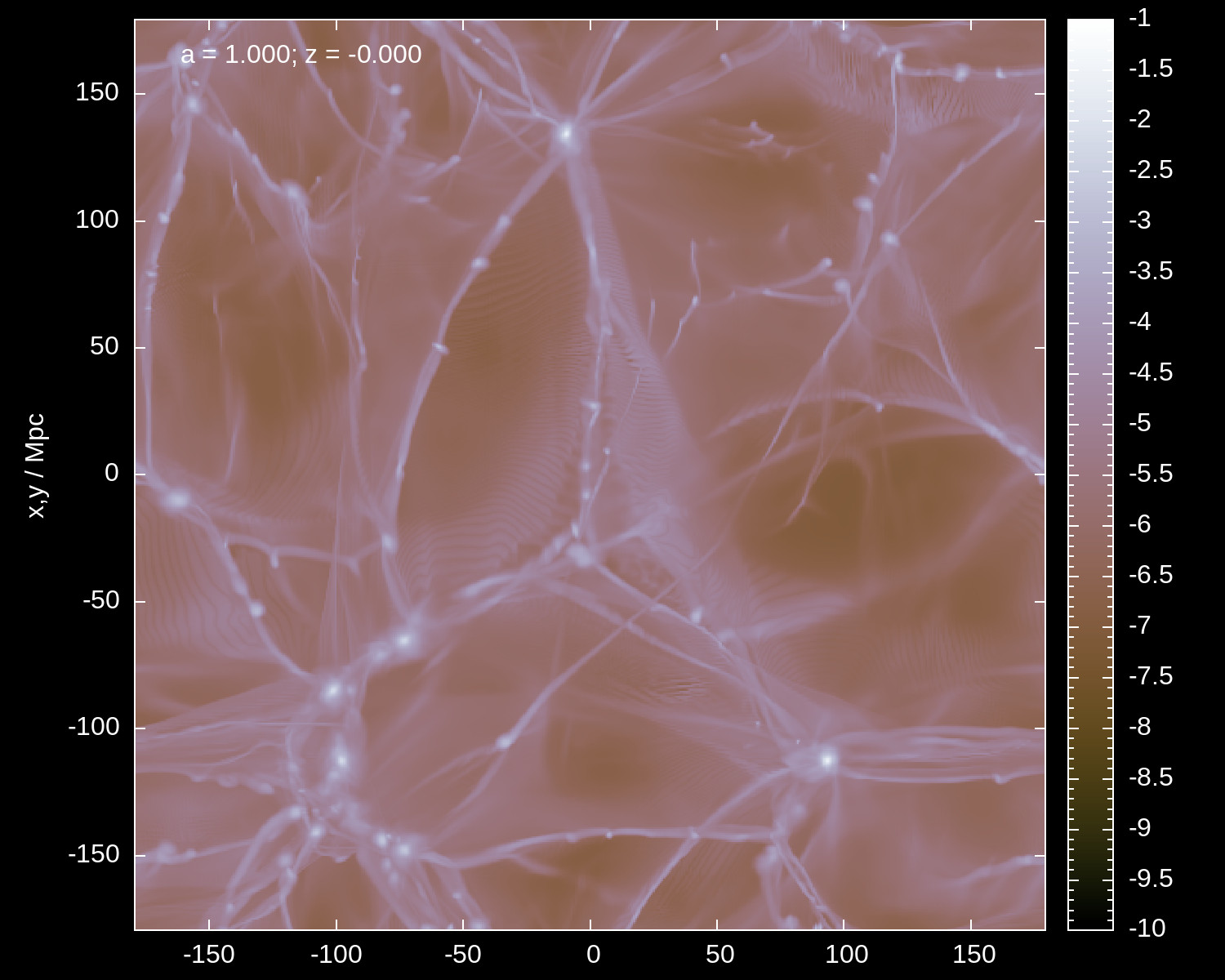}
     \end{subfigure}
         \begin{subfigure}[b]{0.32\textwidth}
         \centering
         \includegraphics[width=\linewidth, height=5cm]{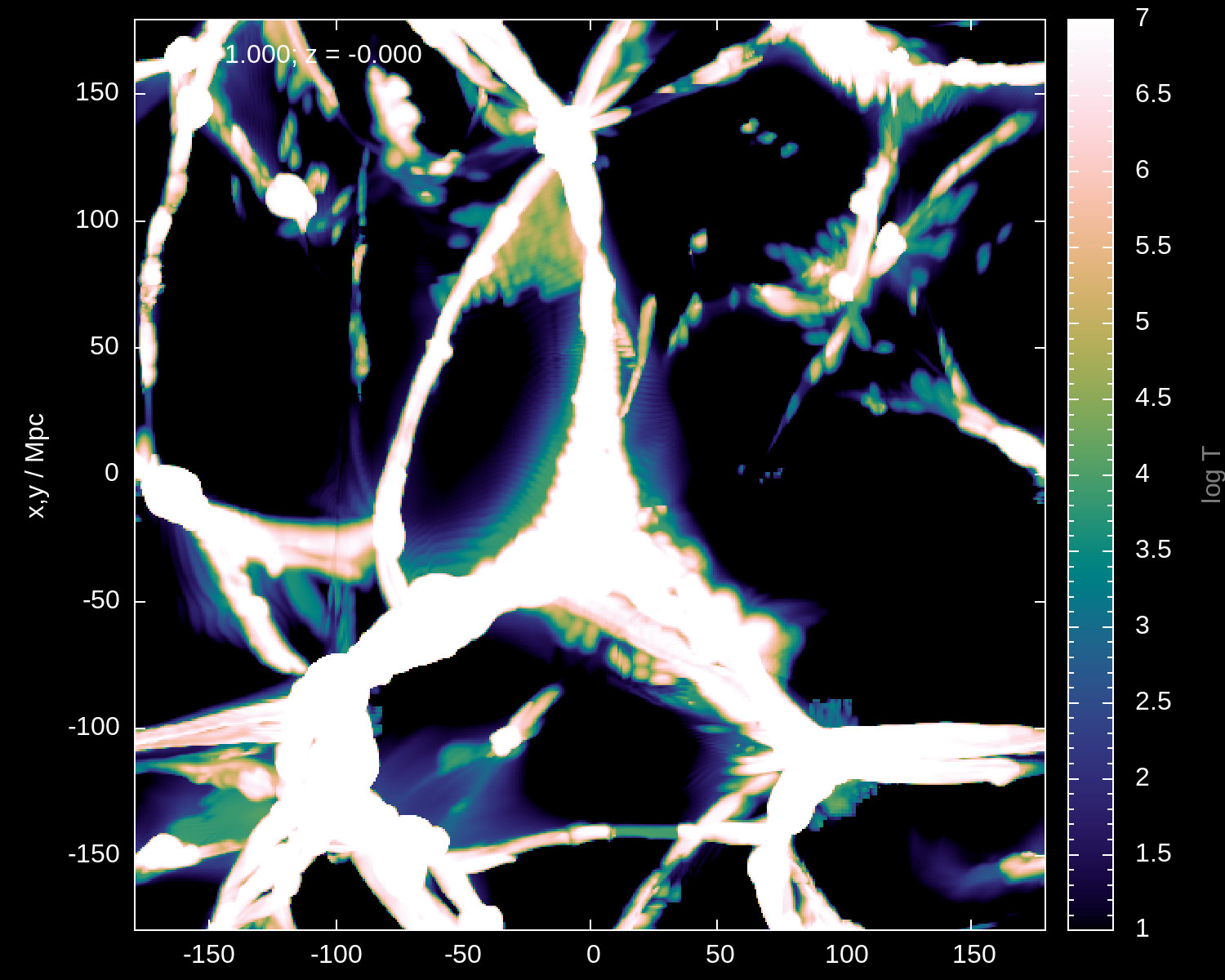}
     \end{subfigure}
         \begin{subfigure}[b]{0.32\textwidth}
         \centering
         \includegraphics[width=\linewidth, height=5cm]{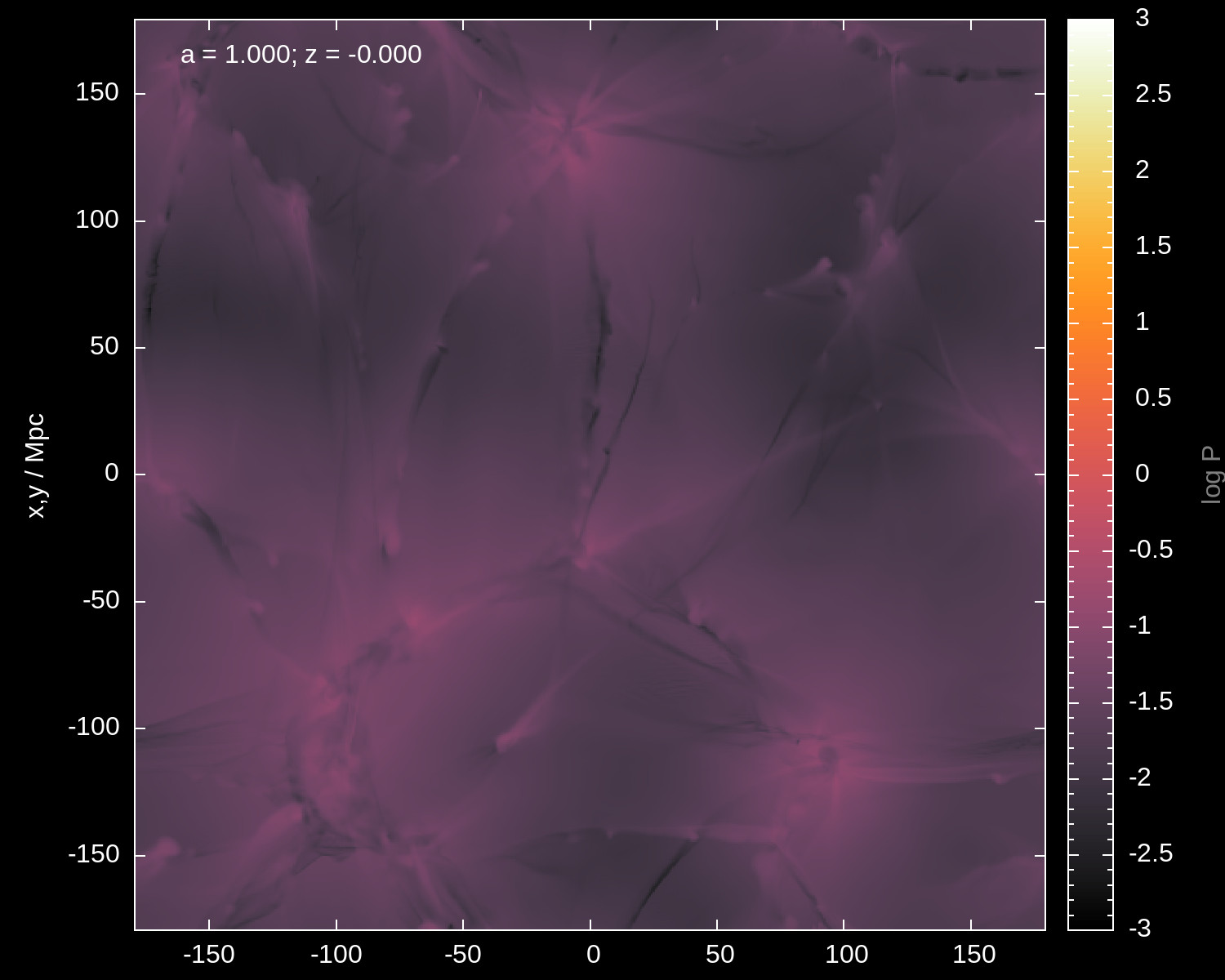}
     \end{subfigure}
    \caption{Density (first column), temperature (second column) and pressure (third column) contrasts for a 200 Mpc/h box (almost 300 Mpc physical size) for $\nu$HDM (upper row) and opt-$\nu$HDM (bottom row) cosmologies at z = 0.}
    \label{nuHDM_snapshots}
\end{figure*}

Simulation snapshots at $z=0$ of the $\nu$HDM model and the opt-$\nu$HDM model for density, temperature and pressure contrasts are presented in Fig. \ref{nuHDM_snapshots}. At a first glance, both cosmologies look very similar, long filaments host big elliptically shaped bound structures. Bridges between them seem to have smaller objects with apparent spins perpendicular to them. The temperature plots show big contrasts of high temperature (T $\approx 10^7$ K) further out of the filamentary structures, enriching the surrounding regions. Pressure gradients appear on the same trend, rising on the over-densities.

\subsubsection{Identifying Milgromian objects}
I have employed the package Amiga Halo Finder \citep[AHF]{2009ApJS..182..608K} for finding gravitationally bound structures. Usually, in the $\Lambda$CDM cosmology, these bound objects are called "halos", but since the $\nu$HDM cosmology obeys Milgromian Dynamics, the paradigm is shifted, halos no longer exist and objects become bound by the principles of Milgromian gravity. 

Firstly, before using the halo finder AHF, RAMSES data must be converted from grid cells into particles with their mass located at the center of the cell. This is done by the conversion tool \textit{ramses2gadget}. The user is asked to specify which type of particles should be converted, sterile neutrinos (\textit{-dm}), baryonic gas (\textit{-g}) or stellar particles (\textit{-s}). The conversion of the data produces binary files as input for the AHF. The user will also notice that the number of cores/threads/tasks that RAMSES has used (with MPI) is the same as the number of the produced binary files (see also my \href{https://github.com/NickSam121}{github} page). 

\begin{figure*} 
    \includegraphics[width=\linewidth]{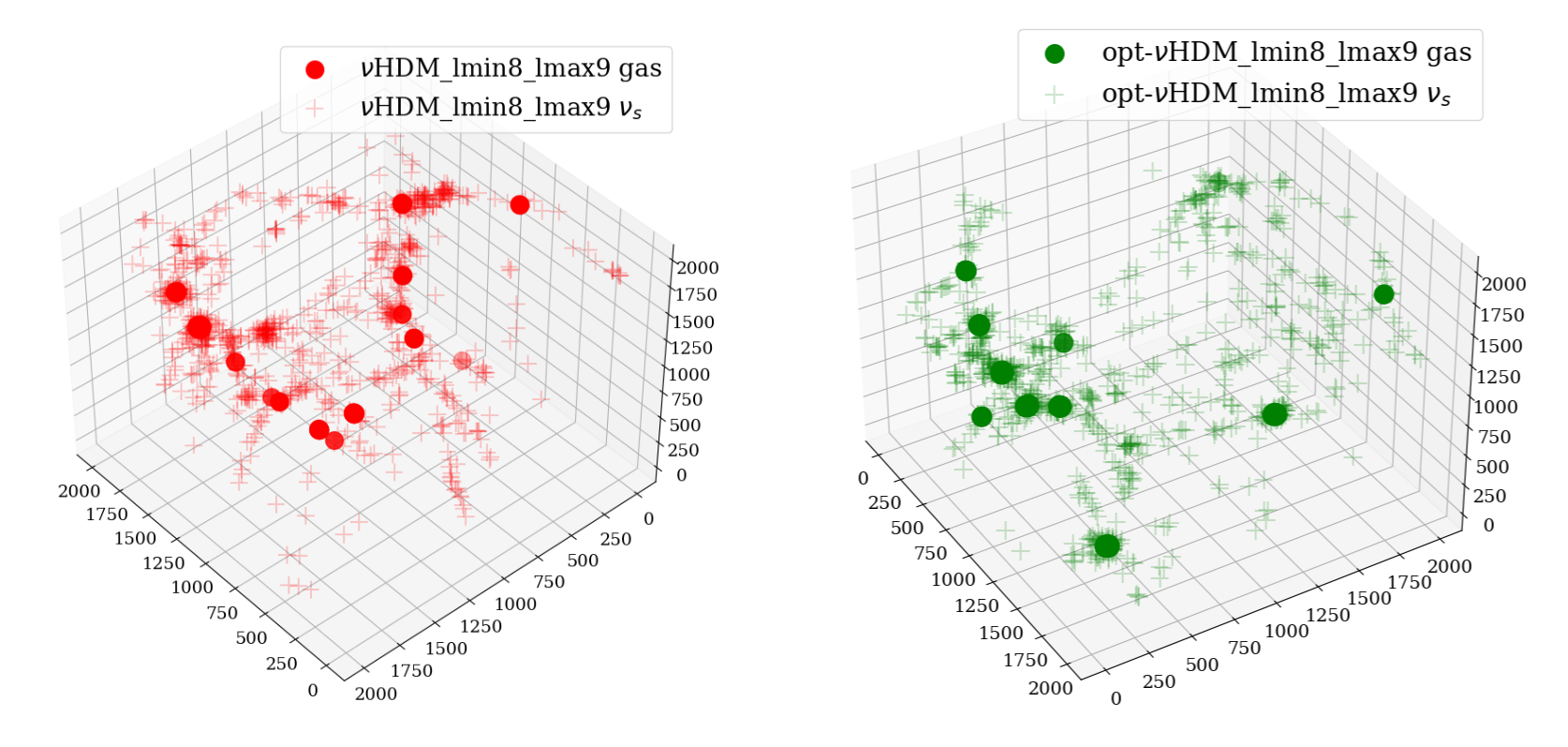}
    \caption{3D distributions of $\nu$HDM (left and red) and opt-$\nu$HDM (right and green) bound structures at $z=0$. The cross symbols represent sterile neutrino assemblies and the circles are gaseous bound structures.}
    \label{3Dpositions}
\end{figure*}

\citet{2023MNRAS.523..453W} had reported that the AHF will be biased for less massive systems, since the algorithm identifies only the bound objects, ignoring the unbound ones. However, under MOND gravity, objects will feel more self-gravity and they will be more stable. Some of the unbound (under Newtonian gravity) objects will be bound under MOND, but are not taken into consideration because of the Newtonian design of the program. Currently, there is no halo finder that conforms with the MOND law. In order to tackle this problem and for ameliorating the status of the MONDian cosmology, I have decided to increase the number of particles from ${2^7}^3$ to ${2^8}^3$. This gives many more structures than reported in the \citet{2023MNRAS.523..453W} study. Namely, \citet[their table. 1]{2023MNRAS.523..453W} had detected 518 bound structures in total (sterile neutrinos and baryons) for a box of 200 Mpc/h. In comparison, my simulations, at the same volume of 200 Mpc/h but with more particles, contain in 1034 opt-$\nu$HDM sterile neutrinos bound structures and 698 opt-$\nu$HDM gaseous ones. Regarding the $\nu$HDM model, there are 377 gaseous and 1092 sterile neutrino structures. In these runs, the refinement level is slightly worse than in the \citet{2023MNRAS.523..453W} study, but this is not necessarily problematic, since I have recovered their results and alleviated some of the known challenges of the model. Before going further into the details, a 3D distribution of the bound structures is presented in Fig. \ref{3Dpositions} for the two $\nu$HDM variant models. The positioning of the gravitationally bound objects is very similar, following MOND gravity. The $\nu$HDM model appears more clumpy, paradoxically counted on the fact that the opt-$\nu$HDM is a heavier universe with $\Omega_M \approx0.49$ and much more sterile-neutrino-dominated. Nevertheless, the volume of the simulation is a significant parameter. For smaller sizes, for instance, of 100 Mpc/h, structures never appear, with the perturbations dying out on the way. On the other extreme, for huge sizes of 3 Gpc/h, the distribution of matter is rather incoherent, with the cosmic web not being developed. Whether this is a result of MOND gravity or the sterile neutrino perturbations in the early Universe is to be investigated.
%or the "drift problem" seen by \citet{2023MNRAS.519.5128N} in their MS study

\begin{figure*} 
    \includegraphics[width=\linewidth]{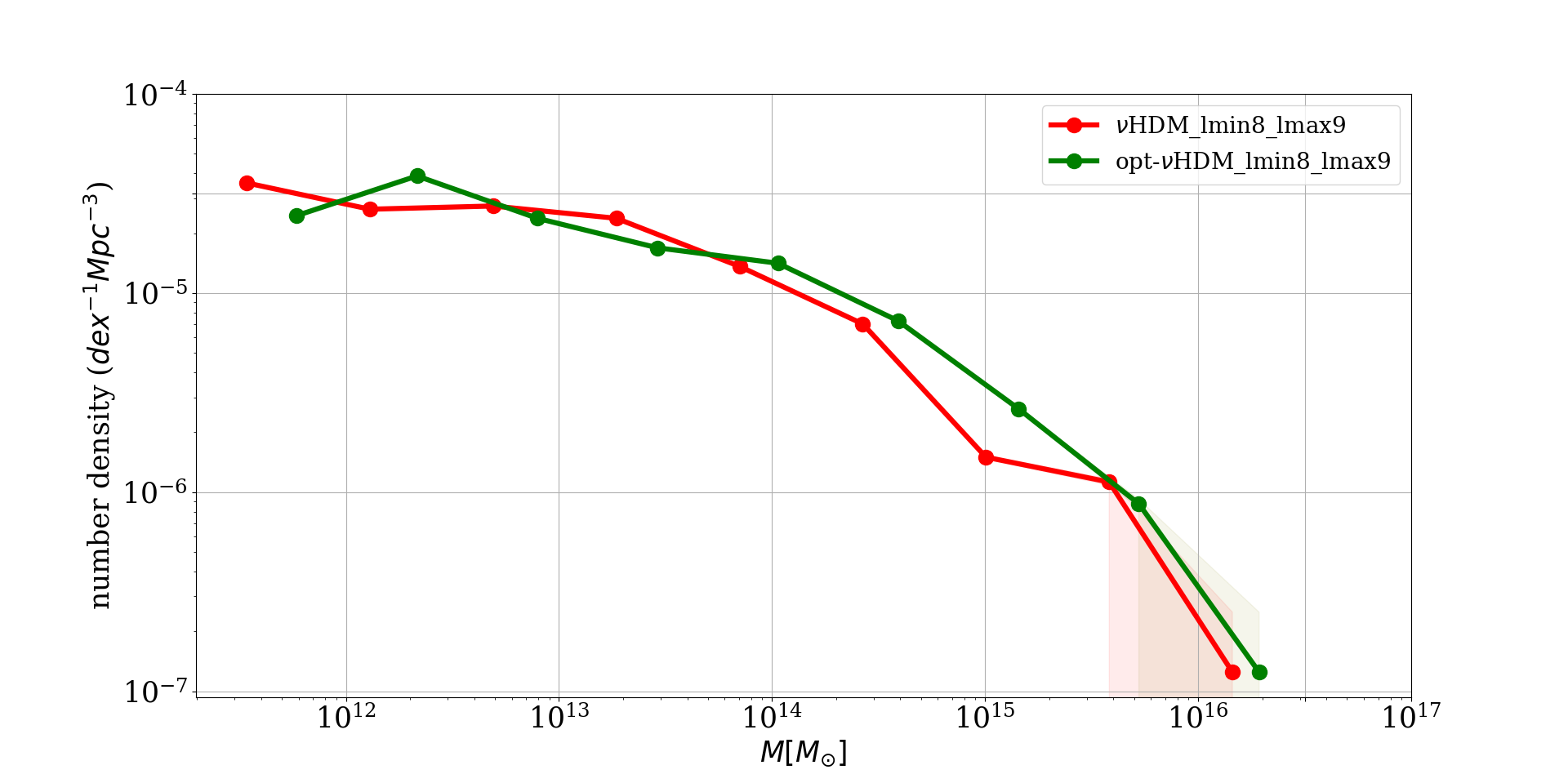}
    \caption{Sterile neutrino only mass function for the two $\nu$HDM variant cosmologies at $z=0$. The shaded areas are Poisson statistics.}
    \label{dm_mass_function}
\end{figure*}

In any case, the late rise of the cosmic web, as \citet{2023MNRAS.523..453W} reported, is, to some degree, eased. The number of particles assisted, such that the bound structures appear at $z\approx5.5$. For a movie of the cosmic evolution of a box of 200 Mpc/h of the opt-$\nu$HDM model, follow this link: \url{https://www.youtube.com/watch?v=XDeCCL_ln-k}. Recent JWST observations \citep{Naidu_2022} have seen galaxies much earlier in the Universe, near $z\approx15$. None of the $\nu$HDM cosmological models seem to approach these earlier times. The $\nu$HDM cosmology may offer some attractive directions to explore scenarios of enhanced structure formation, which are essential to explain the high-redshift galaxies, but hybrid models like these ones seem insufficient. Concerning other combined theoretical frameworks, the delayed cosmogony of the HDM+CDM models is also known since long time \citep{1993ApJ...416....1K, 2024Univ...10...48M}. 

Last, enhancing the number of particles further to ${2^9}^3$ was another idea, but the memory cost is huge, making the simulations unfeasible to complete. SF is also activated in my runs, which have naturally affected the integration but have minor effects at Mpc or Gpc scales. In future efforts, SF will be turned off, capturing less physics, but making the simulation run more effectively. 

A fundamental question of the $\nu$HDM products is the resulting mass function of structures. Although \citet[fig. 6]{2023MNRAS.523..453W} is giving a mass function of structures, consisting of baryons and sterile neutrinos, I separate those objects to understand better their mass distributions. In Fig. \ref{dm_mass_function}, I present the mass function of the sterile neutrino bound objects in the simulation box of 200 Mpc/h, at $z=0$. The two models generate similar sterile neutrino composed gravitational bodies. Note that there are objects of $M <10^{12} M_{\odot}$, which \citet{2023MNRAS.523..453W} could not resolve. On the high-mass end, the over-prediction of vast objects is mitigated. \citet{2010MNRAS.402..395A, 2013ApJ...772...10K} have reached the same conclusion of the $\nu$HDM model generates unphysically massive objects of $M>10^{17}M_{\odot}$. The hydrodynamical simulations \citep{2023MNRAS.523..453W} seem to create better resolved clusters, which probably appeared as unusually massive clusters of mass near $10^{17}M_{\odot}$. The larger number of particles assisted by improving the resolution and therefore, more realistic results are achieved. Overall, the mass distribution of the \citet{2023MNRAS.523..453W} study is recovered and augmented in the low mass end by the calculations I have described above.

\begin{figure*} 
    \includegraphics[width=\linewidth]{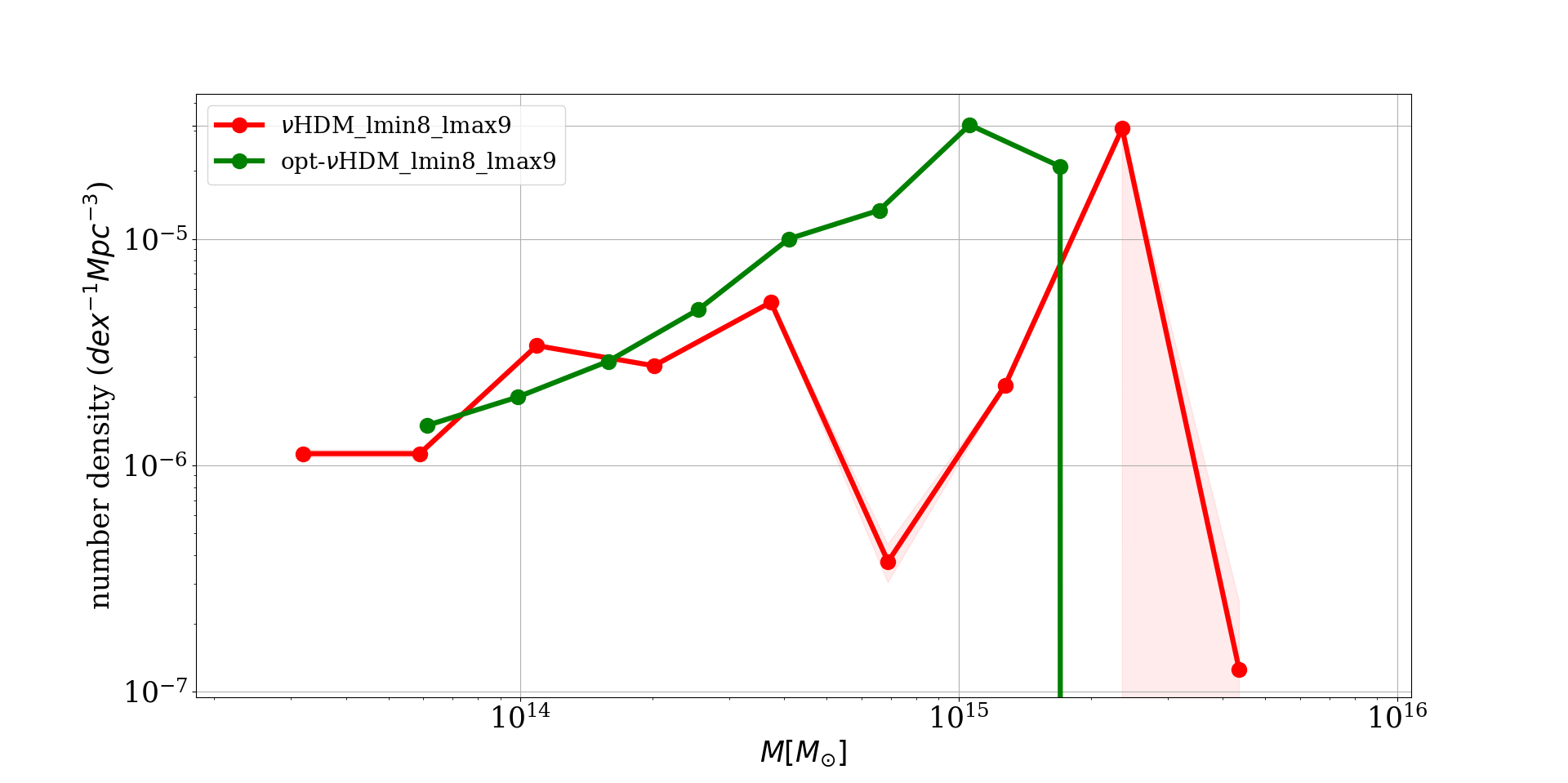}
    \caption{Baryonic gas mass function for the two $\nu$HDM variant cosmologies at $z=0$. The shaded areas are Poisson statistics.}
    \label{gas_mass_function}
\end{figure*}

In addition, the baryonic gas mass function is also presented (Fig. \ref{gas_mass_function}). Apparently, the mass function is "inverted" in this case. Massive gas bound structures seem to be more common than their low-mass counterparts. The gaseous structures may be diminished to the lower mass end, but this does not necessarily mean that they will be sterile-neutrino dominated. It may well be that the sterile neutrino fraction also decreases at the low mass end, as \citet[their fig. 10, lower row, left panel]{2023MNRAS.523..453W} had inspected. This is actually anticipated to occur that objects at this mass range should be purely baryonic and MONDian. The sterile neutrino fraction will be presented in the subsequent paper (Samaras and Kroupa in prep).

The emergence of huge galaxy clusters, like the El Gordo galaxy cluster \citep{2021MNRAS.500.5249A} may be effortless, for the $\nu$HDM variant cosmologies to develop, because of the MOND non-linear gravity enhancement. However, I am not examining, in this work, the cosmic structures at earlier redshifts of $z\approx0.87$.

Overall, this set of simulations seems a nice continuation of the original hydrodynamical work of \citet{2023MNRAS.523..453W}. The emergence of the cosmic web is slightly ameliorated to earlier times of $z\approx 5.5$ from $z\approx 4$. The opt-$\nu$HDM model, although having a non-standard set of cosmological parameters, shares common results with the progenitor $\nu$HDM cosmology. Lastly, none of the models can address currently observations like the KBC void. Going to bigger boxes demands sufficient numerical power and sterile neutrino perturbations in the early Universe do not get materialized easily to groups or clusters of galaxies. 

\subsection{The \texorpdfstring{$h - \Omega_M$}{h-Omega-M} relation}
The last section of this analysis is about a combination of projects, that one from Chapter \ref{banik_paper} concerning the local void solution of the Hubble tension and the previous sections regarding the $\nu$HDM variant cosmological simulations. \citet{2025MNRAS.541..784S} reported a prediction of the age of the opt-$\nu$HDM universe, which was quite old, $t^{\text{opt}-\nu\text{HDM}}_0\approx 14.9$ Gyr. This is far from agreement with the standard model of cosmology $\Lambda$CDM, whose age is $t^{\Lambda\text{CDM}}_0 \approx13.7$ Gyr. I would like to investigate if this is a plausible estimate from the observational point of view. 

\begin{figure*} 
    \includegraphics[width=\linewidth]{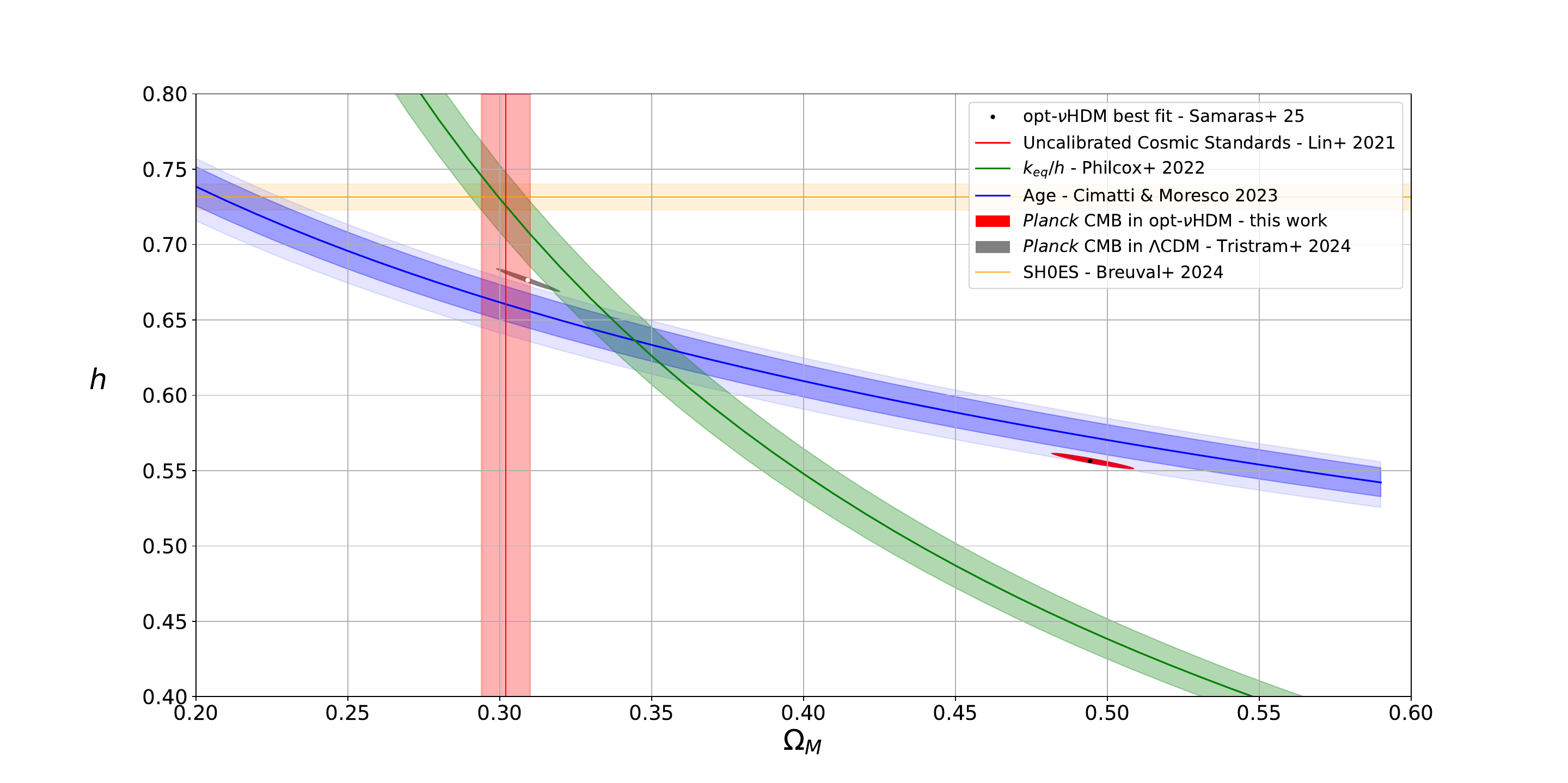}
    \caption{Opt-$\nu$HDM estimate for the relation $h-\Omega_M$, compared to various observational datasets. The prediction $t^{\text{opt}-\nu\text{HDM}}_0\approx 14.9$ Gyr is matched by observations based on the (blue band prediction of) oldest stars of the Milky Way \citep{Cimatti_2023}.}
    \label{h_OmegaM_opt}
\end{figure*}

For this purpose, I have made use of the MCMC chain of the \citet{2025MNRAS.541..784S} investigation. Using a uniform weight for all the $h-\Omega_M$ values, I have over-plotted the \citet{Banik_NS} bands together with the opt-$\nu$HDM ellipse. The nominal values for the opt-$\nu$HDM model is $h-\Omega_M = 0.556-0.49$. As before, the orange band is the SH0ES value of $H_0=73.17$ km/s/Mpc, inferred locally from SNIa data. The red band in the Uncalibrated Cosmic Standards with $\Omega_M = 0.302 \pm0.008$ and the green one is the \citet{2022PhRvD.106f3530P} prediction which yields to $h-\Omega_M = 0.648-0.338$. None of the aforementioned estimates are close to the value of $H^{\text{opt}-\nu\text{HDM}}_0 \approx 55.6$ km/s/Mpc and $\Omega^{\text{opt}-\nu\text{HDM}}_M\approx0.49$.

Strikingly, the age estimate from \citet{Cimatti_2023}, based on the stars in the Galactic disc can be compatible with such an old Universe, like the opt-$\nu$HDM model. The accommodated values from the optimized sterile-neutrino MONDian model is $h-\Omega_M = 0.556-0.49$ which are fitted distinctively at the 1$\sigma$ error band of the estimates. 
Notably, the age of the Universe could potentially be even larger, considering the dust around elliptical galaxies \citep{2025arXiv250504687G}. Further work on contrasting the opt-$\nu$HDM model with distance measurement from DESI \citep{2025JCAP...02..021A} will appear soon (Samaras and Kroupa, in prep).

\section{Conclusions} \label{conclusions} 
The $\Lambda$CDM model has been able to provide answers about the evolution of the Universe from its beginning in the Big Bang until the latest stages of the accelerated expansion due to the unknown DE since its birth \citep{1990Natur.348..705E}. Its substantial role in giving the community the tools to perform calculations is valuable for comprehending Nature. But as $\Lambda$CDM has its merits, so it exhibits its deficiencies. In fact, its shortcomings are also of great importance, hinting at potential resolutions and answers to how the Universe works. Independent tests on the $\Lambda$CDM model falsify it by many $\sigma$ values, but they remain provocatively disregarded. Arguments based on the Chandrasekhar dynamical friction \citep{1943ApJ....97..255C}, like the orbits of the Large and the Small Magellanic Clouds \citep{2024Univ...10..143O} and the galactic bars \citep{2021MNRAS.508..926R}, confirm the lack of observational evidence for dynamical dissipation by extended particle-based dark matter halos. More investigations on the matter distribution of DM models, like the plane of satellites \citep{2013MNRAS.435.2116P} or immense under-dense regions \citep{Keenan_2013} and the big bulk flows \citep{2021A&A...649A.151M}, indicate strong departure from the stochastic merger history and the convergence to homogeneity on $\approx$ 300 Mpc scales.

On the other hand, MOND \citep{2020pama.book.....M} is tremendously under-funded, lagging behind the mainstream bodies of cosmology, which are benefited by the great amounts of investment and human-power. On its own, MOND seems inflexible to overcome the discrepancy of the dynamical mass to the observed mass of gas-rich, X-ray emitting galaxy clusters \citep{Sanders_1999}. Starting from that, sterile neutrinos were hypothesized on top of Milgromian dynamics to explain clusters and address cosmology. The $\nu$HDM cosmology might be a reasonable first response to scrutinize the enhanced structure formation that MOND furnishes. Sterile neutrinos could be better motivated \citep{2022Symm...14.1331B} compared to any other CDM particle, but still that scenario suffers from the aforementioned experiments \citep{2015CaJPh..93..169K}.

Numerically, due to MOND's non-linear nature, any potential analytical or numerical attempts come out  arduous and require intensive and systematic monitoring. N-body MONDian codes \citep{2025A&A...693A.127P} have existed since recently and will tackle unresolved issues, concerning small scale probes of Milgromian Dynamics \citep{2023MNRAS.525.1401H, 2024MNRAS.527.4573B}.

In this sense, this Ph.D. thesis proposed the most advanced cosmological simulations ever made on the $\nu$HDM model. The simulations achieve higher resolution, managing to tackle the late-forming cosmic web, from $z\approx4$ \citep{2023MNRAS.523..453W} to $z\approx 5.5$. The baryonic gas mass function appears on the same trend as previous studies \citep{2013ApJ...772...10K}, but it is now relaxed from unphysically massive clusters of $M\approx10^{17}M_{\odot}$. The rest of this work is under preparation (Samaras and Kroupa) and will elaborate more on the sterile neutrino fraction and the properties of the cosmic $\nu$HDM structures.

%The Bohemian Model of Cosmology is under development (Gjergo and Kroupa, in prep) and does not rely on any of the $\Lambda$CDM assumptions. First-ever simulations, in a universe which is inflation-free, DM-free, DE-free reveals exciting high-z galaxies, natural emergence of El Gordo analogues and KBC-like under dense regions, too. Preliminary results are very promising and new research lines are eventually carved.

\setstretch{1}
\addcontentsline{toc}{section}{References}
%\bibliography{references} 
\bibliography{BIBLIO} 

@ARTICLE{2023MNRAS.523..453W,
       author = {{Wittenburg}, Nils and {Kroupa}, Pavel and {Banik}, Indranil and {Candlish}, Graeme and {Samaras}, Nick},
        title = "{Hydrodynamical structure formation in Milgromian cosmology}",
      journal = {\mnras},
         year = 2023,
        month = jul,
       volume = {523},
       number = {1},
        pages = {453-473},
          doi = {10.1093/mnras/stad1371},
archivePrefix = {arXiv},
       eprint = {2305.05696},
 primaryClass = {astro-ph.CO},
       adsurl = {https://ui.adsabs.harvard.edu/abs/2023MNRAS.523..453W}
}

@BOOK{1990eaun.book.....K,
       author = {{Kolb}, Edward W. and {Turner}, Michael S.},
        title = "{The early universe}",
         year = 1990,
       volume = {69},
       adsurl = {https://ui.adsabs.harvard.edu/abs/1990eaun.book.....K}
}

@ARTICLE{1990Natur.348..705E,
       author = {{Efstathiou}, G. and {Sutherland}, W.J. and {Maddox}, S.-J.},
        title = "{The cosmological constant and cold dark matter}",
      journal = {\nat},
         year = 1990,
        month = dec,
       volume = {348},
       number = {6303},
        pages = {705-707},
          doi = {10.1038/348705a0},
       adsurl = {https://ui.adsabs.harvard.edu/abs/1990Natur.348..705E}
}

@ARTICLE{1995Natur.377..600O,
       author = {{Ostriker}, J.-P. and {Steinhardt}, Paul J.},
        title = "{The observational case for a low-density Universe with a non-zero cosmological constant}",
      journal = {\nat},
         year = 1995,
        month = oct,
       volume = {377},
       number = {6550},
        pages = {600-602},
          doi = {10.1038/377600a0},
       adsurl = {https://ui.adsabs.harvard.edu/abs/1995Natur.377..600O}
}

@article{Riess_1998,
doi = {10.1086/300499},
url = {https://dx.doi.org/10.1086/300499},
year = {1998},
month = sep,
publisher = {},
volume = {116},
number = {3},
pages = {1009},
author = {Riess, Adam G. and Filippenko, Alexei V. and Challis, Peter and Clocchiatti, Alejandro and Diercks, Alan and Garnavich, Peter M. and Gilliland, Ron L. and Hogan, Craig J. and Jha, Saurabh and Kirshner, Robert P. and Leibundgut, B. and Phillips, M. M. and Reiss, David and Schmidt, Brian P. and Schommer, Robert A. and Smith, R. Chris and Spyromilio, J. and Stubbs, Christopher and Suntzeff, Nicholas B. and Tonry, John},
title = {Observational Evidence from Supernovae for an Accelerating Universe and a Cosmological Constant},
journal = {The Astronomical Journal}
}

@article{10.1093/mnras/stx3112,
    author = {Pillepich, Annalisa and Nelson, Dylan and Hernquist, Lars and Springel, Volker and Pakmor, Rüdiger and Torrey, Paul and Weinberger, Rainer and Genel, Shy and Naiman, Jill P and Marinacci, Federico and Vogelsberger, Mark},
    title = "{First results from the IllustrisTNG simulations: the stellar mass content of groups and clusters of galaxies}",
    journal = {Monthly Notices of the Royal Astronomical Society},
    volume = {475},
    number = {1},
    pages = {648-675},
    year = {2017},
    month = {12},
    issn = {0035-8711},
    doi = {10.1093/mnras/stx3112},
    url = {https://doi.org/10.1093/mnras/stx3112},
    eprint = {https://academic.oup.com/mnras/article-pdf/475/1/648/23534158/stx3112.pdf},
}

@article{10.1093/mnras/stz2338,
    author = {Pillepich, Annalisa and Nelson, Dylan and Springel, Volker and Pakmor, Rüdiger and Torrey, Paul and Weinberger, Rainer and Vogelsberger, Mark and Marinacci, Federico and Genel, Shy and vander Wel, Arjen and Hernquist, Lars},
    title = "{First results from the TNG50 simulation: the evolution of stellar and gaseous discs across cosmic time}",
    journal = {Monthly Notices of the Royal Astronomical Society},
    volume = {490},
    number = {3},
    pages = {3196-3233},
    year = {2019},
    month = {09},
    issn = {0035-8711},
    doi = {10.1093/mnras/stz2338},
    url = {https://doi.org/10.1093/mnras/stz2338},
    eprint = {https://academic.oup.com/mnras/article-pdf/490/3/3196/30327933/stz2338.pdf},
}

@ARTICLE{Nel,
       author = {{Nelson}, D. and {Pillepich}, A. and {Genel}, S. and {Vogelsberger}, M. and {Springel}, V. and {Torrey}, P. and {Rodriguez-Gomez}, V. and {Sijacki}, D. and {Snyder}, G.-F. and {Griffen}, B. and {Marinacci}, F. and {Blecha}, L. and {Sales}, L. and {Xu}, D. and {Hernquist}, L.},
        title = "{The illustris simulation: Public data release}",
      journal = {Astronomy and Computing},
         year = 2015,
        month = nov,
       volume = {13},
        pages = {12-37},
          doi = {10.1016/j.ascom.2015.09.003},
archivePrefix = {arXiv},
       eprint = {1504.00362},
 primaryClass = {astro-ph.CO},
       adsurl = {https://ui.adsabs.harvard.edu/abs/2015A&C....13...12N}
}

@ARTICLE{Spri1,
       author = {{Springel}, Volker},
        title = "{E pur si muove: Galilean-invariant cosmological hydrodynamical simulations on a moving mesh}",
      journal = {\mnras},
         year = 2010,
        month = jan,
       volume = {401},
       number = {2},
        pages = {791-851},
          doi = {10.1111/j.1365-2966.2009.15715.x},
archivePrefix = {arXiv},
       eprint = {0901.4107},
 primaryClass = {astro-ph.CO},
       adsurl = {https://ui.adsabs.harvard.edu/abs/2010MNRAS.401..791S}
}

@ARTICLE{Spri2,
       author = {{Springel}, Volker and {White}, Simon D.-M. and {Tormen}, Giuseppe and {Kauffmann}, Guinevere},
        title = "{Populating a cluster of galaxies - I. Results at z=0}",
      journal = {\mnras},
         year = 2001,
        month = dec,
       volume = {328},
       number = {3},
        pages = {726-750},
          doi = {10.1046/j.1365-8711.2001.04912.x},
archivePrefix = {arXiv},
       eprint = {astro-ph/0012055},
 primaryClass = {astro-ph},
       adsurl = {https://ui.adsabs.harvard.edu/abs/2001MNRAS.328..726S}
}

@ARTICLE{2014A&A...571A...1P,
       author = {{Planck Collaboration} and {Ade}, P.A.R. and {Aghanim}, N. and {Alves}, M.-I.-R. and {Armitage-Caplan}, C. and {Arnaud}, M. and {Ashdown}, M. and {Zonca}, A.},
        title = "{Planck 2013 results. I. Overview of products and scientific results}",
      journal = {\aap},
         year = 2014,
        month = nov,
       volume = {571},
          eid = {A1},
        pages = {A1},
          doi = {10.1051/0004-6361/201321529},
archivePrefix = {arXiv},
       eprint = {1303.5062},
 primaryClass = {astro-ph.CO},
       adsurl = {https://ui.adsabs.harvard.edu/abs/2014A&A...571A...1P}
}

@ARTICLE{Davies,
       author = {{Davis}, M. and {Efstathiou}, G. and {Frenk}, C.-S. and {White}, S.-D.-M.},
        title = "{The evolution of large-scale structure in a universe dominated by cold dark matter}",
      journal = {\apj},
         year = 1985,
        month = may,
       volume = {292},
        pages = {371-394},
          doi = {10.1086/163168},
       adsurl = {https://ui.adsabs.harvard.edu/abs/1985ApJ...292..371D}
}

@ARTICLE{Dolag,
       author = {{Dolag}, K. and {Borgani}, S. and {Murante}, G. and {Springel}, V.},
        title = "{Substructures in hydrodynamical cluster simulations}",
      journal = {\mnras},
         year = 2009,
        month = oct,
       volume = {399},
       number = {2},
        pages = {497-514},
          doi = {10.1111/j.1365-2966.2009.15034.x},
archivePrefix = {arXiv},
       eprint = {0808.3401},
 primaryClass = {astro-ph},
       adsurl = {https://ui.adsabs.harvard.edu/abs/2009MNRAS.399..497D}
}

@ARTICLE{2019ComAC...6....2N,
       author = {{Nelson}, Dylan and {Springel}, Volker and {Pillepich}, Annalisa and {Rodriguez-Gomez}, Vicente and {Torrey}, Paul and {Genel}, Shy and {Vogelsberger}, Mark and {Pakmor}, Ruediger and {Marinacci}, Federico and {Weinberger}, Rainer and {Kelley}, Luke and {Lovell}, Mark and {Diemer}, Benedikt and {Hernquist}, Lars},
        title = "{The IllustrisTNG simulations: public data release}",
      journal = {Computational Astrophysics and Cosmology},
         year = 2019,
        month = may,
       volume = {6},
       number = {1},
          eid = {2},
        pages = {2},
          doi = {10.1186/s40668-019-0028-x},
archivePrefix = {arXiv},
       eprint = {1812.05609},
 primaryClass = {astro-ph.GA},
       adsurl = {https://ui.adsabs.harvard.edu/abs/2019ComAC...6....2N}
}

@ARTICLE{Vogel,
       author = {{Vogelsberger}, M. and {Genel}, S. and {Springel}, V. and {Torrey}, P. and {Sijacki}, D. and {Xu}, D. and {Snyder}, G. and {Bird}, S. and {Nelson}, D. and {Hernquist}, L.},
        title = "{Properties of galaxies reproduced by a hydrodynamic simulation}",
      journal = {\nat},
         year = 2014,
        month = may,
       volume = {509},
       number = {7499},
        pages = {177-182},
          doi = {10.1038/nature13316},
archivePrefix = {arXiv},
       eprint = {1405.1418},
 primaryClass = {astro-ph.CO},
       adsurl = {https://ui.adsabs.harvard.edu/abs/2014Natur.509..177V}
}

@article{10.1093/mnras/stx1051,
    author = {Trayford, James W. and Camps, Peter and Theuns, Tom and Baes, Maarten and Bower, Richard G. and Crain, Robert A. and Gunawardhana, Madusha L. P. and Schaller, Matthieu and Schaye, Joop and Frenk, Carlos S.},
    title = "{Optical colours and spectral indices of z=0.1 eagle galaxies with the 3D dust radiative transfer code skirt}",
    journal = {Monthly Notices of the Royal Astronomical Society},
    volume = {470},
    number = {1},
    pages = {771-799},
    year = {2017},
    month = {05},
    issn = {0035-8711},
    doi = {10.1093/mnras/stx1051},
    url = {https://doi.org/10.1093/mnras/stx1051},
    eprint = {https://academic.oup.com/mnras/article-pdf/470/1/771/17872335/stx1051.pdf},
}

@article{Oppen,
    author = {Oppenheimer, Benjamin D. and Davé, Romeel},
    title = "{Cosmological simulations of intergalactic medium enrichment from galactic outflows}",
    journal = {Monthly Notices of the Royal Astronomical Society},
    volume = {373},
    number = {4},
    pages = {1265-1292},
    year = {2006},
    month = {11},
    issn = {0035-8711},
    doi = {10.1111/j.1365-2966.2006.10989.x},
    url = {https://doi.org/10.1111/j.1365-2966.2006.10989.x},
    eprint = {https://academic.oup.com/mnras/article-pdf/373/4/1265/11176562/mnras0373-1265.pdf},
}

@article{10.1093/mnras/stx3040,
    author = {Nelson, Dylan and Pillepich, Annalisa and Springel, Volker and Weinberger, Rainer and Hernquist, Lars and Pakmor, Rüdiger and Genel, Shy and Torrey, Paul and Vogelsberger, Mark and Kauffmann, Guinevere and Marinacci, Federico and Naiman, Jill},
    title = "{First results from the IllustrisTNG simulations: the galaxy colour bimodality}",
    journal = {Monthly Notices of the Royal Astronomical Society},
    volume = {475},
    number = {1},
    pages = {624-647},
    year = {2017},
    month = {11},
    issn = {0035-8711},
    doi = {10.1093/mnras/stx3040},
    url = {https://doi.org/10.1093/mnras/stx3040},
    eprint = {https://academic.oup.com/mnras/article-pdf/475/1/624/23534134/stx3040.pdf},
}

@ARTICLE{2014Natur.509..177V,
       author = {{Vogelsberger}, M. and {Genel}, S. and {Springel}, V. and {Torrey}, P. and {Sijacki}, D. and {Xu}, D. and {Snyder}, G. and {Bird}, S. and {Nelson}, D. and {Hernquist}, L.},
        title = "{Properties of galaxies reproduced by a hydrodynamic simulation}",
      journal = {\nat},
         year = 2014,
        month = may,
       volume = {509},
       number = {7499},
        pages = {177-182},
          doi = {10.1038/nature13316},
archivePrefix = {arXiv},
       eprint = {1405.1418},
 primaryClass = {astro-ph.CO},
       adsurl = {https://ui.adsabs.harvard.edu/abs/2014Natur.509..177V}
}

@article{10.1093/mnras/stx2787,
    author = {Ploeckinger, Sylvia and Sharma, Kuldeep and Schaye, Joop and Crain, Robert A. and Schaller, Matthieu and Barber, Christopher},
    title = "{Tidal dwarf galaxies in cosmological simulations}",
    journal = {Monthly Notices of the Royal Astronomical Society},
    volume = {474},
    number = {1},
    pages = {580-596},
    year = {2017},
    month = {10},
    issn = {0035-8711},
    doi = {10.1093/mnras/stx2787},
    url = {https://doi.org/10.1093/mnras/stx2787},
    eprint = {https://academic.oup.com/mnras/article-pdf/474/1/580/22141251/stx2787.pdf},
}

@article{10.1093/mnras/stx2656,
    author = {Pillepich, Annalisa and Springel, Volker and Nelson, Dylan and Genel, Shy and Naiman, Jill and Pakmor, Rüdiger and Hernquist, Lars and Torrey, Paul and Vogelsberger, Mark and Weinberger, Rainer and Marinacci, Federico},
    title = "{Simulating galaxy formation with the IllustrisTNG model}",
    journal = {Monthly Notices of the Royal Astronomical Society},
    volume = {473},
    number = {3},
    pages = {4077-4106},
    year = {2017},
    month = {10},
    issn = {0035-8711},
    doi = {10.1093/mnras/stx2656},
    url = {https://doi.org/10.1093/mnras/stx2656},
    eprint = {https://academic.oup.com/mnras/article-pdf/473/3/4077/21841785/stx2656.pdf},
}

@article{10.1093/mnras/stu2058,
    author = {Schaye, Joop and Crain, Robert A. and Bower, Richard G. and Furlong, Michelle and Schaller, Matthieu and Theuns, Tom and Dalla Vecchia, Claudio and Frenk, Carlos S. and McCarthy, I. G. and Helly, John C. and Jenkins, Adrian and Rosas-Guevara, Y. M. and White, Simon D. M. and Baes, Maarten and Booth, C. M. and Camps, Peter and Navarro, Julio F. and Qu, Yan and Rahmati, Alireza and Sawala, Till and Thomas, Peter A. and Trayford, James},
    title = "{The EAGLE project: simulating the evolution and assembly of galaxies and their environments}",
    journal = {Monthly Notices of the Royal Astronomical Society},
    volume = {446},
    number = {1},
    pages = {521-554},
    year = {2014},
    month = {11},
    issn = {0035-8711},
    doi = {10.1093/mnras/stu2058},
    url = {https://doi.org/10.1093/mnras/stu2058},
    eprint = {https://academic.oup.com/mnras/article-pdf/446/1/521/4139718/stu2058.pdf},
}

@article{Crain,
    author = {Crain, Robert A. and Schaye, Joop and Bower, Richard G. and Furlong, Michelle and Schaller, Matthieu and Theuns, Tom and Dalla Vecchia, Claudio and Frenk, Carlos S. and McCarthy, Ian G. and Helly, John C. and Jenkins, Adrian and Rosas-Guevara, Yetli M. and White, Simon D. M. and Trayford, James W.},
    title = "{The EAGLE simulations of galaxy formation: calibration of subgrid physics and model variations}",
    journal = {Monthly Notices of the Royal Astronomical Society},
    volume = {450},
    number = {2},
    pages = {1937-1961},
    year = {2015},
    month = {04},
    issn = {0035-8711},
    doi = {10.1093/mnras/stv725},
    url = {https://doi.org/10.1093/mnras/stv725},
    eprint = {https://academic.oup.com/mnras/article-pdf/450/2/1937/3067585/stv725.pdf},
}

@article{Spri3,
    author = {Springel, Volker},
    title = "{The cosmological simulation code gadget-2}",
    journal = {Monthly Notices of the Royal Astronomical Society},
    volume = {364},
    number = {4},
    pages = {1105-1134},
    year = {2005},
    month = {12},
    issn = {0035-8711},
    doi = {10.1111/j.1365-2966.2005.09655.x},
    url = {https://doi.org/10.1111/j.1365-2966.2005.09655.x},
    eprint = {https://academic.oup.com/mnras/article-pdf/364/4/1105/18657201/364-4-1105.pdf},
}

@article{SD,
    author = {Schaye, Joop and Dalla Vecchia, Claudio},
    title = "{On the relation between the Schmidt and Kennicutt Schmidt star formation laws and its implications for numerical simulations}",
    journal = {Monthly Notices of the Royal Astronomical Society},
    volume = {383},
    number = {3},
    pages = {1210-1222},
    year = {2008},
    month = {01},
    issn = {0035-8711},
    doi = {10.1111/j.1365-2966.2007.12639.x},
    url = {https://doi.org/10.1111/j.1365-2966.2007.12639.x},
    eprint = {https://academic.oup.com/mnras/article-pdf/383/3/1210/3497111/mnras0383-1210.pdf},
}

@article{Karachentsev_2013,
doi = {10.1088/0004-6256/146/3/46},
url = {https://dx.doi.org/10.1088/0004-6256/146/3/46},
year = {2013},
month = jul,
publisher = {The American Astronomical Society},
volume = {146},
number = {3},
pages = {46},
author = {Igor D. Karachentsev and Elena I. Kaisina},
title = {STAR FORMATION PROPERTIES IN THE LOCAL VOLUME GALAXIES VIA Hα AND FAR-ULTRAVIOLET FLUXES},
journal = {The Astronomical Journal}
}

@article{Speagle_2014,
doi = {10.1088/0067-0049/214/2/15},
url = {https://dx.doi.org/10.1088/0067-0049/214/2/15},
year = {2014},
month = sep,
publisher = {The American Astronomical Society},
volume = {214},
number = {2},
pages = {15},
author = {J. S. Speagle and C. L. Steinhardt and P. L. Capak and J. D. Silverman},
title = {A HIGHLY CONSISTENT FRAMEWORK FOR THE EVOLUTION OF THE STAR-FORMING “MAIN SEQUENCE” FROM z ∼ 0–6},
journal = {}
}

@article{Kroupa,
	author = {{Kroupa, P.} and {Famaey, B.} and {de Boer, K. S.} and {Dabringhausen, J.} and {Pawlowski, M. S.} and {Boily, C. M.} and {Jerjen, H.} and {Forbes, D.} and {Hensler, G.} and {Metz, M.}},
	title = {Local-Group tests of dark-matter concordance
          cosmology - Towards a new paradigm for structure formation},
	DOI= "10.1051/0004-6361/201014892",
	url= "https://doi.org/10.1051/0004-6361/201014892",
	journal = {A\&A},
	year = 2010,
	volume = 523,
	pages = "A32",
	month = "",
}

@ARTICLE{Lelli2016,
       author = {{Lelli}, Federico and {McGaugh}, Stacy S. and {Schombert}, James M.},
        title = "{SPARC: Mass Models for 175 Disk Galaxies with Spitzer Photometry and Accurate Rotation Curves}",
      journal = {\aj},
         year = 2016,
        month = dec,
       volume = {152},
       number = {6},
          eid = {157},
        pages = {157},
          doi = {10.3847/0004-6256/152/6/157},
archivePrefix = {arXiv},
       eprint = {1606.09251},
 primaryClass = {astro-ph.GA},
       adsurl = {https://ui.adsabs.harvard.edu/abs/2016AJ....152..157L}
}

@ARTICLE{Hasl,
       author = {{Haslbauer}, Moritz and {Kroupa}, Pavel and {Jerabkova}, Tereza},
        title = "{The cosmological star formation history from the Local Cosmological Volume of galaxies and constraints on the matter homogeneity}",
      journal = {\mnras},
         year = 2023,
        month = sep,
       volume = {524},
       number = {3},
        pages = {3252-3262},
          doi = {10.1093/mnras/stad1986},
archivePrefix = {arXiv},
       eprint = {2306.16436},
 primaryClass = {astro-ph.GA},
       adsurl = {https://ui.adsabs.harvard.edu/abs/2023MNRAS.524.3252H}
}

@article{Kroupa2020,
    author = {Kroupa, P and Haslbauer, M and Banik, I and Nagesh, S T and Pflamm-Altenburg, J},
    title = "{Constraints on the star formation histories of galaxies in the Local Cosmological Volume}",
    journal = {Monthly Notices of the Royal Astronomical Society},
    volume = {497},
    number = {1},
    pages = {37-43},
    year = {2020},
    month = {07},
    issn = {0035-8711},
    doi = {10.1093/mnras/staa1851},
    url = {https://doi.org/10.1093/mnras/staa1851},
    eprint = {https://academic.oup.com/mnras/article-pdf/497/1/37/33519932/staa1851.pdf},
}

@INPROCEEDINGS{2023eppg.confE.231K,
       author = {{Kroupa}, P. and {Gjergo}, E. and {Asencio}, E. and {Haslbauer}, M. and {Pflamm-Altenburg}, J. and {Wittenburg}, N. and {Samaras}, N. and {Thies}, I. and {Oehm}, W.},
        title = "{The many tensions with dark-matter based models and implications on the nature of the Universe}",
    booktitle = {School and Workshops on Elementary Particle Physics and Gravity},
         year = 2023,
        month = jan,
          eid = {231},
        pages = {231},
          doi = {10.48550/arXiv.2309.11552},
archivePrefix = {arXiv},
       eprint = {2309.11552},
 primaryClass = {astro-ph.CO},
       adsurl = {https://ui.adsabs.harvard.edu/abs/2023eppg.confE.231K}
}

@ARTICLE{2015CaJPh..93..232L,
       author = {{L{\"u}ghausen}, Fabian and {Famaey}, Benoit and {Kroupa}, Pavel},
        title = "{Phantom of RAMSES (POR): A new Milgromian dynamicsN-body code}",
      journal = {Canadian Journal of Physics},
         year = 2015,
        month = feb,
       volume = {93},
       number = {2},
        pages = {232-241},
          doi = {10.1139/cjp-2014-0168},
archivePrefix = {arXiv},
       eprint = {1405.5963},
 primaryClass = {astro-ph.GA},
       adsurl = {https://ui.adsabs.harvard.edu/abs/2015CaJPh..93..232L}
}

@ARTICLE{2010MNRAS.403..886M,
       author = {{Milgrom}, Mordehai},
        title = "{Quasi-linear formulation of MOND}",
      journal = {\mnras},
         year = 2010,
        month = apr,
       volume = {403},
       number = {2},
        pages = {886-895},
          doi = {10.1111/j.1365-2966.2009.16184.x},
archivePrefix = {arXiv},
       eprint = {0911.5464},
 primaryClass = {astro-ph.CO},
       adsurl = {https://ui.adsabs.harvard.edu/abs/2010MNRAS.403..886M}
}

@article{Naidu_2022,
doi = {10.3847/2041-8213/ac9b22},
url = {https://dx.doi.org/10.3847/2041-8213/ac9b22},
year = {2022},
month = nov,
publisher = {The American Astronomical Society},
volume = {940},
number = {1},
pages = {L14},
author = {Naidu, Rohan P. and Oesch, Pascal A. and Dokkum, Pieter van and Nelson, Erica J. and Suess, Katherine A. and Brammer, Gabriel and Whitaker, Katherine E. and Illingworth, Garth and Bouwens, Rychard and Tacchella, Sandro and Matthee, Jorryt and Allen, Natalie and Bezanson, Rachel and Conroy, Charlie and Labbe, Ivo and Leja, Joel and Leonova, Ecaterina and Magee, Dan and Price, Sedona H. and Setton, David J. and Strait, Victoria and Stefanon, Mauro and Toft, Sune and Weaver, John R. and Weibel, Andrea},
title = {Two Remarkably Luminous Galaxy Candidates at z ≈ 10–12 Revealed by JWST},
journal = {The Astrophysical Journal Letters}
}

@article{Keenan_2013,
doi = {10.1088/0004-637X/775/1/62},
url = {https://dx.doi.org/10.1088/0004-637X/775/1/62},
year = {2013},
month = sep,
publisher = {The American Astronomical Society},
volume = {775},
number = {1},
pages = {62},
author = {R. C. Keenan and A. J. Barger and L. L. Cowie},
title = {EVIDENCE FOR A ∼300 MEGAPARSEC SCALE UNDER-DENSITY IN THE LOCAL GALAXY DISTRIBUTION},
journal = {The Astrophysical Journal}
}

@ARTICLE{2025arXiv250504687G,
       author = {{Gjergo}, Eda and {Kroupa}, Pavel},
        title = "{The Impact of Early Massive Galaxy Formation on the Cosmic Microwave Background}",
      journal = {arXiv e-prints},
         year = 2025,
        month = may,
          eid = {arXiv:2505.04687},
        pages = {arXiv:2505.04687},
          doi = {10.48550/arXiv.2505.04687},
archivePrefix = {arXiv},
       eprint = {2505.04687},
 primaryClass = {astro-ph.GA},
       adsurl = {https://ui.adsabs.harvard.edu/abs/2025arXiv250504687G}
}

@ARTICLE{2021PhRvL.127p1302S,
       author = {{Skordis}, Constantinos and {Z{\l}o{\'s}nik}, Tom},
        title = "{New Relativistic Theory for Modified Newtonian Dynamics}",
      journal = {Phys. Rev. Lett.},
         year = 2021,
        month = oct,
       volume = {127},
       number = {16},
          eid = {161302},
        pages = {161302},
          doi = {10.1103/PhysRevLett.127.161302},
archivePrefix = {arXiv},
       eprint = {2007.00082},
 primaryClass = {astro-ph.CO},
       adsurl = {https://ui.adsabs.harvard.edu/abs/2021PhRvL.127p1302S}
}

@article{10.1093/mnras/stu2713,
    author = {Sparre, Martin and Hayward, Christopher C. and Springel, Volker and Vogelsberger, Mark and Genel, Shy and Torrey, Paul and Nelson, Dylan and Sijacki, Debora and Hernquist, Lars},
    title = "{The star formation main sequence and stellar mass assembly of galaxies in the Illustris simulation}",
    journal = {Monthly Notices of the Royal Astronomical Society},
    volume = {447},
    number = {4},
    pages = {3548-3563},
    year = {2015},
    month = {01},
    issn = {0035-8711},
    doi = {10.1093/mnras/stu2713},
    url = {https://doi.org/10.1093/mnras/stu2713},
    eprint = {https://academic.oup.com/mnras/article-pdf/447/4/3548/5821928/stu2713.pdf},
}

@ARTICLE{2013ApJ...770...57B,
       author = {{Behroozi}, Peter S. and {Wechsler}, Risa H. and {Conroy}, Charlie},
        title = "{The Average Star Formation Histories of Galaxies in Dark Matter Halos from z = 0-8}",
      journal = {\apj},
         year = 2013,
        month = jun,
       volume = {770},
       number = {1},
          eid = {57},
        pages = {57},
          doi = {10.1088/0004-637X/770/1/57},
archivePrefix = {arXiv},
       eprint = {1207.6105},
 primaryClass = {astro-ph.CO},
       adsurl = {https://ui.adsabs.harvard.edu/abs/2013ApJ...770...57B}
}

@ARTICLE{2021MNRAS.508..926R,
       author = {{Roshan}, Mahmood and {Ghafourian}, Neda and {Kashfi}, Tahere and {Banik}, Indranil and {Haslbauer}, Moritz and {Cuomo}, Virginia and {Famaey}, Benoit and {Kroupa}, Pavel},
        title = "{Fast galaxy bars continue to challenge standard cosmology}",
      journal = {\mnras},
         year = 2021,
        month = nov,
       volume = {508},
       number = {1},
        pages = {926-939},
          doi = {10.1093/mnras/stab2553},
archivePrefix = {arXiv},
       eprint = {2106.10304},
 primaryClass = {astro-ph.GA},
       adsurl = {https://ui.adsabs.harvard.edu/abs/2021MNRAS.508..926R}
}

@article{Riess_2022,
doi = {10.3847/2041-8213/ac5c5b},
url = {https://dx.doi.org/10.3847/2041-8213/ac5c5b},
year = {2022},
month = jul,
publisher = {The American Astronomical Society},
volume = {934},
number = {1},
pages = {L7},
author = {Riess, Adam G. and Yuan, Wenlong and Macri, Lucas M. and Scolnic, Dan and Brout, Dillon and Casertano, Stefano and Jones, David O. and Murakami, Yukei and Anand, Gagandeep S. and Breuval, Louise and Brink, Thomas G. and Filippenko, Alexei V. and Hoffmann, Samantha and Jha, Saurabh W. and D’arcy Kenworthy, W. and Mackenty, John and Stahl, Benjamin E. and Zheng, WeiKang},
title = {A Comprehensive Measurement of the Local Value of the Hubble Constant with 1 km s−1 Mpc−1 Uncertainty from the Hubble Space Telescope and the SH0ES Team},
journal = {The Astrophysical Journal Letters}
}

@ARTICLE{CosmoVerse,
       author = {{Di Valentino}, Eleonora and {Levi Said}, Jackson and {Riess}, Adam and {Pollo}, Agnieszka and {Poulin}, Vivian and {G{\'o}mez-Valent}, Adri{\`a} and {Weltman}, Amanda and {Palmese}, Antonella and {Huang}, Caroline D. and {van de Bruck}, Carsten and {Shekhar Saraf}, Chandra and {Kuo}, Cheng-Yu and {Uhlemann}, Cora and {Grand{\'o}n}, Daniela and {Paz}, Dante and {Eckert}, Dominique and {Teixeira}, Elsa M. and {Saridakis}, Emmanuel N. and {Colg{\'a}in}, Eoin {\'O} and {Beutler}, Florian and {Niedermann}, Florian and {Bajardi}, Francesco and {Barenboim}, Gabriela and {Gubitosi}, Giulia and {Musella}, Ilaria and {Banik}, Indranil and {Szapudi}, Istvan and {Singal}, Jack and {Haro Cases}, Jaume and {Chluba}, Jens and {Torrado}, Jes{\'u}s and {Mifsud}, Jurgen and {Jedamzik}, Karsten and {Said}},
        title = "{The CosmoVerse White Paper: Addressing observational tensions in cosmology with systematics and fundamental physics}",
      journal = {arXiv e-prints},
         year = 2025,
        month = apr,
          eid = {arXiv:2504.01669},
        pages = {arXiv:2504.01669},
          doi = {10.48550/arXiv.2504.01669},
archivePrefix = {arXiv},
       eprint = {2504.01669},
 primaryClass = {astro-ph.CO},
       adsurl = {https://ui.adsabs.harvard.edu/abs/2025arXiv250401669D}
}

@ARTICLE{2021PhRvD.103h3533A,
       author = {{Alam}, Shadab and {Aubert}, Marie and {Avila}, Santiago and {Balland}, Christophe and {Bautista}, Julian E. and {Bershady}, Matthew A. and {Bizyaev}, Dmitry and {Blanton}, Michael R. and {Bolton}, Adam S. and {Bovy}, Jo and {Brinkmann}, Jonathan and {Brownstein}, Joel R. and {Burtin}, Etienne and {Chabanier}, Sol{\`e}ne and {Chapman}, Michael J. and {Choi}, Peter Doohyun and {Chuang}, Chia-Hsun and {Comparat}, Johan and {Cousinou}, Marie-Claude and {Cuceu}, Andrei and {Dawson}, Kyle S. and {de la Torre}, Sylvain and {de Mattia}, Arnaud and {Agathe}, Victoria de Sainte and {des Bourboux}, H{\'e}lion du Mas and {Escoffier}, Stephanie and {Etourneau}, Thomas and {Farr}, James and {Font-Ribera}, Andreu and {Frinchaboy}, Peter M. and {Fromenteau}, Sebastien and {Gil-Mar{\'\i}n}, H{\'e}ctor and {Le Goff}, Jean-Marc and {Gonzalez-Morales}, Alma X. and {Gonzalez-Perez}, Violeta and {Grabowski}, Kathleen and {Guy}, Julien and {Hawken}, Adam J. and {Hou}, Jiamin and {Kong}, Hui and {Parker}, James and {Klaene}, Mark and {Kneib}, Jean-Paul and {Lin}, Sicheng and {Long}, Daniel and {Lyke}, Brad W. and {de la Macorra}, Axel and {Martini}, Paul and {Masters}, Karen and {Mohammad}, Faizan G. and {Moon}, Jeongin and {Mueller}, Eva-Maria and {Mu{\-n}oz-Guti{\'e}rrez}, Andrea and {Myers}, Adam D. and {Nadathur}, Seshadri and {Neveux}, Richard and {Newman}, Jeffrey A. and {Noterdaeme}, Pasquier and {Oravetz}, Audrey and {Oravetz}, Daniel and {Palanque-Delabrouille}, Nathalie and {Pan}, Kaike and {Paviot}, Romain and {Percival}, Will J. and {P{\'e}rez-R{\`a}fols}, Ignasi and {Petitjean}, Patrick and {Pieri}, Matthew M. and {Prakash}, Abhishek and {Raichoor}, Anand and {Ravoux}, Corentin and {Rezaie}, Mehdi and {Rich}, James and {Ross}, Ashley J. and {Rossi}, Graziano and {Ruggeri}, Rossana and {Ruhlmann-Kleider}, Vanina and {S{\'a}nchez}, Ariel G. and {S{\'a}nchez}, F. Javier and {S{\'a}nchez-Gallego}, Jos{\'e} R. and {Sayres}, Conor and {Schneider}, Donald P. and {Seo}, Hee-Jong and {Shafieloo}, Arman and {Slosar}, An{\v{z}}e and {Smith}, Alex and {Stermer}, Julianna and {Tamone}, Amelie and {Tinker}, Jeremy L. and {Tojeiro}, Rita and {Vargas-Maga{\-n}a}, Mariana and {Variu}, Andrei and {Wang}, Yuting and {Weaver}, Benjamin A. and {Weijmans}, Anne-Marie and {Y{\`e}che}, Christophe and {Zarrouk}, Pauline and {Zhao}, Cheng and {Zhao}, Gong-Bo and {Zheng}, Zheng},
        title = "{Completed SDSS-IV extended Baryon Oscillation Spectroscopic Survey: Cosmological implications from two decades of spectroscopic surveys at the Apache Point Observatory}",
      journal = {\prd},
         year = 2021,
        month = apr,
       volume = {103},
       number = {8},
          eid = {083533},
        pages = {083533},
          doi = {10.1103/PhysRevD.103.083533},
archivePrefix = {arXiv},
       eprint = {2007.08991},
 primaryClass = {astro-ph.CO},
       adsurl = {https://ui.adsabs.harvard.edu/abs/2021PhRvD.103h3533A}
}

@ARTICLE{2016AJ....151...44D,
       author = {{Dawson}, Kyle S. and {Kneib}, Jean-Paul and {Percival}, Will J. and {Alam}, Shadab and {Albareti}, Franco D. and {Anderson}, Scott F. and {Armengaud}, Eric and {Aubourg}, {\'E}ric and {Bailey}, Stephen and {Bautista}, Julian E. and {Berlind}, Andreas A. and {Bershady}, Matthew A. and {Beutler}, Florian and {Bizyaev}, Dmitry and {Blanton}, Michael R. and {Blomqvist}, Michael and {Bolton}, Adam S. and {Bovy}, Jo and {Brandt}, W.-N. and {Brinkmann}, Jon and {Brownstein}, Joel R. and {Burtin}, Etienne and {Busca}, N.-G. and {Cai}, Zheng and {Chuang}, Chia-Hsun and {Clerc}, Nicolas and {Comparat}, Johan and {Cope}, Frances and {Croft}, Rupert A.-C. and {Cruz-Gonzalez}, Irene and {da Costa}, Luiz N. and {Cousinou}, Marie-Claude and {Darling}, Jeremy and {de la Macorra}, Axel and {de la Torre}, Sylvain and {Delubac}, Timoth{\'e}e and {du Mas des Bourboux}, H{\'e}lion and {Dwelly}, Tom and {Ealet}, Anne and {Eisenstein}, Daniel J. and {Eracleous}, Michael and {Escoffier}, S. and {Fan}, Xiaohui and {Finoguenov}, Alexis and {Font-Ribera}, Andreu and {Frinchaboy}, Peter and {Gaulme}, Patrick and {Georgakakis}, Antonis and {Green}, Paul and {Guo}, Hong and {Guy}, Julien and {Ho}, Shirley and {Holder}, Diana and {Huehnerhoff}, Joe and {Hutchinson}, Timothy and {Jing}, Yipeng and {Jullo}, Eric and {Kamble}, Vikrant and {Kinemuchi}, Karen and {Kirkby}, David and {Kitaura}, Francisco-Shu and {Klaene}, Mark A. and {Laher}, Russ R. and {Lang}, Dustin and {Laurent}, Pierre and {Le Goff}, Jean-Marc and {Li}, Cheng and {Liang}, Yu and {Lima}, Marcos and {Lin}, Qiufan and {Lin}, Weipeng and {Lin}, Yen-Ting and {Long}, Daniel C. and {Lundgren}, Britt and {MacDonald}, Nicholas and {Geimba Maia}, Marcio Antonio and {Malanushenko}, Elena and {Malanushenko}, Viktor and {Mariappan}, Vivek and {McBride}, Cameron K. and {McGreer}, Ian D. and {M{\'e}nard}, Brice and {Merloni}, Andrea and {Meza}, Andres and {Montero-Dorta}, Antonio D. and {Muna}, Demitri and {Myers}, Adam D. and {Nandra}, Kirpal and {Naugle}, Tracy and {Newman}, Jeffrey A. and {Noterdaeme}, Pasquier and {Nugent}, Peter and {Ogando}, Ricardo and {Olmstead}, Matthew D. and {Oravetz}, Audrey and {Oravetz}, Daniel J. and {Padmanabhan}, Nikhil and {Palanque-Delabrouille}, Nathalie and {Pan}, Kaike and {Parejko}, John K. and {P{\^a}ris}, Isabelle and {Peacock}, John A. and {Petitjean}, Patrick and {Pieri}, Matthew M. and {Pisani}, Alice and {Prada}, Francisco and {Prakash}, Abhishek and {Raichoor}, Anand and {Reid}, Beth and {Rich}, James and {Ridl}, Jethro and {Rodriguez-Torres}, Sergio and {Carnero Rosell}, Aurelio and {Ross}, Ashley J. and {Rossi}, Graziano and {Ruan}, John and {Salvato}, Mara and {Sayres}, Conor and {Schneider}, Donald P. and {Schlegel}, David J. and {Seljak}, Uros and {Seo}, Hee-Jong and {Sesar}, Branimir and {Shandera}, Sarah and {Shu}, Yiping and {Slosar}, An{\v{z}}e and {Sobreira}, Flavia and {Streblyanska}, Alina and {Suzuki}, Nao and {Taylor}, Donna and {Tao}, Charling and {Tinker}, Jeremy L. and {Tojeiro}, Rita and {Vargas-Maga{\-n}a}, Mariana and {Wang}, Yuting and {Weaver}, Benjamin A. and {Weinberg}, David H. and {White}, Martin and {Wood-Vasey}, W.-M. and {Yeche}, Christophe and {Zhai}, Zhongxu and {Zhao}, Cheng and {Zhao}, Gong-bo and {Zheng}, Zheng and {Ben Zhu}, Guangtun and {Zou}, Hu},
        title = "{The SDSS-IV Extended Baryon Oscillation Spectroscopic Survey: Overview and Early Data}",
      journal = {\aj},
         year = 2016,
        month = feb,
       volume = {151},
       number = {2},
          eid = {44},
        pages = {44},
          doi = {10.3847/0004-6256/151/2/44},
archivePrefix = {arXiv},
       eprint = {1508.04473},
 primaryClass = {astro-ph.CO},
       adsurl = {https://ui.adsabs.harvard.edu/abs/2016AJ....151...44D}
}

@ARTICLE{2025JCAP...02..021A,
       author = {{Adame}, A.-G. and {Aguilar}, J. and {Ahlen}, S. and {Alam}, S. and {Alexander}, D.-M. and {Alvarez}, M. and {Alves}, O. and {Anand}, A. and {Andrade}, U. and {Armengaud}, E. and {Avila}, S. and {Aviles}, A. and {Awan}, H. and {Bahr-Kalus}, B. and {Bailey}, S. and {Baltay}, C. and {Bault}, A. and {Behera}, J. and {BenZvi}, S. and {Bera}, A. and {Beutler}, F. and {Bianchi}, D. and {Blake}, C. and {Blum}, R. and {Brieden}, S. and {Brodzeller}, A. and {Brooks}, D. and {Buckley-Geer}, E. and {Burtin}, E. and {Calderon}, R. and {Canning}, R. and {Carnero Rosell}, A. and {Cereskaite}, R. and {Cervantes-Cota}, J.-L. and {Chabanier}, S. and {Chaussidon}, E. and {Chaves-Montero}, J. and {Chen}, S. and {Chen}, X. and {Claybaugh}, T. and {Cole}, S. and {Cuceu}, A. and {Davis}, T.-M. and {Dawson}, K. and {de la Macorra}, A. and {de Mattia}, A. and {Deiosso}, N. and {Dey}, A. and {Dey}, B. and {Ding}, Z. and {Doel}, P. and {Edelstein}, J. and {Eftekharzadeh}, S. and {Eisenstein}, D.-J. and {Elliott}, A. and {Fagrelius}, P. and {Fanning}, K. and {Ferraro}, S. and {Ereza}, J. and {Findlay}, N. and {Flaugher}, B. and {Font-Ribera}, A. and {Forero-S{\'a}nchez}, D. and {Forero-Romero}, J.-E. and {Frenk}, C.-S. and {Garcia-Quintero}, C. and {Gazta{\-n}aga}, E. and {Gil-Mar{\'\i}n}, H. and {Gontcho a Gontcho}, S. and {Gonzalez-Morales}, A.-X. and {Gonzalez-Perez}, V. and {Gordon}, C. and {Green}, D. and {Gruen}, D. and {Gsponer}, R. and {Gutierrez}, G. and {Guy}, J. and {Hadzhiyska}, B. and {Hahn}, C. and {Hanif}, M.-M.-S. and {Herrera-Alcantar}, H.-K. and {Honscheid}, K. and {Howlett}, C. and {Huterer}, D. and {Ir{\v{s}}i{\v{c}}}, V. and {Ishak}, M. and {Juneau}, S. and {Kara{\c{c}}ayl{\i}}, N.-G. and {Kehoe}, R. and {Kent}, S. and {Kirkby}, D. and {Kremin}, A. and {Krolewski}, A. and {Lai}, Y. and {Lan}, T. -W. and {Landriau}, M. and {Lang}, D. and {Lasker}, J. and {Le Goff}, J.-M. and {Le Guillou}, L. and {Leauthaud}, A. and {Levi}, M.-E. and {Li}, T.-S. and {Linder}, E. and {Lodha}, K. and {Magneville}, C. and {Manera}, M. and {Margala}, D. and {Martini}, P. and {Maus}, M. and {McDonald}, P. and {Medina-Varela}, L. and {Meisner}, A. and {Mena-Fern{\'a}ndez}, J. and {Miquel}, R. and {Moon}, J. and {Moore}, S. and {Moustakas}, J. and {Mueller}, E. and {Mu{\-n}oz-Guti{\'e}rrez}, A. and {Myers}, A.-D. and {Nadathur}, S. and {Napolitano}, L. and {Neveux}, R. and {Newman}, J.-A. and {Nguyen}, N.-M. and {Nie}, J. and {Niz}, G. and {Noriega}, H.-E. and {Padmanabhan}, N. and {Paillas}, E. and {Palanque-Delabrouille}, N. and {Pan}, J. and {Penmetsa}, S. and {Percival}, W.-J. and {Pieri}, M.-M. and {Pinon}, M. and {Poppett}, C. and {Porredon}, A. and {Prada}, F. and {P{\'e}rez-Fern{\'a}ndez}, A. and {P{\'e}rez-R{\`a}fols}, I. and {Rabinowitz}, D. and {Raichoor}, A. and {Ram{\'\i}rez-P{\'e}rez}, C. and {Ramirez-Solano}, S. and {Rashkovetskyi}, M. and {Ravoux}, C. and {Rezaie}, M. and {Rich}, J. and {Rocher}, A. and {Rockosi}, C. and {Roe}, N.-A. and {Rosado-Marin}, A. and {Ross}, A.-J. and {Rossi}, G. and {Ruggeri}, R. and {Ruhlmann-Kleider}, V. and {Samushia}, L. and {Sanchez}, E. and {Saulder}, C. and {Schlafly}, E.-F. and {Schlegel}, D. and {Schubnell}, M. and {Seo}, H. and {Shafieloo}, A. and {Sharples}, R. and {Silber}, J. and {Slosar}, A. and {Smith}, A. and {Sprayberry}, D. and {Tan}, T. and {Tarl{\'e}}, G. and {Taylor}, P. and {Trusov}, S. and {Ure{\-n}a-L{\'o}pez}, L.-A. and {Vaisakh}, R. and {Valcin}, D. and {Valdes}, F. and {Vargas-Maga{\-n}a}, M. and {Verde}, L. and {Walther}, M. and {Wang}, B. and {Wang}, M.-S. and {Weaver}, B.-A. and {Weaverdyck}, N. and {Wechsler}, R.-H. and {Weinberg}, D.-H. and {White}, M. and {Yu}, J. and {Yu}, Y. and {Yuan}, S. and {Y{\`e}che}, C. and {Zaborowski}, E.-A. and {Zarrouk}, P. and {Zhang}, H. and {Zhao}, C. and {Zhao}, R. and {Zhou}, R. and {Zhuang}, T.},
        title = "{DESI 2024 VI: cosmological constraints from the measurements of baryon acoustic oscillations}",
      journal = {\jcap},
         year = 2025,
        month = feb,
       volume = {2025},
       number = {2},
          eid = {021},
        pages = {021},
          doi = {10.1088/1475-7516/2025/02/021},
archivePrefix = {arXiv},
       eprint = {2404.03002},
 primaryClass = {astro-ph.CO},
       adsurl = {https://ui.adsabs.harvard.edu/abs/2025JCAP...02..021A}
}

@ARTICLE{2022PhR...984....1S,
       author = {{Sch{\"o}neberg}, Nils and {Abell{\'a}n}, Guillermo Franco and {S{\'a}nchez}, Andrea P{\'e}rez and {Witte}, Samuel J. and {Poulin}, Vivian and {Lesgourgues}, Julien},
        title = "{The H$_{0}$ Olympics: A fair ranking of proposed models}",
      journal = {\physrep},
         year = 2022,
        month = oct,
       volume = {984},
        pages = {1-55},
          doi = {10.1016/j.physrep.2022.07.001},
archivePrefix = {arXiv},
       eprint = {2107.10291},
 primaryClass = {astro-ph.CO},
       adsurl = {https://ui.adsabs.harvard.edu/abs/2022PhR...984....1S}
}

@article{Scolnic_2022,
doi = {10.3847/1538-4357/ac8b7a},
url = {https://dx.doi.org/10.3847/1538-4357/ac8b7a},
year = {2022},
month = oct,
publisher = {The American Astronomical Society},
volume = {938},
number = {2},
pages = {113},
author = {Scolnic, Dan and Brout, Dillon and Carr, Anthony and Riess, Adam G. and Davis, Tamara M. and Dwomoh, Arianna and Jones, David O. and Ali, Noor and Charvu, Pranav and Chen, Rebecca and Peterson, Erik R. and Popovic, Brodie and Rose, Benjamin M. and Wood, Charlotte M. and Brown, Peter J. and Chambers, Ken and Coulter, David A. and Dettman, Kyle G. and Dimitriadis, Georgios and Filippenko, Alexei V. and Foley, Ryan J. and Jha, Saurabh W. and Kilpatrick, Charles D. and Kirshner, Robert P. and Pan, Yen-Chen and Rest, Armin and Rojas-Bravo, Cesar and Siebert, Matthew R. and Stahl, Benjamin E. and Zheng, WeiKang},
title = {The Pantheon+ Analysis: The Full Data Set and Light-curve Release},
journal = {The Astrophysical Journal}
}

@ARTICLE{2022ApJ...938..110B,
       author = {{Brout}, Dillon and {Scolnic}, Dan and {Popovic}, Brodie and {Riess}, Adam G. and {Carr}, Anthony and {Zuntz}, Joe and {Kessler}, Rick and {Davis}, Tamara M. and {Hinton}, Samuel and {Jones}, David and {Kenworthy}, W. D'Arcy and {Peterson}, Erik R. and {Said}, Khaled and {Taylor}, Georgie and {Ali}, Noor and {Armstrong}, Patrick and {Charvu}, Pranav and {Dwomoh}, Arianna and {Meldorf}, Cole and {Palmese}, Antonella and {Qu}, Helen and {Rose}, Benjamin M. and {Sanchez}, Bruno and {Stubbs}, Christopher W. and {Vincenzi}, Maria and {Wood}, Charlotte M. and {Brown}, Peter J. and {Chen}, Rebecca and {Chambers}, Ken and {Coulter}, David A. and {Dai}, Mi and {Dimitriadis}, Georgios and {Filippenko}, Alexei V. and {Foley}, Ryan J. and {Jha}, Saurabh W. and {Kelsey}, Lisa and {Kirshner}, Robert P. and {M{\"o}ller}, Anais and {Muir}, Jessie and {Nadathur}, Seshadri and {Pan}, Yen-Chen and {Rest}, Armin and {Rojas-Bravo}, Cesar and {Sako}, Masao and {Siebert}, Matthew R. and {Smith}, Mat and {Stahl}, Benjamin E. and {Wiseman}, Phil},
        title = "{The Pantheon+ Analysis: Cosmological Constraints}",
      journal = {\apj},
         year = 2022,
        month = oct,
       volume = {938},
       number = {2},
          eid = {110},
        pages = {110},
          doi = {10.3847/1538-4357/ac8e04},
archivePrefix = {arXiv},
       eprint = {2202.04077},
 primaryClass = {astro-ph.CO},
       adsurl = {https://ui.adsabs.harvard.edu/abs/2022ApJ...938..110B}
}

@ARTICLE{1990MNRAS.247P...1M,
       author = {{Maddox}, S.-J. and {Sutherland}, W.-J. and {Efstathiou}, G. and {Loveday}, J. and {Peterson}, B.-A.},
        title = "{Galaxy Evolution at Low Redshift}",
      journal = {\mnras},
         year = 1990,
        month = nov,
       volume = {247},
        pages = {1P},
       adsurl = {https://ui.adsabs.harvard.edu/abs/1990MNRAS.247P...1M}
}

@article{10.1093/mnras/stt2024,
    author = {Whitbourn, J. R. and Shanks, T.},
    title = {The local hole revealed by galaxy counts and redshifts},
    journal = {Monthly Notices of the Royal Astronomical Society},
    volume = {437},
    number = {3},
    pages = {2146-2162},
    year = {2013},
    month = {11},
    issn = {0035-8711},
    doi = {10.1093/mnras/stt2024},
    url = {https://doi.org/10.1093/mnras/stt2024},
    eprint = {https://academic.oup.com/mnras/article-pdf/437/3/2146/18455902/stt2024.pdf},
}

@ARTICLE{2013A&A...555A.117R,
       author = {{Rubart}, M. and {Schwarz}, D.-J.},
        title = "{Cosmic radio dipole from NVSS and WENSS}",
      journal = {\aap},
         year = 2013,
        month = jul,
       volume = {555},
          eid = {A117},
        pages = {A117},
          doi = {10.1051/0004-6361/201321215},
archivePrefix = {arXiv},
       eprint = {1301.5559},
 primaryClass = {astro-ph.CO},
       adsurl = {https://ui.adsabs.harvard.edu/abs/2013A&A...555A.117R}
}

@ARTICLE{2015A&A...574A..26B,
       author = {{B{\"o}hringer}, Hans and {Chon}, Gayoung and {Bristow}, Martyn and {Collins}, Chris A.},
        title = "{The extended ROSAT-ESO Flux-Limited X-ray Galaxy Cluster Survey (REFLEX II). V. Exploring a local underdensity in the southern sky}",
      journal = {\aap},
         year = 2015,
        month = feb,
       volume = {574},
          eid = {A26},
        pages = {A26},
          doi = {10.1051/0004-6361/201424817},
archivePrefix = {arXiv},
       eprint = {1410.2172},
 primaryClass = {astro-ph.CO},
       adsurl = {https://ui.adsabs.harvard.edu/abs/2015A&A...574A..26B}
}

@ARTICLE{2023arXiv231001473K,
       author = {{Kroupa}, Pavel},
        title = "{A modern view of galaxies and their stellar populations}",
      journal = {arXiv e-prints},
         year = 2023,
        month = oct,
          eid = {arXiv:2310.01473},
        pages = {arXiv:2310.01473},
          doi = {10.48550/arXiv.2310.01473},
archivePrefix = {arXiv},
       eprint = {2310.01473},
 primaryClass = {astro-ph.GA},
       adsurl = {https://ui.adsabs.harvard.edu/abs/2023arXiv231001473K}
}

@ARTICLE{2016A&A...594A..16P,
       author = {{Planck Collaboration}},
        title = "{Planck 2015 results. XVI. Isotropy and statistics of the CMB}",
      journal = {\aap},
         year = 2016,
        month = sep,
       volume = {594},
          eid = {A16},
        pages = {A16},
          doi = {10.1051/0004-6361/201526681},
archivePrefix = {arXiv},
       eprint = {1506.07135},
 primaryClass = {astro-ph.CO},
       adsurl = {https://ui.adsabs.harvard.edu/abs/2016A&A...594A..16P}
}

@ARTICLE{Sergij2,
       author = {{Mazurenko}, Sergij and {Banik}, Indranil and {Kroupa}, Pavel},
        title = "{The redshift dependence of the inferred H$_{0}$ in a local void solution to the Hubble tension}",
      journal = {\mnras},
         year = 2025,
        month = feb,
       volume = {536},
       number = {4},
        pages = {3232-3241},
          doi = {10.1093/mnras/stae2758},
archivePrefix = {arXiv},
       eprint = {2412.12245},
 primaryClass = {astro-ph.CO},
       adsurl = {https://ui.adsabs.harvard.edu/abs/2025MNRAS.536.3232M}
}

@ARTICLE{Sergij1,
       author = {{Mazurenko}, Sergij and {Banik}, Indranil and {Kroupa}, Pavel and {Haslbauer}, Moritz},
        title = "{A simultaneous solution to the Hubble tension and observed bulk flow within 250 h$^{-1}$ Mpc}",
      journal = {\mnras},
         year = 2024,
        month = jan,
       volume = {527},
       number = {3},
        pages = {4388-4396},
          doi = {10.1093/mnras/stad3357},
archivePrefix = {arXiv},
       eprint = {2311.17988},
 primaryClass = {astro-ph.CO},
       adsurl = {https://ui.adsabs.harvard.edu/abs/2024MNRAS.527.4388M}
}

@ARTICLE{Banik_NS,
       author = {{Banik}, Indranil and {Samaras}, Nick},
        title = "{Constraints on the Hubble and matter density parameters with and without modelling the CMB anisotropies}",
      journal = {arXiv e-prints},
         year = 2024,
        month = oct,
          eid = {arXiv:2410.00804},
        pages = {arXiv:2410.00804},
          doi = {10.48550/arXiv.2410.00804},
archivePrefix = {arXiv},
       eprint = {2410.00804},
 primaryClass = {astro-ph.CO},
       adsurl = {https://ui.adsabs.harvard.edu/abs/2024arXiv241000804B}
}

@article{Haslbauer_KBC,
    author = {Haslbauer, Moritz and Banik, Indranil and Kroupa, Pavel},
    title = {The KBC void and Hubble tension contradict ΛCDM on a Gpc scale − Milgromian dynamics as a possible solution},
    journal = {Monthly Notices of the Royal Astronomical Society},
    volume = {499},
    number = {2},
    pages = {2845-2883},
    year = {2020},
    month = {10},
    issn = {0035-8711},
    doi = {10.1093/mnras/staa2348},
    url = {https://doi.org/10.1093/mnras/staa2348},
    eprint = {https://academic.oup.com/mnras/article-pdf/499/2/2845/34031754/staa2348.pdf},
}

@ARTICLE{2023MNRAS.524.1885W,
       author = {{Watkins}, Richard and {Allen}, Trey and {Bradford}, Collin James and {Ramon}, Albert and {Walker}, Alexandra and {Feldman}, Hume A. and {Cionitti}, Rachel and {Al-Shorman}, Yara and {Kourkchi}, Ehsan and {Tully}, R. Brent},
        title = "{Analysing the large-scale bulk flow using cosmicflows4: increasing tension with the standard cosmological model}",
      journal = {\mnras},
         year = 2023,
        month = sep,
       volume = {524},
       number = {2},
        pages = {1885-1892},
          doi = {10.1093/mnras/stad1984},
archivePrefix = {arXiv},
       eprint = {2302.02028},
 primaryClass = {astro-ph.CO},
       adsurl = {https://ui.adsabs.harvard.edu/abs/2023MNRAS.524.1885W}
}

@ARTICLE{Banik_desmond,
       author = {{Banik}, Indranil and {Desmond}, Harry and {Samaras}, Nick},
        title = "{Challenges to a sharp change in G as a solution to the Hubble tension}",
      journal = {\mnras},
         year = 2025,
        month = may,
       volume = {539},
       number = {2},
        pages = {1553-1561},
          doi = {10.1093/mnras/staf567},
archivePrefix = {arXiv},
       eprint = {2411.15301},
 primaryClass = {astro-ph.CO},
       adsurl = {https://ui.adsabs.harvard.edu/abs/2025MNRAS.539.1553B}
}

@article{Lin_2021,
doi = {10.3847/1538-4357/ac12cf},
url = {https://dx.doi.org/10.3847/1538-4357/ac12cf},
year = {2021},
month = {10},
publisher = {The American Astronomical Society},
volume = {920},
number = {2},
pages = {159},
author = {Lin, Weikang and Chen, Xingang and Mack, Katherine J.},
title = {Early Universe Physics Insensitive and Uncalibrated Cosmic Standards: Constraints on Ωm and Implications for the Hubble Tension},
journal = {The Astrophysical Journal}
}

@ARTICLE{2022PhRvD.106f3530P,
       author = {{Philcox}, Oliver H.~E. and {Farren}, Gerrit S. and {Sherwin}, Blake D. and {Baxter}, Eric J. and {Brout}, Dillon J.},
        title = "{Determining the Hubble constant without the sound horizon: A 3.6 \% constraint on H$_{0}$ from galaxy surveys, CMB lensing, and supernovae}",
      journal = {\prd},
         year = 2022,
        month = sep,
       volume = {106},
       number = {6},
          eid = {063530},
        pages = {063530},
          doi = {10.1103/PhysRevD.106.063530},
archivePrefix = {arXiv},
       eprint = {2204.02984},
 primaryClass = {astro-ph.CO},
       adsurl = {https://ui.adsabs.harvard.edu/abs/2022PhRvD.106f3530P}
}

@article{Eisenstein_1998,
doi = {10.1086/305424},
url = {https://dx.doi.org/10.1086/305424},
year = {1998},
month = {4},
publisher = {},
volume = {496},
number = {2},
pages = {605},
author = {Eisenstein, Daniel J. and Hu, Wayne},
title = {Baryonic Features in the Matter Transfer Function},
journal = {The Astrophysical Journal}
}

@article{Cimatti_2023,
doi = {10.3847/1538-4357/ace439},
url = {https://dx.doi.org/10.3847/1538-4357/ace439},
year = {2023},
month = {8},
publisher = {The American Astronomical Society},
volume = {953},
number = {2},
pages = {149},
author = {Cimatti, Andrea and Moresco, Michele},
title = {Revisiting the Oldest Stars as Cosmological Probes: New Constraints on the Hubble Constant},
journal = {The Astrophysical Journal}
}

@ARTICLE{1996A&A...312..345D,
       author = {{degl'Innocenti}, S. and {Fiorentini}, G. and {Raffelt}, G.-G. and {Ricci}, B. and {Weiss}, A.},
        title = "{Time-variation of Newton's constant and the age of globular clusters.}",
      journal = {\aap},
         year = 1996,
        month = 8,
       volume = {312},
        pages = {345-352},
          doi = {10.48550/arXiv.astro-ph/9509090},
archivePrefix = {arXiv},
       eprint = {astro-ph/9509090},
 primaryClass = {astro-ph},
       adsurl = {https://ui.adsabs.harvard.edu/abs/1996A&A...312..345D}
}

@ARTICLE{2012PhRvD..85l3006D,
       author = {{Davis}, Anne-Christine and {Lim}, Eugene A. and {Sakstein}, Jeremy and {Shaw}, Douglas J.},
        title = "{Modified gravity makes galaxies brighter}",
      journal = {\prd},
         year = 2012,
        month = jun,
       volume = {85},
       number = {12},
          eid = {123006},
        pages = {123006},
          doi = {10.1103/PhysRevD.85.123006},
archivePrefix = {arXiv},
       eprint = {1102.5278},
 primaryClass = {astro-ph.CO},
       adsurl = {https://ui.adsabs.harvard.edu/abs/2012PhRvD..85l3006D}
}

@ARTICLE{2020RSPSA.47600303A,
       author = {{Arnscheidt}, Constantin W. and {Rothman}, Daniel H.},
        title = "{Routes to global glaciation}",
      journal = {Proceedings of the Royal Society of London Series A},
         year = 2020,
        month = jul,
       volume = {476},
       number = {2239},
          eid = {20200303},
        pages = {20200303},
          doi = {10.1098/rspa.2020.0303},
       adsurl = {https://ui.adsabs.harvard.edu/abs/2020RSPSA.47600303A}
}

@ARTICLE{2012Sci...338..651C,
       author = {{Connelly}, James N. and {Bizzarro}, Martin and {Krot}, Alexander N. and {Nordlund}, {\r{A}}ke and {Wielandt}, Daniel and {Ivanova}, Marina A.},
        title = "{The Absolute Chronology and Thermal Processing of Solids in the Solar Protoplanetary Disk}",
      journal = {Science},
         year = 2012,
        month = nov,
       volume = {338},
       number = {6107},
        pages = {651},
          doi = {10.1126/science.1226919},
       adsurl = {https://ui.adsabs.harvard.edu/abs/2012Sci...338..651C}
}

@ARTICLE{2022ApJ...938...36R,
       author = {{Riess}, Adam G. and {Breuval}, Louise and {Yuan}, Wenlong and {Casertano}, Stefano and {Macri}, Lucas M. and {Bowers}, J. Bradley and {Scolnic}, Dan and {Cantat-Gaudin}, Tristan and {Anderson}, Richard I. and {Cruz Reyes}, Mauricio},
        title = "{Cluster Cepheids with High Precision Gaia Parallaxes, Low Zero-point Uncertainties, and Hubble Space Telescope Photometry}",
      journal = {\apj},
         year = 2022,
        month = oct,
       volume = {938},
       number = {1},
          eid = {36},
        pages = {36},
          doi = {10.3847/1538-4357/ac8f24},
archivePrefix = {arXiv},
       eprint = {2208.01045},
 primaryClass = {astro-ph.CO},
       adsurl = {https://ui.adsabs.harvard.edu/abs/2022ApJ...938...36R}
}

@ARTICLE{2025A&A...697A.109L,
       author = {{Lamine}, B. and {Ozdalkiran}, Y. and {Mirouze}, L. and {Erdogan}, F. and {Ilic}, S. and {Tutusaus}, I. and {Kou}, R. and {Blanchard}, A.},
        title = "{Cosmological measurement of the gravitational constant G using the CMB, BAO, and BBN}",
      journal = {\aap},
         year = 2025,
        month = may,
       volume = {697},
          eid = {A109},
        pages = {A109},
          doi = {10.1051/0004-6361/202451602},
archivePrefix = {arXiv},
       eprint = {2407.15553},
 primaryClass = {astro-ph.CO},
       adsurl = {https://ui.adsabs.harvard.edu/abs/2025A&A...697A.109L}
}

@article{PhysRevD.100.043537,
  title = {Local resolution of the Hubble tension: The impact of screened fifth forces on the cosmic distance ladder},
  author = {Desmond, Harry and Jain, Bhuvnesh and Sakstein, Jeremy},
  journal = {Phys. Rev. D},
  volume = {100},
  issue = {4},
  pages = {043537},
  numpages = {20},
  year = {2019},
  month = Aug,
  publisher = {American Physical Society},
  doi = {10.1103/PhysRevD.100.043537},
  url = {https://link.aps.org/doi/10.1103/PhysRevD.100.043537}
}

@ARTICLE{2018PhRvD..97h3505W,
       author = {{Wright}, Bill S. and {Li}, Baojiu},
        title = "{Type Ia supernovae, standardizable candles, and gravity}",
      journal = {\prd},
         year = 2018,
        month = apr,
       volume = {97},
       number = {8},
          eid = {083505},
        pages = {083505},
          doi = {10.1103/PhysRevD.97.083505},
archivePrefix = {arXiv},
       eprint = {1710.07018},
 primaryClass = {astro-ph.CO},
       adsurl = {https://ui.adsabs.harvard.edu/abs/2018PhRvD..97h3505W}
}

@ARTICLE{2015CaJPh..93..250M,
       author = {{McGaugh}, Stacy S.},
        title = "{A tale of two paradigms: the mutual incommensurability of {\ensuremath{\Lambda}}CDM and MOND}",
      journal = {Canadian Journal of Physics},
         year = 2015,
        month = feb,
       volume = {93},
       number = {2},
        pages = {250-259},
          doi = {10.1139/cjp-2014-0203},
archivePrefix = {arXiv},
       eprint = {1404.7525},
 primaryClass = {astro-ph.CO},
       adsurl = {https://ui.adsabs.harvard.edu/abs/2015CaJPh..93..250M}
}

@article{Bulbul_2014,
doi = {10.1088/0004-637X/789/1/13},
url = {https://dx.doi.org/10.1088/0004-637X/789/1/13},
year = {2014},
month = jun,
publisher = {The American Astronomical Society},
volume = {789},
number = {1},
pages = {13},
author = {Bulbul, Esra and Markevitch, Maxim and Foster, Adam and Smith, Randall K. and Loewenstein, Michael and Randall, Scott W.},
title = {DETECTION OF AN UNIDENTIFIED EMISSION LINE IN THE STACKED X-RAY SPECTRUM OF GALAXY CLUSTERS},
journal = {The Astrophysical Journal}
}

@ARTICLE{2017arXiv170208430M,
       author = {{Merle}, Alexander},
        title = "{keV sterile neutrino Dark Matter}",
      journal = {arXiv e-prints},
         year = 2017,
        month = feb,
          eid = {arXiv:1702.08430},
        pages = {arXiv:1702.08430},
          doi = {10.48550/arXiv.1702.08430},
archivePrefix = {arXiv},
       eprint = {1702.08430},
 primaryClass = {hep-ph},
       adsurl = {https://ui.adsabs.harvard.edu/abs/2017arXiv170208430M}
}

@article{Langacker:1988fp,
    author = "Langacker, Paul",
    title = "{Is the Standard Model Unique?}",
    reportNumber = "Print-88-0672 (PENN), UPR-0362T",
    journal = "Comments Nucl. Part. Phys.",
    volume = "19",
    number = "1",
    pages = "1--23",
    year = "1989"
}

@article{PhysRevD.88.043502,
  title = {Warm dark matter as a solution to the small scale crisis: New constraints from high redshift Lyman-$\ensuremath{\alpha}$ forest data},
  author = {Viel, Matteo and Becker, George D. and Bolton, James S. and Haehnelt, Martin G.},
  journal = {Phys. Rev. D},
  volume = {88},
  issue = {4},
  pages = {043502},
  numpages = {19},
  year = {2013},
  month = Aug,
  publisher = {American Physical Society},
  doi = {10.1103/PhysRevD.88.043502},
  url = {https://link.aps.org/doi/10.1103/PhysRevD.88.043502}
}

@article{König_2016,
doi = {10.1088/1475-7516/2016/11/038},
url = {https://dx.doi.org/10.1088/1475-7516/2016/11/038},
year = {2016},
month = nov,
publisher = {},
volume = {2016},
number = {11},
pages = {038},
author = {Johannes König and Alexander Merle and Maximilian Totzauer},
title = {keV sterile neutrino dark matter from singlet scalar decays: the most general case},
journal = {Journal of Cosmology and Astroparticle Physics}
}

@ARTICLE{1994PhRvL..72...17D,
       author = {{Dodelson}, Scott and {Widrow}, Lawrence M.},
        title = "{Sterile neutrinos as dark matter}",
      journal = {\prl},
         year = 1994,
        month = jan,
       volume = {72},
       number = {1},
        pages = {17-20},
          doi = {10.1103/PhysRevLett.72.17},
archivePrefix = {arXiv},
       eprint = {hep-ph/9303287},
 primaryClass = {hep-ph},
       adsurl = {https://ui.adsabs.harvard.edu/abs/1994PhRvL..72...17D}
}

@article{PhysRevLett.134.081804,
  title = {Dual-Baseline Search for Active-to-Sterile Neutrino Oscillations in NOvA},
  author = {Acero, M. A. and Acharya, B. and Adamson, P. and Anfimov, N. and Antoshkin, A. and Arrieta-Diaz, E. and Asquith, L. and Aurisano, A. and Back, A. and Balashov, N. and Baldi, P. and Bambah, B. A. and Bannister, E. F. and Barros, A. and Bat, A. and Bays, K. and Bernstein, R. and Bezerra, T. J. C. and Bhatnagar, V. and Bhattarai, D. and Bhuyan, B. and Bian, J. and Booth, A. C. and Bowles, R. and Brahma, B. and Bromberg, C. and Buchanan, N. and Butkevich, A. and Calvez, S. and Carroll, T. J. and Catano-Mur, E. and Cesar, J. P. and Chatla, A. and Chirco, R. and Choudhary, B. C. and Christensen, A. and Cicala, M. F. and Coan, T. E. and Cooleybeck, A. and Cortes-Parra, C. and Coveyou, D. and Cremonesi, L. and Davies, G. S. and Derwent, P. F. and Ding, P. and Djurcic, Z. and Dobbs, K. and Dolce, M. and Doyle, D. and Tonguino, D. Due\~nas and Dukes, E. C. and Dye, A. and Ehrlich, R. and Ewart, E. and Filip, P. and Frank, M. J. and Gallagher, H. R. and Gao, F. and Giri, A. and Gomes, R. A. and Goodman, M. C. and Groh, M. and Group, R. and Habig, A. and Hakl, F. and Hartnell, J. and Hatcher, R. and Hausner, H. and He, M. and Heller, K. and Hewes, V. and Himmel, A. and Horoho, T. and Ivanova, A. and Jargowsky, B. and Jarosz, J. and Judah, M. and Kakorin, I. and Kalitkina, A. and Kaplan, D. M. and Kirezli-Ozdemir, B. and Kleykamp, J. and Klimov, O. and Koerner, L. W. and Kolupaeva, L. and Kralik, R. and Kumar, A. and Kus, V. and Lackey, T. and Lang, K. and Lesmeister, J. and Lister, A. and Liu, J. and Lock, J. A. and Lokajicek, M. and MacMahon, M. and Magill, S. and Mann, W. A. and Manoharan, M. T. and Plata, M. Manrique and Marshak, M. L. and Martinez-Casales, M. and Matveev, V. and Mehta, B. and Messier, M. D. and Meyer, H. and Miao, T. and Mikola, V. and Miller, W. H. and Mishra, S. and Mishra, S. R. and Mislivec, A. and Mohanta, R. and Moren, A. and Morozova, A. and Mu, W. and Mualem, L. and Muether, M. and Myers, D. and Naples, D. and Nath, A. and Nelleri, S. and Nelson, J. K. and Nichol, R. and Niner, E. and Norman, A. and Norrick, A. and Oh, H. and Olshevskiy, A. and Olson, T. and Ozkaynak, M. and Pal, A. and Paley, J. and Panda, L. and Patterson, R. B. and Pawloski, G. and Petti, R. and Plunkett, R. K. and Prais, L. R. and Rabelhofer, M. and Rafique, A. and Raj, V. and Rajaoalisoa, M. and Ramson, B. and Rebel, B. and Roy, P. and Samoylov, O. and Sanchez, M. C. and Falero, S. S\'anchez and Shanahan, P. and Sharma, P. and Sheshukov, A. and Shmakov, A. and Shivam and Shorrock, W. and Shukla, S. and Singha, D. K. and Singh, I. and Singh, P. and Singh, V. and Smith, E. and Smolik, J. and Snopok, P. and Solomey, N. and Sousa, A. and Soustruznik, K. and Strait, M. and Suter, L. and Sutton, A. and Sutton, K. and Swain, S. and Sweeney, C. and Sztuc, A. and Oregui, B. Tapia and Talukdar, N. and Tas, P. and Thakore, T. and Thomas, J. and Tiras, E. and Titus, M. and Torun, Y. and Tran, D. and Tripathi, J. and Trokan-Tenorio, J. and Urheim, J. and Vahle, P. and Vallari, Z. and Villamil, J. D. and Vockerodt, K. J. and Wallbank, M. and Weber, C. and Wetstein, M. and Whittington, D. and Wickremasinghe, D. A. and Wieber, T. and Wolcott, J. and Wrobel, M. and Wu, S. and Wu, W. and Wu, W. and Xiao, Y. and Yaeggy, B. and Yahaya, A. and Yankelevich, A. and Yonehara, K. and Zadorozhnyy, S. and Zalesak, J. and Zwaska, R.},
  collaboration = {NOvA Collaboration},
  journal = {Phys. Rev. Lett.},
  volume = {134},
  issue = {8},
  pages = {081804},
  numpages = {8},
  year = {2025},
  month = Feb,
  publisher = {American Physical Society},
  doi = {10.1103/PhysRevLett.134.081804},
  url = {https://link.aps.org/doi/10.1103/PhysRevLett.134.081804}
}

@article{PhysRevD.64.112007,
  title = {Evidence for neutrino oscillations from the observation of ${\overline{\ensuremath{\nu}}}_{e}$ appearance in a ${\overline{\ensuremath{\nu}}}_{\ensuremath{\mu}}$ beam},
  author = {Aguilar, A. and Auerbach, L. B. and Burman, R. L. and Caldwell, D. O. and Church, E. D. and Cochran, A. K. and Donahue, J. B. and Fazely, A. and Garvey, G. T. and Gunasingha, R. M. and Imlay, R. and Louis, W. C. and Majkic, R. and Malik, A. and Metcalf, W. and Mills, G. B. and Sandberg, V. and Smith, D. and Stancu, I. and Sung, M. and Tayloe, R. and VanDalen, G. J. and Vernon, W. and Wadia, N. and White, D. H. and Yellin, S.},
  collaboration = {LSND Collaboration},
  journal = {Phys. Rev. D},
  volume = {64},
  issue = {11},
  pages = {112007},
  numpages = {22},
  year = {2001},
  month = Nov,
  publisher = {American Physical Society},
  doi = {10.1103/PhysRevD.64.112007},
  url = {https://link.aps.org/doi/10.1103/PhysRevD.64.112007}
}

@article{Mohanty:2023wl,
  author = "Mohanty, Shailaja",
  title = "{Search for a light sterile neutrino with KATRIN}",
  doi = "10.22323/1.449.0164",
  journal = "PoS",
  year = 2023,
  volume = "EPS-HEP2023",
  pages = "164"
}

@ARTICLE{1993ApJ...416....1K,
       author = {{Klypin}, Anatoly and {Holtzman}, Jon and {Primack}, Joel and {Regos}, Eniko},
        title = "{Structure Formation with Cold plus Hot Dark Matter}",
      journal = {\apj},
         year = 1993,
        month = oct,
       volume = {416},
        pages = {1},
          doi = {10.1086/173210},
archivePrefix = {arXiv},
       eprint = {astro-ph/9305011},
 primaryClass = {astro-ph},
       adsurl = {https://ui.adsabs.harvard.edu/abs/1993ApJ...416....1K}
}

@ARTICLE{1983ApJ...270..365M,
       author = {{Milgrom}, M.},
        title = "{A modification of the Newtonian dynamics as a possible alternative to the hidden mass hypothesis.}",
      journal = {\apj},
         year = 1983,
        month = jul,
       volume = {270},
        pages = {365-370},
          doi = {10.1086/161130},
       adsurl = {https://ui.adsabs.harvard.edu/abs/1983ApJ...270..365M}
}

@ARTICLE{2025A&A...693A.127P,
       author = {{Pflamm-Altenburg}, J.},
        title = "{Asymmetry in the tidal tails of open star clusters from direct N-body integrations in Milgrom-law dynamics}",
      journal = {\aap},
         year = 2025,
        month = jan,
       volume = {693},
          eid = {A127},
        pages = {A127},
          doi = {10.1051/0004-6361/202347796},
archivePrefix = {arXiv},
       eprint = {2411.13675},
 primaryClass = {astro-ph.GA},
       adsurl = {https://ui.adsabs.harvard.edu/abs/2025A&A...693A.127P}
}

@ARTICLE{2016arXiv160604942T,
       author = {{Thies}, Ingo and {Kroupa}, Pavel and {Famaey}, Benoit},
        title = "{Simulating disk galaxies and interactions in Milgromian dynamics}",
      journal = {arXiv e-prints},
         year = 2016,
        month = jun,
          eid = {arXiv:1606.04942},
        pages = {arXiv:1606.04942},
          doi = {10.48550/arXiv.1606.04942},
archivePrefix = {arXiv},
       eprint = {1606.04942},
 primaryClass = {astro-ph.GA},
       adsurl = {https://ui.adsabs.harvard.edu/abs/2016arXiv160604942T}
}

@ARTICLE{2023MNRAS.519.5128N,
       author = {{Nagesh}, Srikanth T. and {Kroupa}, Pavel and {Banik}, Indranil and {Famaey}, Benoit and {Ghafourian}, Neda and {Roshan}, Mahmood and {Thies}, Ingo and {Zhao}, Hongsheng and {Wittenburg}, Nils},
        title = "{Simulations of star-forming main-sequence galaxies in Milgromian gravity}",
      journal = {\mnras},
         year = 2023,
        month = mar,
       volume = {519},
       number = {4},
        pages = {5128-5148},
          doi = {10.1093/mnras/stac3645},
archivePrefix = {arXiv},
       eprint = {2212.07447},
 primaryClass = {astro-ph.GA},
       adsurl = {https://ui.adsabs.harvard.edu/abs/2023MNRAS.519.5128N}
}

@article{Sanders_1999,
doi = {10.1086/311865},
url = {https://dx.doi.org/10.1086/311865},
year = {1998},
month = dec,
publisher = {},
volume = {512},
number = {1},
pages = {L23},
author = {Sanders, R. H.},
title = {The Virial Discrepancy in Clusters of Galaxies in the Context of Modified Newtonian Dynamics},
journal = {The Astrophysical Journal}
}

@ARTICLE{2010MNRAS.402..395A,
       author = {{Angus}, G.~W. and {Famaey}, B. and {Diaferio}, A.},
        title = "{Equilibrium configurations of 11 eV sterile neutrinos in MONDian galaxy clusters}",
      journal = {\mnras},
         year = 2010,
        month = feb,
       volume = {402},
       number = {1},
        pages = {395-408},
          doi = {10.1111/j.1365-2966.2009.15895.x},
archivePrefix = {arXiv},
       eprint = {0906.3322},
 primaryClass = {astro-ph.CO},
       adsurl = {https://ui.adsabs.harvard.edu/abs/2010MNRAS.402..395A}
}

@ARTICLE{2013ApJ...772...10K,
       author = {{Katz}, Harley and {McGaugh}, Stacy and {Teuben}, Peter and {Angus}, G.~W.},
        title = "{Galaxy Cluster Bulk Flows and Collision Velocities in QUMOND}",
      journal = {\apj},
         year = 2013,
        month = jul,
       volume = {772},
       number = {1},
          eid = {10},
        pages = {10},
          doi = {10.1088/0004-637X/772/1/10},
archivePrefix = {arXiv},
       eprint = {1305.3651},
 primaryClass = {astro-ph.CO},
       adsurl = {https://ui.adsabs.harvard.edu/abs/2013ApJ...772...10K}
}

@ARTICLE{2012LRR....15...10F,
       author = {{Famaey}, Beno{\^\i}t and {McGaugh}, Stacy S.},
        title = "{Modified Newtonian Dynamics (MOND): Observational Phenomenology and Relativistic Extensions}",
      journal = {Living Reviews in Relativity},
         year = 2012,
        month = dec,
       volume = {15},
       number = {1},
          eid = {10},
        pages = {10},
          doi = {10.12942/lrr-2012-10},
archivePrefix = {arXiv},
       eprint = {1112.3960},
 primaryClass = {astro-ph.CO},
       adsurl = {https://ui.adsabs.harvard.edu/abs/2012LRR....15...10F}
}

@ARTICLE{2009ApJS..182..608K,
       author = {{Knollmann}, Steffen R. and {Knebe}, Alexander},
        title = "{AHF: Amiga's Halo Finder}",
      journal = {\apjs},
         year = 2009,
        month = jun,
       volume = {182},
       number = {2},
        pages = {608-624},
          doi = {10.1088/0067-0049/182/2/608},
archivePrefix = {arXiv},
       eprint = {0904.3662},
 primaryClass = {astro-ph.CO},
       adsurl = {https://ui.adsabs.harvard.edu/abs/2009ApJS..182..608K}
}

@article{Lewis_2000,
doi = {10.1086/309179},
url = {https://dx.doi.org/10.1086/309179},
year = {2000},
month = aug,
publisher = {},
volume = {538},
number = {2},
pages = {473},
author = {Lewis, Antony and Challinor, Anthony and Lasenby, Anthony},
title = {Efficient Computation of Cosmic Microwave Background Anisotropies in
Closed Friedmann-Robertson-Walker Models},
journal = {The Astrophysical Journal}
}

@article{10.1111/j.1365-2966.2011.18820.x,
    author = {Hahn, Oliver and Abel, Tom},
    title = {Multi-scale initial conditions for cosmological simulations},
    journal = {Monthly Notices of the Royal Astronomical Society},
    volume = {415},
    number = {3},
    pages = {2101-2121},
    year = {2011},
    month = {08},
    issn = {0035-8711},
    doi = {10.1111/j.1365-2966.2011.18820.x},
    url = {https://doi.org/10.1111/j.1365-2966.2011.18820.x},
    eprint = {https://academic.oup.com/mnras/article-pdf/415/3/2101/5967972/mnras0415-2101.pdf},
}

@ARTICLE{2020A&A...641A...1P,
       author = {{Planck Collaboration} and {Aghanim}, N. and {Akrami}, Y. and {Arroja}, F. and {Ashdown}, M. and {Aumont}, J. and {Baccigalupi}, C. and {Ballardini}, M. and {Banday}, A.~J. and {Barreiro}, R.~B. and {Bartolo}, N. and {Basak}, S. and {Battye}, R. and {Benabed}, K. and {Bernard}, J. -P. and {Bersanelli}, M. and {Bielewicz}, P. and {Bock}, J.~J. and {Bond}, J.~R. and {Borrill}, J. and {Bouchet}, F.~R. and {Boulanger}, F. and {Bucher}, M. and {Burigana}, C. and {Butler}, R.~C. and {Calabrese}, E. and {Cardoso}, J. -F. and {Carron}, J. and {Casaponsa}, B. and {Challinor}, A. and {Chiang}, H.~C. and {Colombo}, L.~P.~L. and {Combet}, C. and {Contreras}, D. and {Crill}, B.~P. and {Cuttaia}, F. and {de Bernardis}, P. and {de Zotti}, G. and {Delabrouille}, J. and {Delouis}, J. -M. and {D{\'e}sert}, F. -X. and {Di Valentino}, E. and {Dickinson}, C. and {Diego}, J.~M. and {Donzelli}, S. and {Dor{\'e}}, O. and {Douspis}, M. and {Ducout}, A. and {Dupac}, X. and {Efstathiou}, G. and {Elsner}, F. and {En{\ss}lin}, T.~A. and {Eriksen}, H.~K. and {Falgarone}, E. and {Fantaye}, Y. and {Fergusson}, J. and {Fernandez-Cobos}, R. and {Finelli}, F. and {Forastieri}, F. and {Frailis}, M. and {Franceschi}, E. and {Frolov}, A. and {Galeotta}, S. and {Galli}, S. and {Ganga}, K. and {G{\'e}nova-Santos}, R.~T. and {Gerbino}, M. and {Ghosh}, T. and {Gonz{\'a}lez-Nuevo}, J. and {G{\'o}rski}, K.~M. and {Gratton}, S. and {Gruppuso}, A. and {Gudmundsson}, J.~E. and {Hamann}, J. and {Handley}, W. and {Hansen}, F.~K. and {Helou}, G. and {Herranz}, D. and {Hildebrandt}, S.~R. and {Hivon}, E. and {Huang}, Z. and {Jaffe}, A.~H. and {Jones}, W.~C. and {Karakci}, A. and {Keih{\"a}nen}, E. and {Keskitalo}, R. and {Kiiveri}, K. and {Kim}, J. and {Kisner}, T.~S. and {Knox}, L. and {Krachmalnicoff}, N. and {Kunz}, M. and {Kurki-Suonio}, H. and {Lagache}, G. and {Lamarre}, J. -M. and {Langer}, M. and {Lasenby}, A. and {Lattanzi}, M. and {Lawrence}, C.~R. and {Le Jeune}, M. and {Leahy}, J.~P. and {Lesgourgues}, J. and {Levrier}, F. and {Lewis}, A. and {Liguori}, M. and {Lilje}, P.~B. and {Lilley}, M. and {Lindholm}, V. and {L{\'o}pez-Caniego}, M. and {Lubin}, P.~M. and {Ma}, Y. -Z. and {Mac{\'\i}as-P{\'e}rez}, J.~F. and {Maggio}, G. and {Maino}, D. and {Mandolesi}, N. and {Mangilli}, A. and {Marcos-Caballero}, A. and {Maris}, M. and {Martin}, P.~G. and {Martinelli}, M. and {Mart{\'\i}nez-Gonz{\'a}lez}, E. and {Matarrese}, S. and {Mauri}, N. and {McEwen}, J.~D. and {Meerburg}, P.~D. and {Meinhold}, P.~R. and {Melchiorri}, A. and {Mennella}, A. and {Migliaccio}, M. and {Millea}, M. and {Mitra}, S. and {Miville-Desch{\^e}nes}, M. -A. and {Molinari}, D. and {Moneti}, A. and {Montier}, L. and {Morgante}, G. and {Moss}, A. and {Mottet}, S. and {M{\"u}nchmeyer}, M. and {Natoli}, P. and {N{\o}rgaard-Nielsen}, H.~U. and {Oxborrow}, C.~A. and {Pagano}, L. and {Paoletti}, D. and {Partridge}, B. and {Patanchon}, G. and {Pearson}, T.~J. and {Peel}, M. and {Peiris}, H.~V. and {Perrotta}, F. and {Pettorino}, V. and {Piacentini}, F. and {Polastri}, L. and {Polenta}, G. and {Puget}, J. -L. and {Rachen}, J.~P. and {Reinecke}, M. and {Remazeilles}, M. and {Renault}, C. and {Renzi}, A. and {Rocha}, G. and {Rosset}, C. and {Roudier}, G. and {Rubi{\~n}o-Mart{\'\i}n}, J.~A. and {Ruiz-Granados}, B. and {Salvati}, L. and {Sandri}, M. and {Savelainen}, M. and {Scott}, D. and {Shellard}, E.~P.~S. and {Shiraishi}, M. and {Sirignano}, C. and {Sirri}, G. and {Spencer}, L.~D. and {Sunyaev}, R. and {Suur-Uski}, A. -S. and {Tauber}, J.~A. and {Tavagnacco}, D. and {Tenti}, M. and {Terenzi}, L. and {Toffolatti}, L. and {Tomasi}, M. and {Trombetti}, T. and {Valiviita}, J. and {Van Tent}, B. and {Vibert}, L. and {Vielva}, P. and {Villa}, F. and {Vittorio}, N. and {Wandelt}, B.~D. and {Wehus}, I.~K. and {White}, M. and {White}, S.~D.~M. and {Zacchei}, A. and {Zonca}, A.},
        title = "{Planck 2018 results. I. Overview and the cosmological legacy of Planck}",
      journal = {\aap},
         year = 2020,
        month = sep,
       volume = {641},
          eid = {A1},
        pages = {A1},
          doi = {10.1051/0004-6361/201833880},
archivePrefix = {arXiv},
       eprint = {1807.06205},
 primaryClass = {astro-ph.CO},
       adsurl = {https://ui.adsabs.harvard.edu/abs/2020A&A...641A...1P}
}

@ARTICLE{2011JCAP...07..034B,
       author = {{Blas}, Diego and {Lesgourgues}, Julien and {Tram}, Thomas},
        title = "{The Cosmic Linear Anisotropy Solving System (CLASS). Part II: Approximation schemes}",
      journal = {\jcap},
         year = 2011,
        month = jul,
       volume = {2011},
       number = {7},
          eid = {034},
        pages = {034},
          doi = {10.1088/1475-7516/2011/07/034},
archivePrefix = {arXiv},
       eprint = {1104.2933},
 primaryClass = {astro-ph.CO},
       adsurl = {https://ui.adsabs.harvard.edu/abs/2011JCAP...07..034B}
}

@ARTICLE{2021CaJPh..99..607N,
       author = {{Nagesh}, S.~T. and {Banik}, I. and {Thies}, I. and {Kroupa}, P. and {Famaey}, B. and {Wittenburg}, N. and {Parziale}, R. and {Haslbauer}, M.},
        title = "{The Phantom of RAMSES user guide for galaxy simulations using Milgromian and Newtonian gravity}",
      journal = {Canadian Journal of Physics},
         year = 2021,
        month = aug,
       volume = {99},
       number = {8},
        pages = {607-613},
          doi = {10.1139/cjp-2020-0624},
archivePrefix = {arXiv},
       eprint = {2101.11011},
 primaryClass = {astro-ph.IM},
       adsurl = {https://ui.adsabs.harvard.edu/abs/2021CaJPh..99..607N}
}

@article{10.1046/j.1365-8711.1998.01459.x,
    author = {Sanders, R.H.},
    title = {Cosmology with modified Newtonian dynamics (MOND)},
    journal = {Monthly Notices of the Royal Astronomical Society},
    volume = {296},
    number = {4},
    pages = {1009-1018},
    year = {1998},
    month = {06},
    issn = {0035-8711},
    doi = {10.1046/j.1365-8711.1998.01459.x},
    url = {https://doi.org/10.1046/j.1365-8711.1998.01459.x},
    eprint = {https://academic.oup.com/mnras/article-pdf/296/4/1009/18408341/296-4-1009.pdf},
}

@ARTICLE{2002A&A...385..337T,
       author = {{Teyssier}, R.},
        title = "{Cosmological hydrodynamics with adaptive mesh refinement. A new high resolution code called RAMSES}",
      journal = {\aap},
         year = 2002,
        month = apr,
       volume = {385},
        pages = {337-364},
          doi = {10.1051/0004-6361:20011817},
archivePrefix = {arXiv},
       eprint = {astro-ph/0111367},
 primaryClass = {astro-ph},
       adsurl = {https://ui.adsabs.harvard.edu/abs/2002A&A...385..337T}
}

@ARTICLE{2022Symm...14.1331B,
       author = {{Banik}, Indranil and {Zhao}, Hongsheng},
        title = "{From Galactic Bars to the Hubble Tension: Weighing Up the Astrophysical Evidence for Milgromian Gravity}",
      journal = {Symmetry},
         year = 2022,
        month = jun,
       volume = {14},
       number = {7},
          eid = {1331},
        pages = {1331},
          doi = {10.3390/sym14071331},
archivePrefix = {arXiv},
       eprint = {2110.06936},
 primaryClass = {astro-ph.CO},
       adsurl = {https://ui.adsabs.harvard.edu/abs/2022Symm...14.1331B}
}

@ARTICLE{1984ApJ...286....7B,
       author = {{Bekenstein}, J. and {Milgrom}, M.},
        title = "{Does the missing mass problem signal the breakdown of Newtonian gravity?}",
      journal = {\apj},
         year = 1984,
        month = nov,
       volume = {286},
        pages = {7-14},
          doi = {10.1086/162570},
       adsurl = {https://ui.adsabs.harvard.edu/abs/1984ApJ...286....7B}
}

@ARTICLE{1999PhLA..253..273M,
       author = {{Milgrom}, Mordehai},
        title = "{The modified dynamics as a vacuum effect}",
      journal = {Physics Letters A},
         year = 1999,
        month = mar,
       volume = {253},
       number = {5-6},
        pages = {273-279},
          doi = {10.1016/S0375-9601(99)00077-8},
archivePrefix = {arXiv},
       eprint = {astro-ph/9805346},
 primaryClass = {astro-ph},
       adsurl = {https://ui.adsabs.harvard.edu/abs/1999PhLA..253..273M}
}

@ARTICLE{2017ApJ...836..152L,
       author = {{Lelli}, Federico and {McGaugh}, Stacy S. and {Schombert}, James M. and {Pawlowski}, Marcel S.},
        title = "{One Law to Rule Them All: The Radial Acceleration Relation of Galaxies}",
      journal = {\apj},
         year = 2017,
        month = feb,
       volume = {836},
       number = {2},
          eid = {152},
        pages = {152},
          doi = {10.3847/1538-4357/836/2/152},
archivePrefix = {arXiv},
       eprint = {1610.08981},
 primaryClass = {astro-ph.GA},
       adsurl = {https://ui.adsabs.harvard.edu/abs/2017ApJ...836..152L}
}

@ARTICLE{1991MNRAS.249..523B,
       author = {{Begeman}, K.~G. and {Broeils}, A.~H. and {Sanders}, R.~H.},
        title = "{Extended rotation curves of spiral galaxies : dark haloes and modified dynamics.}",
      journal = {\mnras},
         year = 1991,
        month = apr,
       volume = {249},
        pages = {523},
          doi = {10.1093/mnras/249.3.523},
       adsurl = {https://ui.adsabs.harvard.edu/abs/1991MNRAS.249..523B}
}

@ARTICLE{2024Univ...10..143O,
       author = {{Oehm}, Wolfgang and {Kroupa}, Pavel},
        title = "{The Relevance of Dynamical Friction for the MW/LMC/SMC Triple System}",
      journal = {Universe},
         year = 2024,
        month = mar,
       volume = {10},
       number = {3},
          eid = {143},
        pages = {143},
          doi = {10.3390/universe10030143},
archivePrefix = {arXiv},
       eprint = {2403.17999},
 primaryClass = {astro-ph.GA},
       adsurl = {https://ui.adsabs.harvard.edu/abs/2024Univ...10..143O}
}

@ARTICLE{planck_overview,
       author = {{Planck Collaboration} and {Aghanim}, N. and {Akrami}, Y. and {Arroja}, F. and {Ashdown}, M. and {Aumont}, J. and {Baccigalupi}, C. and {Ballardini}, M. and {Banday}, A.~J. and {Barreiro}, R.~B. and {Bartolo}, N. and {Basak}, S. and {Battye}, R. and {Benabed}, K. and {Bernard}, J. -P. and {Bersanelli}, M. and {Bielewicz}, P. and {Bock}, J.~J. and {Bond}, J.~R. and {Borrill}, J. and {Bouchet}, F.~R. and {Boulanger}, F. and {Bucher}, M. and {Burigana}, C. and {Butler}, R.~C. and {Calabrese}, E. and {Cardoso}, J. -F. and {Carron}, J. and {Casaponsa}, B. and {Challinor}, A. and {Chiang}, H.~C. and {Colombo}, L.~P.~L. and {Combet}, C. and {Contreras}, D. and {Crill}, B.~P. and {Cuttaia}, F. and {de Bernardis}, P. and {de Zotti}, G. and {Delabrouille}, J. and {Delouis}, J. -M. and {D{\'e}sert}, F. -X. and {Di Valentino}, E. and {Dickinson}, C. and {Diego}, J.~M. and {Donzelli}, S. and {Dor{\'e}}, O. and {Douspis}, M. and {Ducout}, A. and {Dupac}, X. and {Efstathiou}, G. and {Elsner}, F. and {En{\ss}lin}, T.~A. and {Eriksen}, H.~K. and {Falgarone}, E. and {Fantaye}, Y. and {Fergusson}, J. and {Fernandez-Cobos}, R. and {Finelli}, F. and {Forastieri}, F. and {Frailis}, M. and {Franceschi}, E. and {Frolov}, A. and {Galeotta}, S. and {Galli}, S. and {Ganga}, K. and {G{\'e}nova-Santos}, R.~T. and {Gerbino}, M. and {Ghosh}, T. and {Gonz{\'a}lez-Nuevo}, J. and {G{\'o}rski}, K.~M. and {Gratton}, S. and {Gruppuso}, A. and {Gudmundsson}, J.~E. and {Hamann}, J. and {Handley}, W. and {Hansen}, F.~K. and {Helou}, G. and {Herranz}, D. and {Hildebrandt}, S.~R. and {Hivon}, E. and {Huang}, Z. and {Jaffe}, A.~H. and {Jones}, W.~C. and {Karakci}, A. and {Keih{\"a}nen}, E. and {Keskitalo}, R. and {Kiiveri}, K. and {Kim}, J. and {Kisner}, T.~S. and {Knox}, L. and {Krachmalnicoff}, N. and {Kunz}, M. and {Kurki-Suonio}, H. and {Lagache}, G. and {Lamarre}, J. -M. and {Langer}, M. and {Lasenby}, A. and {Lattanzi}, M. and {Lawrence}, C.~R. and {Le Jeune}, M. and {Leahy}, J.~P. and {Lesgourgues}, J. and {Levrier}, F. and {Lewis}, A. and {Liguori}, M. and {Lilje}, P.~B. and {Lilley}, M. and {Lindholm}, V. and {L{\'o}pez-Caniego}, M. and {Lubin}, P.~M. and {Ma}, Y. -Z. and {Mac{\'\i}as-P{\'e}rez}, J.~F. and {Maggio}, G. and {Maino}, D. and {Mandolesi}, N. and {Mangilli}, A. and {Marcos-Caballero}, A. and {Maris}, M. and {Martin}, P.~G. and {Martinelli}, M. and {Mart{\'\i}nez-Gonz{\'a}lez}, E. and {Matarrese}, S. and {Mauri}, N. and {McEwen}, J.~D. and {Meerburg}, P.~D. and {Meinhold}, P.~R. and {Melchiorri}, A. and {Mennella}, A. and {Migliaccio}, M. and {Millea}, M. and {Mitra}, S. and {Miville-Desch{\^e}nes}, M. -A. and {Molinari}, D. and {Moneti}, A. and {Montier}, L. and {Morgante}, G. and {Moss}, A. and {Mottet}, S. and {M{\"u}nchmeyer}, M. and {Natoli}, P. and {N{\o}rgaard-Nielsen}, H.~U. and {Oxborrow}, C.~A. and {Pagano}, L. and {Paoletti}, D. and {Partridge}, B. and {Patanchon}, G. and {Pearson}, T.~J. and {Peel}, M. and {Peiris}, H.~V. and {Perrotta}, F. and {Pettorino}, V. and {Piacentini}, F. and {Polastri}, L. and {Polenta}, G. and {Puget}, J. -L. and {Rachen}, J.~P. and {Reinecke}, M. and {Remazeilles}, M. and {Renault}, C. and {Renzi}, A. and {Rocha}, G. and {Rosset}, C. and {Roudier}, G. and {Rubi{\~n}o-Mart{\'\i}n}, J.~A. and {Ruiz-Granados}, B. and {Salvati}, L. and {Sandri}, M. and {Savelainen}, M. and {Scott}, D. and {Shellard}, E.~P.~S. and {Shiraishi}, M. and {Sirignano}, C. and {Sirri}, G. and {Spencer}, L.~D. and {Sunyaev}, R. and {Suur-Uski}, A. -S. and {Tauber}, J.~A. and {Tavagnacco}, D. and {Tenti}, M. and {Terenzi}, L. and {Toffolatti}, L. and {Tomasi}, M. and {Trombetti}, T. and {Valiviita}, J. and {Van Tent}, B. and {Vibert}, L. and {Vielva}, P. and {Villa}, F. and {Vittorio}, N. and {Wandelt}, B.~D. and {Wehus}, I.~K. and {White}, M. and {White}, S.~D.~M. and {Zacchei}, A. and {Zonca}, A.},
        title = "{Planck 2018 results. I. Overview and the cosmological legacy of Planck}",
      journal = {\aap},
         year = 2020,
        month = sep,
       volume = {641},
          eid = {A1},
        pages = {A1},
          doi = {10.1051/0004-6361/201833880},
archivePrefix = {arXiv},
       eprint = {1807.06205},
 primaryClass = {astro-ph.CO},
       adsurl = {https://ui.adsabs.harvard.edu/abs/2020A&A...641A...1P}
}

@ARTICLE{2024ApJ...973...30B,
       author = {{Breuval}, Louise and {Riess}, Adam G. and {Casertano}, Stefano and {Yuan}, Wenlong and {Macri}, Lucas M. and {Romaniello}, Martino and {Murakami}, Yukei S. and {Scolnic}, Daniel and {Anand}, Gagandeep S. and {Soszy{\'n}ski}, Igor},
        title = "{Small Magellanic Cloud Cepheids Observed with the Hubble Space Telescope Provide a New Anchor for the SH0ES Distance Ladder}",
      journal = {\apj},
         year = 2024,
        month = sep,
       volume = {973},
       number = {1},
          eid = {30},
        pages = {30},
          doi = {10.3847/1538-4357/ad630e},
archivePrefix = {arXiv},
       eprint = {2404.08038},
 primaryClass = {astro-ph.CO},
       adsurl = {https://ui.adsabs.harvard.edu/abs/2024ApJ...973...30B}
}

@ARTICLE{2020A&A...633A..19B,
       author = {{B{\"o}hringer}, Hans and {Chon}, Gayoung and {Collins}, Chris A.},
        title = "{Observational evidence for a local underdensity in the Universe and its effect on the measurement of the Hubble constant}",
      journal = {\aap},
         year = 2020,
        month = jan,
       volume = {633},
          eid = {A19},
        pages = {A19},
          doi = {10.1051/0004-6361/201936400},
archivePrefix = {arXiv},
       eprint = {1907.12402},
 primaryClass = {astro-ph.CO},
       adsurl = {https://ui.adsabs.harvard.edu/abs/2020A&A...633A..19B}
}

@ARTICLE{2022MNRAS.511.5742W,
       author = {{Wong}, Jonathan H.~W. and {Shanks}, T. and {Metcalfe}, N. and {Whitbourn}, J.~R.},
        title = "{The local hole: a galaxy underdensity covering 90 per cent of sky to {\ensuremath{\approx}}200 Mpc}",
      journal = {\mnras},
         year = 2022,
        month = apr,
       volume = {511},
       number = {4},
        pages = {5742-5755},
          doi = {10.1093/mnras/stac396},
archivePrefix = {arXiv},
       eprint = {2107.08505},
 primaryClass = {astro-ph.CO},
       adsurl = {https://ui.adsabs.harvard.edu/abs/2022MNRAS.511.5742W}
}

@article{PhysRevD.104.L021303,
  title = {Rapid transition of ${G}_{\mathrm{eff}}$ at ${z}_{t}\ensuremath{\simeq}0.01$ as a possible solution of the Hubble and growth tensions},
  author = {Marra, Valerio and Perivolaropoulos, Leandros},
  journal = {Phys. Rev. D},
  volume = {104},
  issue = {2},
  pages = {L021303},
  numpages = {6},
  year = {2021},
  month = Jul,
  publisher = {American Physical Society},
  doi = {10.1103/PhysRevD.104.L021303},
  url = {https://link.aps.org/doi/10.1103/PhysRevD.104.L021303}
}

@ARTICLE{2021PhRvD.104l3511P,
       author = {{Perivolaropoulos}, Leandros and {Skara}, Foteini},
        title = "{Hubble tension or a transition of the Cepheid SnIa calibrator parameters?}",
      journal = {\prd},
         year = 2021,
        month = dec,
       volume = {104},
       number = {12},
          eid = {123511},
        pages = {123511},
          doi = {10.1103/PhysRevD.104.123511},
archivePrefix = {arXiv},
       eprint = {2109.04406},
 primaryClass = {astro-ph.CO},
       adsurl = {https://ui.adsabs.harvard.edu/abs/2021PhRvD.104l3511P}
}

@ARTICLE{2011MNRAS.415.2101H,
       author = {{Hahn}, Oliver and {Abel}, Tom},
        title = "{Multi-scale initial conditions for cosmological simulations}",
      journal = {\mnras},
         year = 2011,
        month = aug,
       volume = {415},
       number = {3},
        pages = {2101-2121},
          doi = {10.1111/j.1365-2966.2011.18820.x},
archivePrefix = {arXiv},
       eprint = {1103.6031},
 primaryClass = {astro-ph.CO},
       adsurl = {https://ui.adsabs.harvard.edu/abs/2011MNRAS.415.2101H}
}

@ARTICLE{2020A&A...641A...5P,
       author = {{Planck Collaboration} and {Aghanim}, N. and {Akrami}, Y. and {Ashdown}, M. and {Aumont}, J. and {Baccigalupi}, C. and {Ballardini}, M. and {Banday}, A.~J. and {Barreiro}, R.~B. and {Bartolo}, N. and {Basak}, S. and {Benabed}, K. and {Bernard}, J. -P. and {Bersanelli}, M. and {Bielewicz}, P. and {Bock}, J.~J. and {Bond}, J.~R. and {Borrill}, J. and {Bouchet}, F.~R. and {Boulanger}, F. and {Bucher}, M. and {Burigana}, C. and {Butler}, R.~C. and {Calabrese}, E. and {Cardoso}, J. -F. and {Carron}, J. and {Casaponsa}, B. and {Challinor}, A. and {Chiang}, H.~C. and {Colombo}, L.~P.~L. and {Combet}, C. and {Crill}, B.~P. and {Cuttaia}, F. and {de Bernardis}, P. and {de Rosa}, A. and {de Zotti}, G. and {Delabrouille}, J. and {Delouis}, J. -M. and {Di Valentino}, E. and {Diego}, J.~M. and {Dor{\'e}}, O. and {Douspis}, M. and {Ducout}, A. and {Dupac}, X. and {Dusini}, S. and {Efstathiou}, G. and {Elsner}, F. and {En{\ss}lin}, T.~A. and {Eriksen}, H.~K. and {Fantaye}, Y. and {Fernandez-Cobos}, R. and {Finelli}, F. and {Frailis}, M. and {Fraisse}, A.~A. and {Franceschi}, E. and {Frolov}, A. and {Galeotta}, S. and {Galli}, S. and {Ganga}, K. and {G{\'e}nova-Santos}, R.~T. and {Gerbino}, M. and {Ghosh}, T. and {Giraud-H{\'e}raud}, Y. and {Gonz{\'a}lez-Nuevo}, J. and {G{\'o}rski}, K.~M. and {Gratton}, S. and {Gruppuso}, A. and {Gudmundsson}, J.~E. and {Hamann}, J. and {Handley}, W. and {Hansen}, F.~K. and {Herranz}, D. and {Hivon}, E. and {Huang}, Z. and {Jaffe}, A.~H. and {Jones}, W.~C. and {Keih{\"a}nen}, E. and {Keskitalo}, R. and {Kiiveri}, K. and {Kim}, J. and {Kisner}, T.~S. and {Krachmalnicoff}, N. and {Kunz}, M. and {Kurki-Suonio}, H. and {Lagache}, G. and {Lamarre}, J. -M. and {Lasenby}, A. and {Lattanzi}, M. and {Lawrence}, C.~R. and {Le Jeune}, M. and {Levrier}, F. and {Lewis}, A. and {Liguori}, M. and {Lilje}, P.~B. and {Lilley}, M. and {Lindholm}, V. and {L{\'o}pez-Caniego}, M. and {Lubin}, P.~M. and {Ma}, Y. -Z. and {Mac{\'\i}as-P{\'e}rez}, J.~F. and {Maggio}, G. and {Maino}, D. and {Mandolesi}, N. and {Mangilli}, A. and {Marcos-Caballero}, A. and {Maris}, M. and {Martin}, P.~G. and {Mart{\'\i}nez-Gonz{\'a}lez}, E. and {Matarrese}, S. and {Mauri}, N. and {McEwen}, J.~D. and {Meinhold}, P.~R. and {Melchiorri}, A. and {Mennella}, A. and {Migliaccio}, M. and {Millea}, M. and {Miville-Desch{\^e}nes}, M. -A. and {Molinari}, D. and {Moneti}, A. and {Montier}, L. and {Morgante}, G. and {Moss}, A. and {Natoli}, P. and {N{\o}rgaard-Nielsen}, H.~U. and {Pagano}, L. and {Paoletti}, D. and {Partridge}, B. and {Patanchon}, G. and {Peiris}, H.~V. and {Perrotta}, F. and {Pettorino}, V. and {Piacentini}, F. and {Polenta}, G. and {Puget}, J. -L. and {Rachen}, J.~P. and {Reinecke}, M. and {Remazeilles}, M. and {Renzi}, A. and {Rocha}, G. and {Rosset}, C. and {Roudier}, G. and {Rubi{\~n}o-Mart{\'\i}n}, J.~A. and {Ruiz-Granados}, B. and {Salvati}, L. and {Sandri}, M. and {Savelainen}, M. and {Scott}, D. and {Shellard}, E.~P.~S. and {Sirignano}, C. and {Sirri}, G. and {Spencer}, L.~D. and {Sunyaev}, R. and {Suur-Uski}, A. -S. and {Tauber}, J.~A. and {Tavagnacco}, D. and {Tenti}, M. and {Toffolatti}, L. and {Tomasi}, M. and {Trombetti}, T. and {Valiviita}, J. and {Van Tent}, B. and {Vielva}, P. and {Villa}, F. and {Vittorio}, N. and {Wandelt}, B.~D. and {Wehus}, I.~K. and {Zacchei}, A. and {Zonca}, A.},
        title = "{Planck 2018 results. V. CMB power spectra and likelihoods}",
      journal = {\aap},
         year = 2020,
        month = sep,
       volume = {641},
          eid = {A5},
        pages = {A5},
          doi = {10.1051/0004-6361/201936386},
archivePrefix = {arXiv},
       eprint = {1907.12875},
 primaryClass = {astro-ph.CO},
       adsurl = {https://ui.adsabs.harvard.edu/abs/2020A&A...641A...5P}
}

@ARTICLE{2005MNRAS.363..603F,
       author = {{Famaey}, Benoit and {Binney}, James},
        title = "{Modified Newtonian dynamics in the Milky Way}",
      journal = {\mnras},
         year = 2005,
        month = oct,
       volume = {363},
       number = {2},
        pages = {603-608},
          doi = {10.1111/j.1365-2966.2005.09474.x},
archivePrefix = {arXiv},
       eprint = {astro-ph/0506723},
 primaryClass = {astro-ph},
       adsurl = {https://ui.adsabs.harvard.edu/abs/2005MNRAS.363..603F}
}

@INPROCEEDINGS{2023APS..APRG13003A,
       author = {{Abazajian}, Kevork},
        title = "{Astrophysical Constraints on Warm Dark Matter}",
    booktitle = {APS April Meeting Abstracts},
         year = 2023,
       series = {APS Meeting Abstracts},
       volume = {2023},
        month = jan,
          eid = {G13.003},
        pages = {G13.003},
       adsurl = {https://ui.adsabs.harvard.edu/abs/2023APS..APRG13003A}
}

@ARTICLE{2009ApJS..180..306D,
       author = {{Dunkley}, J. and {Komatsu}, E. and {Nolta}, M.~R. and {Spergel}, D.~N. and {Larson}, D. and {Hinshaw}, G. and {Page}, L. and {Bennett}, C.~L. and {Gold}, B. and {Jarosik}, N. and {Weiland}, J.~L. and {Halpern}, M. and {Hill}, R.~S. and {Kogut}, A. and {Limon}, M. and {Meyer}, S.~S. and {Tucker}, G.~S. and {Wollack}, E. and {Wright}, E.~L.},
        title = "{Five-Year Wilkinson Microwave Anisotropy Probe Observations: Likelihoods and Parameters from the WMAP Data}",
      journal = {\apjs},
         year = 2009,
        month = feb,
       volume = {180},
       number = {2},
        pages = {306-329},
          doi = {10.1088/0067-0049/180/2/306},
archivePrefix = {arXiv},
       eprint = {0803.0586},
 primaryClass = {astro-ph},
       adsurl = {https://ui.adsabs.harvard.edu/abs/2009ApJS..180..306D}
}

@ARTICLE{2021MNRAS.500.5249A,
       author = {{Asencio}, Elena and {Banik}, Indranil and {Kroupa}, Pavel},
        title = "{A massive blow for {\ensuremath{\Lambda}}CDM - the high redshift, mass, and collision velocity of the interacting galaxy cluster El Gordo contradicts concordance cosmology}",
      journal = {\mnras},
         year = 2021,
        month = jan,
       volume = {500},
       number = {4},
        pages = {5249-5267},
          doi = {10.1093/mnras/staa3441},
archivePrefix = {arXiv},
       eprint = {2012.03950},
 primaryClass = {astro-ph.CO},
       adsurl = {https://ui.adsabs.harvard.edu/abs/2021MNRAS.500.5249A}
}

@ARTICLE{2025MNRAS.541..784S,
       author = {{Samaras}, Nick and {Grandis}, Sebastian and {Kroupa}, Pavel},
        title = "{On the initial conditions of the {\ensuremath{\nu}}HDM cosmological model}",
      journal = {\mnras},
         year = 2025,
        month = aug,
       volume = {541},
       number = {2},
        pages = {784-797},
          doi = {10.1093/mnras/staf1041},
archivePrefix = {arXiv},
       eprint = {2506.19196},
 primaryClass = {astro-ph.CO},
       adsurl = {https://ui.adsabs.harvard.edu/abs/2025MNRAS.541..784S}
}

@ARTICLE{2022MNRAS.517.3613K,
       author = {{Kroupa}, Pavel and {Jerabkova}, Tereza and {Thies}, Ingo and {Pflamm-Altenburg}, Jan and {Famaey}, Benoit and {Boffin}, Henri M.~J. and {Dabringhausen}, J{\"o}rg and {Beccari}, Giacomo and {Prusti}, Timo and {Boily}, Christian and {Haghi}, Hosein and {Wu}, Xufen and {Haas}, Jaroslav and {Zonoozi}, Akram Hasani and {Thomas}, Guillaume and {{\v{S}}ubr}, Ladislav and {Aarseth}, Sverre J.},
        title = "{Asymmetrical tidal tails of open star clusters: stars crossing their cluster's pr{\'a}h$^{{\dagger}}$ challenge Newtonian gravitation}",
      journal = {\mnras},
         year = 2022,
        month = dec,
       volume = {517},
       number = {3},
        pages = {3613-3639},
          doi = {10.1093/mnras/stac2563},
archivePrefix = {arXiv},
       eprint = {2210.13472},
 primaryClass = {astro-ph.GA},
       adsurl = {https://ui.adsabs.harvard.edu/abs/2022MNRAS.517.3613K}
}

@ARTICLE{2024ApJ...970...94K,
       author = {{Kroupa}, Pavel and {Pflamm-Altenburg}, Jan and {Mazurenko}, Sergij and {Wu}, Wenjie and {Thies}, Ingo and {Jadhav}, Vikrant and {Jerabkova}, Tereza},
        title = "{Open Star Clusters and Their Asymmetrical Tidal Tails}",
      journal = {\apj},
         year = 2024,
        month = jul,
       volume = {970},
       number = {1},
          eid = {94},
        pages = {94},
          doi = {10.3847/1538-4357/ad4c66},
archivePrefix = {arXiv},
       eprint = {2405.09609},
 primaryClass = {astro-ph.GA},
       adsurl = {https://ui.adsabs.harvard.edu/abs/2024ApJ...970...94K}
}

@ARTICLE{2024Univ...10...48M,
       author = {{McGaugh}, Stacy S.},
        title = "{Discord in Concordance Cosmology and Anomalously Massive Early Galaxies}",
      journal = {Universe},
         year = 2024,
        month = jan,
       volume = {10},
       number = {1},
          eid = {48},
        pages = {48},
          doi = {10.3390/universe10010048},
archivePrefix = {arXiv},
       eprint = {2312.03127},
 primaryClass = {astro-ph.CO},
       adsurl = {https://ui.adsabs.harvard.edu/abs/2024Univ...10...48M}
}

@ARTICLE{1943ApJ....97..255C,
       author = {{Chandrasekhar}, S.},
        title = "{Dynamical Friction. I. General Considerations: the Coefficient of Dynamical Friction.}",
      journal = {\apj},
         year = 1943,
        month = mar,
       volume = {97},
        pages = {255},
          doi = {10.1086/144517},
       adsurl = {https://ui.adsabs.harvard.edu/abs/1943ApJ....97..255C}
}

@ARTICLE{2013MNRAS.435.2116P,
       author = {{Pawlowski}, Marcel S. and {Kroupa}, Pavel},
        title = "{The rotationally stabilized VPOS and predicted proper motions of the Milky Way satellite galaxies}",
      journal = {\mnras},
         year = 2013,
        month = nov,
       volume = {435},
       number = {3},
        pages = {2116-2131},
          doi = {10.1093/mnras/stt1429},
archivePrefix = {arXiv},
       eprint = {1309.1159},
 primaryClass = {astro-ph.CO},
       adsurl = {https://ui.adsabs.harvard.edu/abs/2013MNRAS.435.2116P}
}

@ARTICLE{2021A&A...649A.151M,
       author = {{Migkas}, K. and {Pacaud}, F. and {Schellenberger}, G. and {Erler}, J. and {Nguyen-Dang}, N.~T. and {Reiprich}, T.~H. and {Ramos-Ceja}, M.~E. and {Lovisari}, L.},
        title = "{Cosmological implications of the anisotropy of ten galaxy cluster scaling relations}",
      journal = {\aap},
         year = 2021,
        month = may,
       volume = {649},
          eid = {A151},
        pages = {A151},
          doi = {10.1051/0004-6361/202140296},
archivePrefix = {arXiv},
       eprint = {2103.13904},
 primaryClass = {astro-ph.CO},
       adsurl = {https://ui.adsabs.harvard.edu/abs/2021A&A...649A.151M}
}

@ARTICLE{2024MNRAS.527.4573B,
       author = {{Banik}, Indranil and {Pittordis}, Charalambos and {Sutherland}, Will and {Famaey}, Benoit and {Ibata}, Rodrigo and {Mieske}, Steffen and {Zhao}, Hongsheng},
        title = "{Strong constraints on the gravitational law from Gaia DR3 wide binaries}",
      journal = {\mnras},
         year = 2024,
        month = jan,
       volume = {527},
       number = {3},
        pages = {4573-4615},
          doi = {10.1093/mnras/stad3393},
archivePrefix = {arXiv},
       eprint = {2311.03436},
 primaryClass = {astro-ph.SR},
       adsurl = {https://ui.adsabs.harvard.edu/abs/2024MNRAS.527.4573B}
}

@BOOK{2020pama.book.....M,
       author = {{Merritt}, David},
        title = "{A Philosophical Approach to MOND: Assessing the Milgromian Research Program in Cosmology}",
         year = 2020,
       adsurl = {https://ui.adsabs.harvard.edu/abs/2020pama.book.....M}
}

@ARTICLE{2015CaJPh..93..169K,
       author = {{Kroupa}, Pavel},
        title = "{Galaxies as simple dynamical systems: observational data disfavor dark matter and stochastic star formation}",
      journal = {Canadian Journal of Physics},
         year = 2015,
        month = feb,
       volume = {93},
       number = {2},
        pages = {169-202},
          doi = {10.1139/cjp-2014-0179},
archivePrefix = {arXiv},
       eprint = {1406.4860},
 primaryClass = {astro-ph.GA},
       adsurl = {https://ui.adsabs.harvard.edu/abs/2015CaJPh..93..169K}
}

@ARTICLE{2023MNRAS.525.1401H,
       author = {{Hernandez}, X.},
        title = "{Internal kinematics of Gaia DR3 wide binaries: anomalous behaviour in the low acceleration regime}",
      journal = {\mnras},
         year = 2023,
        month = oct,
       volume = {525},
       number = {1},
        pages = {1401-1415},
          doi = {10.1093/mnras/stad2306},
archivePrefix = {arXiv},
       eprint = {2304.07322},
 primaryClass = {astro-ph.GA},
       adsurl = {https://ui.adsabs.harvard.edu/abs/2023MNRAS.525.1401H}
}

@ARTICLE{2024MNRAS.533..729H,
       author = {{Hernandez}, X. and {Chae}, Kyu-Hyun and {Aguayo-Ortiz}, A.},
        title = "{A critical review of recent Gaia wide binary gravity tests}",
      journal = {\mnras},
         year = 2024,
        month = sep,
       volume = {533},
       number = {1},
        pages = {729-742},
          doi = {10.1093/mnras/stae1823},
archivePrefix = {arXiv},
       eprint = {2312.03162},
 primaryClass = {astro-ph.GA},
       adsurl = {https://ui.adsabs.harvard.edu/abs/2024MNRAS.533..729H}
}
%%% Attachments - uncomment (delete %) these 4 lines if you have attachments
\clearpage
%\pagenumbering{arabic}
\addcontentsline{toc}{section}{Publications}
% Attachments go here (Uncomment 4 lines at the end of 0_main.tex)
% Adding a star distables the numbering
% PDF files can be added by \includepdf[pages=-]{yourfile.pdf}
\section*{Publications}\label{publications}

\begin{itemize}
  \item \textit{On the initial conditions of the $\nu$HDM cosmological model} - \\  \textbf{Samaras, Nick}, Grandis S., Kroupa, P. \\ https://doi.org/10.1093/mnras/staf1041
  \item \textit{Hydrodynamical structure formation in Milgromian cosmology} - \\ 
  Witenburg, N., Kroupa, P., Banik, I., Candlish, G., \textbf{Samaras, Nick}.\\ https://doi.org/10.1093/mnras/stad1371
  \item \textit{Challenges to a sharp change in G as a solution to the Hubble tension} - \\ 
  Banik, I., Desmond, H., \textbf{Samaras, Nick}. \\ https://doi.org/10.1093/mnras/staf567
  \item \textit{Constraints on the Hubble and matter density parameters with and without modelling the CMB anisotropies} - \\ Banik, I., \textbf{Samaras, Nick}. \\ https://arxiv.org/abs/2410.00804 (minor revision subm.)
  \item \textit{The many tensions with dark-matter based models and implications on the nature of the Universe} - \\ Kroupa, P. , Gjergo, E., Asencio, E., Haslbauer, M., Pflamm-Altenburg, J., Wittenburg, N., \textbf{Samaras, Nick}, Thies, I., Oehm, W.\\
  https://pos.sissa.it/436/231/
  \item \textit{The CosmoVerse White Paper: Addressing observational tensions in cosmology with systematics and fundamental physics} - \\ Di Valentino, E. Levi Said Jackson, Riess, A. ..., \textbf{Samaras, Nick}, ...\\
  https://arxiv.org/abs/2504.01669
\end{itemize}

Work in progress

\begin{itemize}
    \item Investigating $\nu$HDM cosmogony (in prep.)
    \item MOND workshop + White Paper in Leiden (Sep 2025)
\end{itemize}

\end{document}